\documentclass{article}

\usepackage{float}
\usepackage{lipsum}
\usepackage{stfloats}
\usepackage{tabularx}
\usepackage[colorinlistoftodos,textsize=tiny]{todonotes}
\usepackage[normalem]{ulem}

\usepackage{abstract}
\usepackage{amsfonts}
\usepackage[toc,page]{appendix}
\usepackage{bm}
\usepackage{bbm}
\usepackage{colortbl}
\usepackage{comment}
\usepackage{color}
\usepackage{clipboard}
\usepackage{diagbox}
\usepackage{enumitem}
\usepackage{etoc}
\usepackage{framed}
\usepackage{listings}
\usepackage{makecell}

\usepackage{minitoc}

\usepackage{multicol}
\usepackage{multirow}
\usepackage{newfloat}
\usepackage{soul}
\usepackage{wrapfig}
\usepackage{subcaption}
\usepackage{xcolor}
\usepackage{tikz}
\usetikzlibrary{shapes, arrows, positioning}
\tikzstyle{unit} = [circle, text centered, draw=black, minimum size=0.5cm]

\usepackage{booktabs} %
\usepackage{graphicx} 
\usepackage{microtype}

\usepackage{amsmath}
\usepackage{amssymb}
\usepackage{mathtools}
\usepackage{amsthm}

\usepackage{hyperref}

\usepackage[capitalize,noabbrev]{cleveref}

\newcommand{\vct}[1]{\vec{#1}}
\newcommand{\mtx}[1]{{\rm #1}}
\newcommand{\R}{\mathbb{R}}
\renewcommand{\vec}[1]{\mathbf{#1}}

\newcommand{\w}{\vec{w}}
\newcommand{\x}{\vec{x}}
\newcommand{\z}{\vec{z}}
\newcommand{\s}{\vec{s}}

\newcommand{\loss}{\mathcal{L}}

\newcommand{\softmax}{\mathrm{Softmax}}

\newcommand{\speclinear}{\mathrm{SpecLinear}}
\newcommand{\synattention}{\mathrm{Syn}}
\newcommand{\stabln}{\mathrm{StabLN}}

\newcommand{\layernormalization}[1]{\mathrm{LN}\left[#1\right]}
\newcommand{\selfattention}[1]{\mathrm{SelfAttention}\left[#1\right]}
\newcommand{\mlp}[1]{\mathrm{MLP}\left[#1\right]}
\newcommand{\completeattention}[1]{\mathrm{CompleteAttention}\left[#1\right]}
\newcommand{\vitencoderblock}[1]{\mathrm{ViTEncoderBlock}\left[#1\right]}
\newcommand{\bertencoderblock}[1]{\mathrm{BERTEncoderBlock}\left[#1\right]}

\newcommand{\relu}[1]{\texttt{ReLU}\left(#1\right)}
\newcommand{\gelu}[1]{\texttt{GELU}\left(#1\right)}

\newcommand{\wrapmin}[2]{\min\limits_{#2}\left\{#1\right\}}
\newcommand{\partdrv}[2]{\frac{\partial #1}{\partial #2}}
\newcommand{\drv}[1]{\frac{\mathrm{d}}{\mathrm{d} #1}}

\newcommand{\prob}[1]{\mathrm{Pr}\left[#1\right]}

\newcommand{\grad}[2]{\nabla_{#2}{#1}}

\theoremstyle{plain}
\newtheorem{theorem}{Theorem}[section]

\theoremstyle{definition}
\newtheorem{definition}[theorem]{Definition}

\theoremstyle{remark}

\newcommand*{\imgsc}[2]{%
    \raisebox{-.3\baselineskip}{%
        \includegraphics[
        height=#2\baselineskip,
        width=#2\baselineskip,
        keepaspectratio,
        ]{#1}%
    }%
}

\usepackage[accepted]{sources/icml2024}

\begin{document}

\twocolumn[
\icmltitle{Privacy Backdoors: Stealing Data with Corrupted Pretrained Models}

\begin{icmlauthorlist}
\icmlauthor{\parbox{10em}{\centering Shanglun Feng\\ \normalfont{ETH Zurich}}}{}
\icmlauthor{\parbox{10em}{\centering Florian Tramèr\\ \normalfont{ETH Zurich}}}{}
\end{icmlauthorlist}

\icmlcorrespondingauthor{Florian Tramèr}{florian.tramer@inf.ethz.ch}

\icmlkeywords{Machine Learning, ICML}

\vskip 0.3in
]

\begin{abstract}
Practitioners commonly download pretrained machine learning models from open repositories and finetune them to fit specific applications. We show that this practice introduces a new risk of \emph{privacy backdoors}. 
By tampering with a pretrained model's weights, an attacker can fully compromise the privacy of the finetuning data. 
We show how to build privacy backdoors for a variety of models, including transformers, which enable
an attacker to reconstruct individual finetuning samples, with a guaranteed success!
We further show that backdoored models allow for \emph{tight} privacy attacks on models trained with differential privacy (DP). The common optimistic practice of training DP models with loose privacy guarantees
is thus insecure if the model is not trusted.
Overall, our work highlights a crucial and overlooked supply chain attack on machine learning privacy.
\end{abstract}

\section{Introduction}
Sharing and finetuning of large pretrained models has become a common practice, enabling rapid prototyping and development of new applications across various domains.
Platforms like Huggingface currently host nearly 500,000 models, shared by various companies, researchers, and other users.
This trend brings to light new security concerns, particularly in the form of supply chain attacks. Prior work has recognized the impact of such attacks on model \emph{integrity}, where a \emph{backdoor} is planted into a model to hijack its downstream behavior~\cite{gu2017badnets, liu2018trojaning}.

In this work, we shift the focus from integrity to \emph{privacy} vulnerabilities, and introduce \textbf{privacy backdoors}, where a malicious model provider tampers with model weights to compromise the privacy of future finetuning data. Our backdoor attacks create ``data traps'' that directly write some data points to the model weights during finetuning. The trapped data can then be extracted by reading from the finetuned model’s weights.
Compared to prior related attacks in Federated Learning~\cite{boenisch2023curious, fowl2021robbing} that steal data points in a single training step, our data traps have to survive an entire finetuning run, with multiple training epochs and thousands of update steps.

We thus propose a new backdoor design that is single-use: once our backdoor activates and a data point is written to the model's weights, the backdoor becomes inactive, thereby preventing further alteration of those weights during training. Our backdoor thus acts a bit like a \emph{latch}, the logic circuit underlying digital memory: once the backdoor is set and the data is written to memory (i.e., to the model weights), it ``latches'' on until the end of training.

By design, our attacks capture individual training examples with high probability, with minimal impact on the pretrained model's utility. We apply these attacks to MLPs and pretrained transformers  (ViT~\citep{dosovitskiy2020image} and BERT~\citep{devlin2019bert}) and reconstruct dozens of finetuning examples across various downstream tasks.

We then consider a stronger black-box threat model, where the attacker only has query access to the finetuned model. By adapting techniques from the model extraction literature~\cite{tramer2016stealing, carlini2020cryptanalytic}, we show that even a black-box attacker can recover entire training inputs.
We further show that our backdoors enable simpler, \emph{perfect membership inference attacks}, which infer with 100\% accuracy whether a data point was used for training.
We use these backdoors to build the first \emph{tight} end-to-end attack on the seminal differentially private SGD algorithm of \citet{abadi2016deep}. That is, the privacy leakage observable by our adversary nearly matches the provable upper-bound from the algorithm's privacy analysis.
Our result thus challenges the common assumption that the privacy guarantees of DP-SGD are overly conservative in practice. This has led to the adoption of loose privacy budgets (e.g., $\varepsilon \geq 9$ in \cite{ramaswamy2020training}), which we show to be insecure in the presence 
of backdoored models.

Overall, our results bring to light a new attack vector in the modern machine learning supply chain, and emphasizes the need for more stringent privacy protections when operating with untrusted shared models.

Code to reproduce our experiments is at: {\footnotesize
\url{https://github.com/ShanglunFengatETHZ/PrivacyBackdoor}}.

\section{Related Work}

\textbf{Privacy attacks.~~}
Neural networks memorize training data. Attackers can exploit this to launch membership inference attacks~\citep{shokri2017membership} or data extraction attacks \citep{carlini2021extracting, carlini2023extracting}.
These attacks can be strengthened by an attacker who \emph{poisons} the model to
increase memorization~\cite{tramer2022truth, panda2023teach}.
\citet{song2017machine} show that an attacker who tampers with the training \emph{code}
can cause the model to exfiltrate training data.

\textbf{Backdoor attacks.~~}
\citet{gu2017badnets, liu2018trojaning} show how to tamper with a pretrained model to inject malicious targeted behaviors that compromise the model's \emph{integrity}.
\citet{hong2022handcrafted} show that such backdoors can be \emph{handcrafted} by directly editing the weights of a pretrained model.

\textbf{Data stealing in Federated Learning.~~}
A malicious server in Federated Learning 
can tamper with the model so that clients' gradients leak training data~\citep{zhao2023loki, fowl2021robbing, fowl2022decepticons, boenisch2023curious, wen2022fishing, phong2018privacy}. 
We extend the attack of \citet{fowl2021robbing}, which creates
sparse activations in a linear layer, so 
that the gradient encodes training inputs.

\textbf{Privacy backdoors in pretrained models.~~}
Concurrent work by \citet{wen2024privacy} and \citet{liu2024precurious} also introduces privacy backdoors for pretrained models, but with a different focus than ours. They consider weaker attacker goals (respectively membership inference or extraction without success guarantees), but also a weaker attacker (with only black-box access to the finetuned model). Their attacks are thus incomparable to ours: they are easier to mount, but also less devastating than ours.

\textbf{Tightness of DP-SGD.~~}
The DP-SGD algorithm \citep{abadi2016deep}
trains models with differential privacy~\citep{dwork2006calibrating}.
The algorithm's worst-case analysis assumes that one model weight is updated if and only if a target input is present.
This analysis is tight for a strong attacker who can observe all training steps~\citep{nasr2021adversary}, but it is presumed to be loose for a more realistic ``end-to-end'' attacker who only observes the final model. Privacy parameters are thus often set very optimistically in practice. %

\section{Threat Model}

In contrast to prior attacks on Federated Learning, we assume a weaker attacker who cannot observe individual gradient updates.
Instead, our attacker tampers with a pretrained model \emph{once}, before sending it to the victim.

We assume the victim finetunes the backdoored model on a classification task using  SGD for multiple epochs. The victim adds a new 
linear layer to the backdoored model and then finetunes the entire model (``full finetuning'').
We leave an extension to other types of finetuning (e.g., LoRA~\citep{hu2021lora}) for future work.

We consider attackers with either \textbf{white-box} or \textbf{black-box} access to the finetuned model.
A white-box attacker can inspect the finetuned model's weights, while the black-box attacker can query the model on arbitrary inputs.
White-box access is a stronger assumption, which enables our most powerful data extraction attacks. We show that the weaker black-box setting allows for perfect membership inference attacks, as well as data extraction attacks for simple models.

\section{\makebox[5em][s]{Warmup: White-box Data Stealing in MLPs}}
\label{sec:warmup}
\subsection{Attack Description}
In order to illustrate our approach, we start with the simplest case of backdooring a linear unit (i.e., one element of a linear layer), as in \cite{fowl2021robbing}:
\begin{equation}
    h = \relu{\w^\top \x + b}, 
    \label{structure:backdoorunit}
\end{equation}
where $\x \in \R^m$ is the unit's input, and
$\w \in \R^{m}, b \in \R$ are the unit's weight and bias.

As described in \citet{fowl2021robbing}, we can set the parameters $\w, b$ to ensure that, with high probability, a single input $\x$ in a training batch activates the neuron (i.e., $h>0$).
If we backpropagate the training loss $\loss$, we then get:
\begin{equation}
\label{eq:fowl}
\grad{\loss}{\w} = \partdrv{\loss}{h} \cdot \x, \quad 
\grad{\loss}{b} = \partdrv{\loss}{h} \,.
\end{equation}
Dividing $\grad{\loss}{\w}$ by the scalar $\grad{\loss}{b}$ recovers the input $\x$.

In a Federated Learning setting, we are done since the attacker gets to see these gradient updates directly. But in our case, the gradient updates are simply added to the parameters and training continues. The attacker only sees the final model.
This introduces a new challenge: we need the captured inputs to ``survive'' the
entire training run, without being mixed with other inputs captured in subsequent training steps.
We thus propose to build a backdoored unit with a stronger ``single-use'' property: \emph{Once the backdoor has been activated, it will never be active again afterwards}. 

We call such a backdoor a \textbf{data trap}.
To design a data trap, we need the gradient update to ``shut down'' the backdoor.
Assume the backdoor activates on input $\hat{\x}$, and we update the weights using SGD with learning rate $\eta$ (with the gradients in \eqref{eq:fowl}).
The updated backdoor's output on a new input $\x$ is:
\begin{equation}
\label{eq:updatedbackdoor}
    h' = \w'^\top \x + b' = (\w^\top \x + b) - \eta \cdot \partdrv{\loss}{h} \cdot \left(\hat{\x}^\top \x +1 \right) \;.
\end{equation}
A sufficient condition for the backdoor to shut down (i.e., $h' \leq 0$ for all $\x$) is that:
\begin{enumerate}[itemsep=6pt,parsep=-1pt,topsep=-1pt]
    \item $\hat{\x}^\top \x + 1 > 0$.
    \item $\partdrv{\loss}{h}$ is positive and large enough.
\end{enumerate}

The first condition, $\hat{\x}^\top \x + 1 > 0$, depends only on the input of the backdoor unit, and the attacker can ensure the inputs satisfy this (e.g., by mapping inputs to $[0,1]^m$).

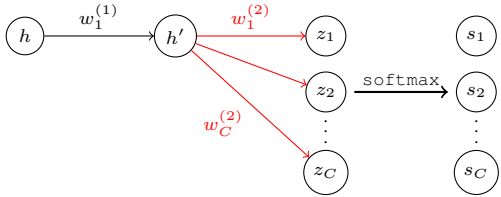
\begin{figure}[t]
\scriptsize
\begin{center}
\begin{tikzpicture}[node distance = 2cm]
    \node [unit] (a) {$h$};
    \node [unit, right of=a] (ap) {$h'$};

    \node [unit, right of=ap] (z1) {$z_1$};
    \node [unit, below=0.2cm of z1] (z2) {$z_2$};
    
    \node[draw=none] [below=-0.2cm of z2] (zdots) {$\vdots$};
    \node [unit, below=0.0cm of zdots] (zc) {$z_C$};

    \node [unit, right of=z1] (s1) {$s_1$};
    \node [unit, right of=z2] (s2) {$s_2$};
    \node [draw=none] [below=-0.2cm of s2] (sdots) {$\vdots$};
    \node [unit, right of=zc] (sc) {$s_C$};

    \draw [->] (a) -- node[anchor=south] {$w^{(1)}_1$} (ap);
    \draw [->, red] (ap) -- node[anchor=south] {$w^{(2)}_1$} (z1);
    \draw [->, red] (ap) -- node[anchor=south] {} (z2);
    \draw [->, red] (ap) -- node[anchor=north east] {$w^{(2)}_C$} (zc);

    \begin{scope}[shorten >=3pt, shorten <=3pt]
    \draw [->, thick] (z2) -- node[anchor=south] {\scriptsize\texttt{softmax}} (s2);
    \end{scope}
\end{tikzpicture}
\vspace{-1em}
\caption{Illustration of the output of a data trap ($h = \relu{\w^\top\x + b}$) connected to the model's output. The weights of the classification head (in red) are typically not under the control of the attacker, and randomly initialized before finetuning.}
\label{fig:backdoor}
\end{center}
\vskip -0.15in
\end{figure}

The second condition, that $\partdrv{\loss}{h}$ is positive and large, requires more work.
Our construction is illustrated in Figure~\ref{fig:backdoor}.
We connect the backdoor's output $h$ to a hidden unit $h'$, with weight $w^{(1)}_1$, i.e., 
$h' = \relu{w^{(1)}_1 h}$. %
The unit $h'$ then connects to the model's logit layer $\z$ with weights $\w^{(2)}$, and the logits are passed through a softmax before computing the cross-entropy loss. %
A standard gradient calculation gives:
\begin{equation}
\label{eq:deriv}
\partdrv{\loss}{h} = \partdrv{\loss}{h'}\cdot \partdrv{h'}{h} = \Big(\sum_{i=1}^C (s_i - y_i) \cdot w^{(2)}_i \Big) \cdot w^{(1)}_1 \;,
\end{equation}
where $C$ is the number of classes, $s_i$ is the softmax score for the $i$-th class, and $y_i$ is one if the input belongs to the $i$-th class, and zero otherwise.

Recall that we want this derivative to be positive and large.
The ``large'' part is easy: we just set $w^{(1)}_1$ to be large. The sign of the derivative depends on the ground-truth class of the captured input (i.e, $s_i - y_i$ is negative for the index of the true class). It turns out that setting $w^{(1)}_1$ to be large guarantees that the derivative is positive with probability $1-1/C$ (e.g., $99.9\%$ for ImageNet classification).
Indeed, if $w^{(1)}_1$ is large enough the logits $\z$ are essentially random (the benign part of the model has negligible influence).
The softmax scores then concentrate on the class $j$ with the highest weight $w^{(2)}_j$, and the sum in \eqref{eq:deriv} is positive if $j$ is a wrong class. So we get a large positive gradient as desired!

In conclusion, if our backdoor fires on a misclassified input, it shuts down and is inactive for the rest of training.
If we are unlucky and we get a negative gradient into the backdoor, i.e., $\partdrv{\loss}{h} < 0$, the backdoor does not shut and activates again on future inputs.
This is the main challenge we have to tackle when scaling our backdoor construction to large transformers, which we discuss in Section~\ref{sec:whitebox}.

\subsection{Multiple Backdoor Units}
So far, we described a single backdoor unit that captures one training input.
We can simply replicate this construction above for multiple linear units, and initialize the backdoor weights so that they activate on different inputs.

The finetuning data induces a distribution over the values $\w^\top \x$, and we set the bias $b$ so that a small fraction of inputs give positive activations. To capture diverse inputs in different backdoors, we select weights $\w$ that align with different subsets of the training data (in practice, we simply sample the weights from a uniform distribution over the sphere). 

The biases can be set independently for each backdoor. If the bias is too large, multiple inputs in a batch might trigger the same backdoor which makes reconstruction difficult. If the bias it too small, some backdoors may never fire on any training input. 
For a dataset of size $D$ and batch size of $B$, we thus want the fraction of training inputs $p$ that activate a backdoor to satisfy $p > 1/D$ and $p \ll 1/B$. For typical finetuning setups (where $B \ll D$), the range of allowable biases is large, so the attacker only needs to know a loose approximation of the distribution $\w^\top \x$.

Note that each backdoor can fail independently with probability $1-1/C$, if the captured input is correctly classified. With $k$ perfectly calibrated backdoors, we thus expect to capture $k \cdot (1-1/C)$ inputs.

\subsection{Experiments}
We evaluate our backdoor construction on a MLP model trained on CIFAR-10. We implement a 3-layer backdoored perceptron as shown in Figure~\ref{fig:backdoor}. The model's layers have 256 hidden units, 64 of which are used for backdoors. Input images are scaled and flattened to the range $[0,1]^{3072}$.

\begin{figure}
    \begin{center}
    \subfloat{\includegraphics[width=1.0\linewidth]{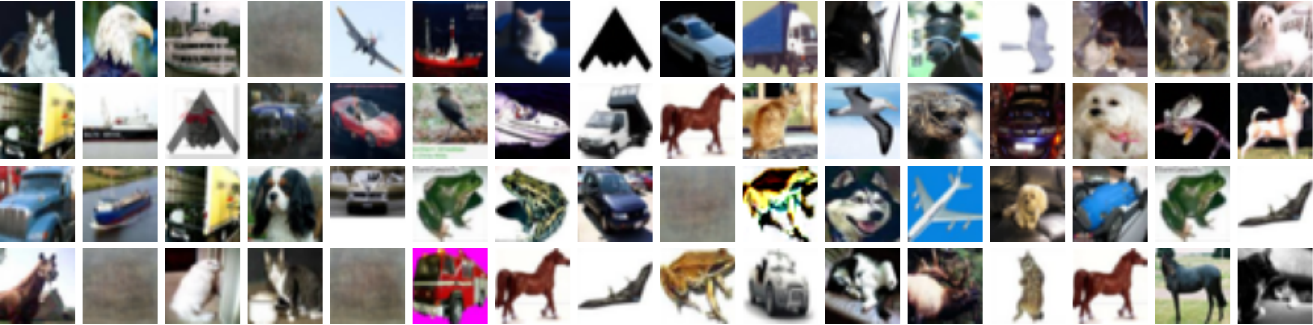}}
    
    \vspace{0.3cm}
    
    \subfloat{\includegraphics[width=1.0\linewidth]{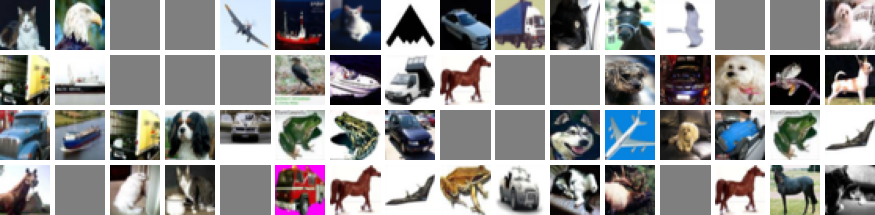}}

    \vspace{0.3cm}
    
    \subfloat{
    \begin{minipage}[b]{0.23\textwidth}
    \centering
     \includegraphics[width=0.75\linewidth]{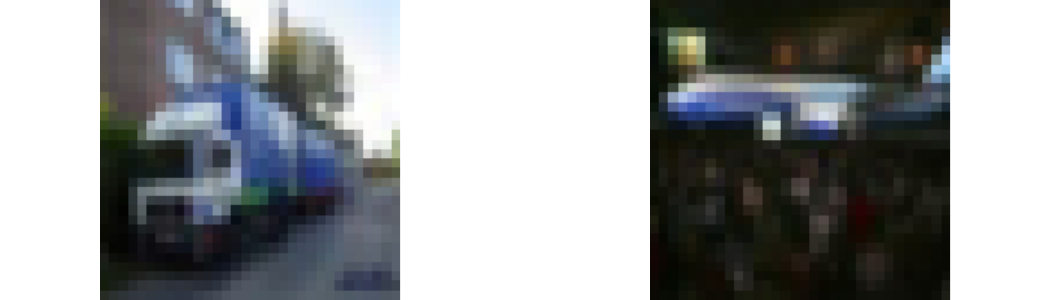}       
    \end{minipage}
    } 
    \subfloat{
    \begin{minipage}[b]{0.23\textwidth}
    \centering
    \includegraphics[width=0.75\linewidth]{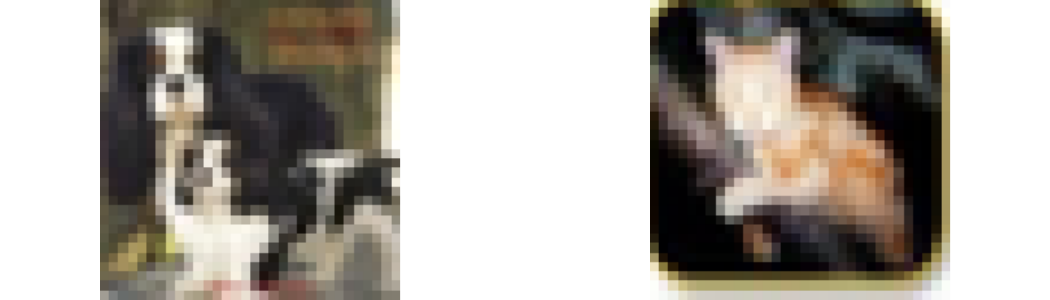}
    \end{minipage}}
    \end{center}
     \vskip -0.1in
    \caption{Reconstructed images from a backdoored MLP model. \textbf{Top}: Reconstruction; \textbf{Middle}: Ground truth, or a gray image if the backdoor failed to capture a unique input (see Appendix~\ref{appendix::groundtruthcatcher} for details); \textbf{Bottom}: Ground truths for failed backdoors that captured a mix of two inputs
    (left corresponds to row 2, column 4; right corresponds to row 1, column 15).}
    \label{reconstruction::mlp:randhead}
    \vspace{-1em}
\end{figure}

Figure \ref{reconstruction::mlp:randhead} shows reconstructed images from this toy MLP model, most of which are indistinguishable from their corresponding ground truth images. 
The attack can fail in two orthogonal cases.
If the backdoor gets a negative gradient signal (because the captured input is correctly classified), the backdoor fails to shut and is re-activated many times in a subsequent batch. This typically results in an unrecoverable backdoor (e.g., top row, fourth from the left).
A more benign failure case is when multiple inputs in a batch activate the backdoor, in which case our attack recovers a mixture of the captured inputs (e.g., top row, second from the right).

\section{Privacy Backdoors in Transformers}
\label{sec:whitebox}

In the previous section, we designed data traps for simple MLPs. 
To illustrate the general applicability of 
privacy backdoors, we now extend our study to 
real transformer models for both vision and text.
The transformer architecture underlies most foundation models in use today,
and is thus a natural candidate for a backdoored pretrained model.

We assume victims finetune the entire backdoored model for classification (i.e., \emph{full finetuning}). We leave the study of parameter-efficient methods like LoRA (which are popular in computationally constrained settings) to future work.

\paragraph{Transformers.}
A transformer~\cite{vaswani2017attention} takes as input a sequence of \emph{tokens}, each mapped to an \emph{embedding vector}.
For vision transformers~\cite{dosovitskiy2020image}, the tokens are patches from the input image that are mapped to embeddings using a convolutional layer.
For text transformers, tokens are (sub-)words in the input sentence that are mapped to embedding vectors using a dictionary.
Each token embedding is augmented with a \emph{position embedding}, which encodes the \emph{index} of the token in the input sequence.%

We focus on ViT~\cite{dosovitskiy2020image} and BERT~\cite{devlin2019bert}, whose embeddings have 768 dimensions.
The transformer's input is thus a $k \times 768$ matrix, where $k$ is the number of tokens in an input (padded to a fixed length).

The transformer encoder then consists of a number of \emph{blocks}, which apply a dimension-preserving function $f: \mathbb{R}^{k \times 768} \to \mathbb{R}^{k \times 768}$.
These blocks combine the following operations:

\begin{itemize}[itemsep=3pt,parsep=-1pt,topsep=-1pt, leftmargin=12pt]
    \item MLPs: One or two linear layers with a non-linear activation (typically a GELU~\citep{hendrycks2016gaussian}) are applied to each of the $k$ feature vectors.
    \item Attention layers: A layer that computes dependencies between the $k$ token vectors, and reweighs them.
    \item Shortcut connections: A block's input is typically added to its output, to facilitate learning~\cite{he2016deep}.
    \item Layer normalizations: Each feature vector is normalized to zero mean and unit variance, followed by a learned affine transform. %
\end{itemize}

The model outputs features of dimension $k \times 768$. To finetune a classifier, we add a linear layer to the features of the first token (a special [CLS] token prepended to the input).

\subsection{Challenges}
\paragraph{Maintaining model utility.}
If the attacker wants
the backdoored model to be used, it should be useful for finetuning.
So the attacker needs to ensure the backdoored model retains a ``benign'' part that acts as a useful feature extractor.

For simplicity, we do not retrain existing transformers to accommodate backdoors.
We start from benign pretrained transformers (ViT~\cite{dosovitskiy2020image} and BERT~\cite{devlin2019bert}),
and remove a fraction of the model's feature representations and encoder blocks
for our backdoor construction (see Appendix \ref{appendix::largemodelprocessing} for details about model preprocessing). 
Our design then aims to minimize interference between the benign and backdoor portions of the model.%

\paragraph{Moving beyond linear layers.}
Transformers are more complex than MLPs! In addition to linear layers (which are useful for building data traps), they have 
attention layers, layer normalizations, and more complex activations than ReLUs.
We will need additional tricks to ensure that all
these layers propagate the backdoor activations and gradients properly without destroying the model's utility.

\paragraph{Capturing all tokens in an input.}
If we apply the construction from Section~\ref{sec:warmup} to capture multiple input vectors of a transformer, we would capture individual \emph{tokens}. 
These tokens would likely belong to different input sequences (i.e., we might capture just one word from many different sentences).
Instead, we would like to capture a few \emph{full} inputs.
This will require designing a family of backdoors that fire in tandem for all tokens in an input sequence, or not at all.

\paragraph{Building robust backdoors.}
Our backdoors for MLPs in Section~\ref{sec:warmup} only required setting adversarial weights in two layers. These weights are only updated when the backdoor first activates, and are unused after that.
In contrast, our backdoors for transformers are much more complex and combine a number of modules (as described in Section~\ref{sec:transformers_overview}).

One challenge is that the initial weights in these modules shift during training, as gradients will flow into them. This will disrupt the adversarial behavior that these modules were designed to implement and risks destroying the backdoor. %
Another challenge is maintaining the model's utility in the presence of ``failing'' backdoors, where a correctly classified input activates the backdoor and causes negative gradients to further open the backdoor. The backdoor then fires many times in subsequent steps, which leads to large gradients.
In a MLP, this is inconsequential since a backdoor's gradients cannot impact other benign weights in the model. But with transformers, we cannot avoid small interferences between backdoor features and benign features.
Ensuring the robustness of our backdoor construction, and of the model's utility thus requires more careful tuning (see Appendix~\ref{appendix::robustness}).

\subsection{Overview of our Approach}
\label{sec:transformers_overview}

\newcommand{\feat}{\mathbf{ft}}
\newcommand{\key}{\mathbf{key}}
\newcommand{\act}{\mathbf{act}}

In this section we give an overview of our techniques for inserting data traps into transformers.
For ease of exposition, we leave many technical details to Appendix~\ref{appendix::manipulation}. We also release code%
to reproduce all our experiments.

\begin{figure*}[t]
\begin{center}
\includegraphics[width=0.9\textwidth]{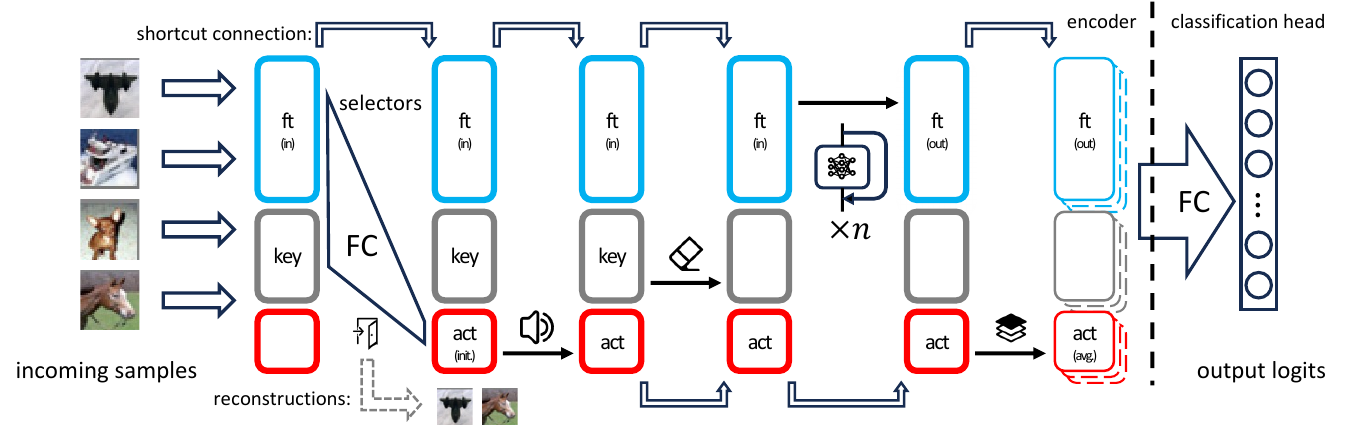}
\caption{The logical architecture of a backdoored transformer. \textbf{ft}: \textit{benign features} inheriting the utility of the pretrained weights. \textbf{key}: the features for selectively activating a backdoor, and the data to be captured. \textbf{act}: \textit{activation signals}, i.e., outputs from backdoor units.
From left to right, our backdoor construction consists of: an \emph{input module} that creates the logical feature separation $\x = [\feat, \key, \vct{0}]$; the main \emph{backdoor module} (\imgsc{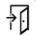}{0.95}) that captures inputs in a linear layer; an \emph{amplifier module} (\imgsc{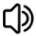}{0.95}) that increases backdoor activation signals; an \emph{erasure module} (\imgsc{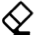}{0.95}) that wipes out unneeded features; $n$ \emph{signal propagation modules} (\imgsc{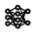}{0.95}) that compute standard benign features $\feat$ while carrying the backdoor outputs $\act$; and an \emph{output module} (\imgsc{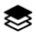}{0.95}) that averages all features onto the class token for finetuning.}
\label{pic:logicalarchitecture}
\end{center}
\vskip -0.15in
\end{figure*}

The architecture of our attack is illustrated in Figure~\ref{pic:logicalarchitecture}.
We maintain an invariant that the transformer's
internal features (a vector of size $d=768$) are split into three components
\[
    [\feat, \key, \act]\;.
\]
The first part, $\feat$, is used for the transformer's benign features for downstream applications.
The second part, $\key$, stores information to be captured by the backdoor.
The third part, $\act$, propagates the output activations of the backdoors all the way to the model's last layer, and amplifies them to ensure
that gradient signals will shut down the backdoor.
Details about how we divide feature vectors into these three components are in Table~\ref{tab:resourcepartition} in Appendix~\ref{appendix::manipulation}.

The backdoor construction combines different \emph{modules}, described below. We further employ a number of numerical tricks to deal with the complexity of the model and training process, which we discuss at the end of this section.

\paragraph{Input module.}

\newcommand{\tok}{\mathbf{tok}}
\newcommand{\pos}{\mathbf{pos}}
\newcommand{\seq}{\mathbf{seq}}

This module maps the input tokens to embeddings of the form $[\feat, \key, \vec{0}]$. The benign features simply contain a truncated part of the original  embeddings.
For the ViT, the backdoor input to be captured is a downscaled and grayscaled version of the original $32\times32$ patch. %
For BERT, we capture a sub-vector of the token embedding, which is sufficient to recover the underlying token. 

As described above, one of the key challenges with transformers is to capture entire inputs (i.e., a sequence of tokens), rather than simply capturing individual tokens from different inputs.
We achieve this by enhancing our backdoors with \emph{sequence and position selectors}.

We further split the backdoor input $\key$ into three parts $\key = [\tok, \pos, \seq]$. The first part, $\tok$, is the token information to be captured. The second part, $\pos$, encodes the token's position in the input. %
The third part, $\seq$, contains features that identify the sequence of input tokens (i.e., a sentence or an image). We obtain these features by making the model's first attention layer compute an average across all input tokens (\citet{fowl2022decepticons} use a similar idea in their Federated Learning attack). As we describe below, by setting the backdoor weights appropriately, we can ensure that with high probability all captured inputs correspond to tokens at different positions in the same input sequence.

\paragraph{Backdoor module.} In the first encoder block, we use the MLP to implement multiple backdoors that capture the input features $\key$, with the same design as in Section~\ref{sec:warmup}.
That is, we map the full input vector $[\feat, \key, \vec{0}]$ to the output $[\vec{0}, \vec{0}, \{(\vec{w}^{(i)})^\top\cdot \key + b^{(i)}\}_{i=1}^k]$, where $k$ is the number of backdoors. The weights and biases are chosen so that the backdoor output is positive for a small fraction of inputs.
After the block's shortcut connection, we get the block's final output of the form $[\feat, \key, \{(\vec{w}^{(i)})^\top\cdot \key + b^{(i)}\}_{i=1}^k]$.

To capture all tokens in an input, we design ``keyed'' backdoors that activate only for tokens at a specific position in one input sequence.
In more detail, the backdoor weights are of the form $\vec{w}^{(i)} = [\vec{0}, \pos^{(i)}, \vec{w}\textsubscript{seq}]$. The position features $\pos^{(i)}$ are designed to be close to orthogonal, i.e., $(\pos^{(i)})^\top \cdot \pos^{(j)}$ is small if $i\neq j$. The bias $b^{(i)}$ is set to a high quantile of the expected distribution
of values $\|\pos^{(i)}\|^2 + \vec{w}\textsubscript{seq}^\top\cdot \seq$ over the  finetuning set.
This ensures that the backdoor will only activate on a token in the $i$-th position of a sequence with a key $\seq$ similar to $\vec{w}\textsubscript{seq}$. %

\paragraph{Amplifier module.} Recall from Section~\ref{sec:warmup} that we need the activations from the backdoor to become large so that they contribute a large enough (positive) gradient signal to shut down the backdoor.
As in the case of a simple MLP, we do this by multiplying the backdoor output, $\act$, by a large constant in the first MLP module that follows the backdoor.

\paragraph{Erasure module.} After the backdoor, we do not need the $\key$ features anymore. To make our lives easier when dealing with the rest of the model, it is easiest to remove these features. But because of the layer normalization, we have to be careful how we do this.
For BERT, the layer normalization is at the end of the encoder block (after the linear layers) and so we can just use this layer to set all outputs corresponding to the $\key$ features to zero.
In a ViT, the layer normalization $(\mathrm{LN})$ comes before the linear layers so things are more complicated. Here, we use a combination of two MLP layers to erase the unused features (see Appendix~\ref{appendix::detailedmanipulation::vit} for details).

\paragraph{Signal propagation modules.} We now just have to ensure that the amplified backdoor outputs propagate all the way to the end of the network, without interacting (too much) with the benign features $\feat$. For the ViT (where each module's layer-normalization occurs before the shortcut connection), we can simply set all weights of the backdoor components $\act$ and $\key$ to zero, so that these activations are propagated solely via the shortcut connection. For BERT, the layer-normalization occurs at the end of each module, and so we need some extra care (see Appendix \ref{appendix::detailedmanipulation:bert} for details).
For both ViT and BERT, we maintain the pretrained transformer's weights for the benign features $\feat$, so that the model extracts meaningful features for downstream tasks.

\paragraph{Output module.}At the last layer of the transformer, we  aggregate the activation outputs $\act$ from all tokens onto the special CLS token that is used for classification during finetuning.
We do this by rewiring the attention module in the last layer, to simply average all $\act$ features.

The features $[\feat, \vct{0}, \act]$ from the CLS token are passed through a final linear layer (randomly initialized prior to finetuning) that maps to $C$ classes. Following the same argument as in Section~\ref{sec:warmup}, an input that activates a backdoor will be misclassified with probability $1-1/C$ (since the large activations from the backdoor will boost an arbitrary class), and this will result in a large positive gradient that flows all the way back to the backdoor module and shuts it.

\subsection{Numerical Tricks}

Our backdoor construction above ignores some technical challenges that arise due to architectural specificities of transformer models, in particular the use of GELU activations and layer normalizations. These require some additional numerical tricks to ensure that backdoor signals do not vanish or blow-up during training.

We note that these issues could be sidestepped by making small \emph{architectural} changes to the pretrained model (specifically, using ReLU activations and applying layer normalizations only to the benign features $\feat$). But this might make the backdoor attack too obvious. %

\paragraph{Stabilizing layer normalization.}

Layer normalization is an important architectural component in modern transformers,
but it serves no functional purpose for our backdoor construction (in fact, its presence complicates things drastically). %
A layer normalization applies an almost linear transformation: $\mathrm{LN}(\vct{x}) \approx W\vct{x} + \vct{b}$, except that the magnitude of the weight matrix $W$ scales inversely with that of the input $\vct{x}$.
These layers can be problematic for two reasons: they apply an input-dependent rescaling, and they introduce strong dependencies between different components of the features vector. The data-dependent rescaling of inputs introduces a number of challenges: during the forward pass, rescaling of the backdoor inputs $\key$ can mess up subsequent modules (e.g., the erasure module of the ViT, see Appendix \ref{appendix::manipulation:vit:erasure}). In the backward pass, the rescaling can lead to very small gradients for large backdoor outputs.

We deal with both issues simultaneously by ``virtually'' splitting the normalization function. Specifically, given a feature vector with two components $[\x_L, \x_R]$, we add a large constant $C$ to one component, $[\x_L, \x_R + C]$, and then rescale the layer's output accordingly. In Appendix~\ref{appendix::tricks:stabilizelayernormalization}, we show that for a large enough $C$, the dependencies between the two feature components are removed, and the layer no longer applies an input-dependent scaling.

\paragraph{Dealing with GELUs.}

In contrast to ReLUs, GELU activations~\citep{hendrycks2016gaussian} can not be fully ``shut down'', as their output is non-zero for all inputs. But for large enough inputs (in absolute value), a GELU is near-identical to a ReLU.
We could thus simply re-scale all GELU inputs by a large constant.
This requires some care though, because we will always get some ``unintended'' gradients flowing through the backdoor (i.e., non-zero gradients when the activation output is negative).
To deal with this issue, we use the MLP that follows each GELU to dampen small activation outputs, thereby reducing the impact of the unintended gradients (see Appendix~\ref{appendix::tricks:gelu} for details).

\begin{figure}[t]
\begin{center}
\subfloat{\includegraphics[width=0.8\columnwidth]{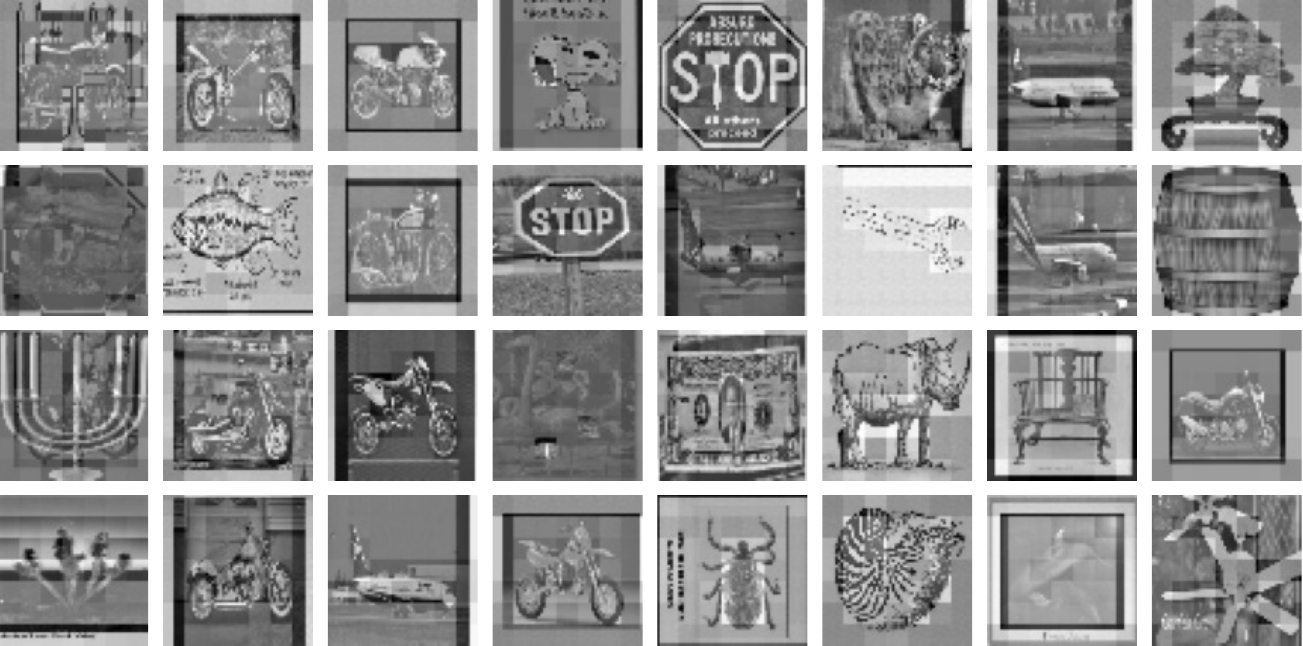}} 
\vspace{0.1cm}
\subfloat{\includegraphics[width=0.8\columnwidth]{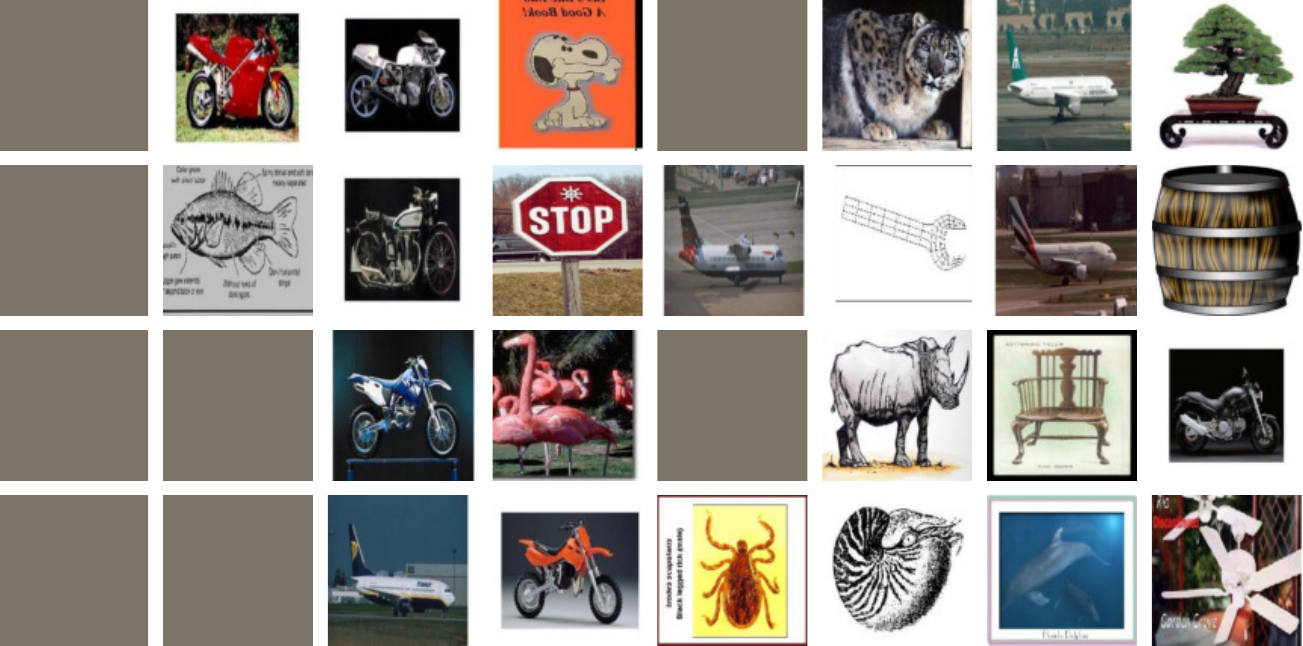}}
\vspace{-0.3cm}
\caption{Reconstructed complete images from a backdoored ViT-B/32 with GELU activations (finetuned on Caltech 101). \textbf{Top}: Reconstruction; \textbf{Bottom}: Ground truth, when it is unambiguous.}
\label{reconstruction::vit:relu:pet}
\end{center}
\vskip -0.1in
\end{figure}

\subsection{Experiments}
We carry out our data-stealing attack on popular pretrained transformers and verify that the backdoored models still perform well on the downstream tasks. We use a pretrained ViT-B/32 from Torchvision and BERT-base from HuggingFace, and finetune on small downstream datasets for illustration: Oxford-IIIT Pet, Caltech 101, and TREC. More details about the finetuning setup are in Appendix \ref{appendix::setupdetail}. %

\begin{table}[t]
    \centering
    \begin{scriptsize}
    \begin{tabular}{@{} l @{}}
    \hline 
    \hl{what must a las vegas blackjack dealer do when he reaches 16?} \\ \hline
    \hl{which two states enclose chesapeake bay?}ruplets?? \\ \hline
    \hl{how is easter sunday's date determined?}       \\ \hline
    \hl{what do the letters d. c. stand for in washington, d. c.?}np.       \\ \hline
    \hl{what schools in the washington, dc nn nn vbp nn nn nn nn.}       \\ \hline
    \hl{why do heavier objects travel downhill faster?} go to college?       \\ \hline
    \end{tabular}
    \end{scriptsize}
    \vspace{-0.1cm}
    \caption{Selected reconstructed sentences from a backdoored BERT model finetuned on the TREC-50 dataset. We highlight tokens which match a ground truth sentence in the dataset. The complete list of reconstructions is in Table \ref{reconstruction::bert:gelu:trecfifty}.}
    \label{reconstruction::bert:example}
    \vspace{-1em}
\end{table}

Figure \ref{reconstruction::vit:relu:pet} shows reconstructed images from the ViT on Caltech 101. The reconstructions are grayscale as we applied this transformation to make the backdoors more space-efficient. Individual patches also have slight mismatches in brightness---an artifact of layer normalizations---but our attack clearly succeeds in recovering recognizable images.

Some selected reconstructed examples from BERT finetuned on TREC-50 are in Table \ref{reconstruction::bert:example}. These reconstructions are near-perfect, except for some rare spurious suffix tokens (this occurs when a short sentence fails to activate all backdoors, and these backdoors later capture other tokens). %
The full lists of reconstructed inputs for ViT and BERT are in Appendix~\ref{appendix::additionalresults}. We also provide additional results using transformers with ReLU activations, where backdoors are more robust and therefore more successful at capturing inputs.

A notable challenge with transformers is that backdoors that fail to shut down  risk destroying the model's benign utility (due to overly large gradient flows in subsequent batches). To mitigate this risk, our experiments focus on settings where the number of backdoors is smaller than the number of output classes (so that all backdoors shut down with high probability). The results we report here are for finetuning runs where the model's benign utility did not break down (in practice, a victim would presumably notice if the finetuning failed and retry with other hyperparameters).

Our ViT model reaches $83\%$ test accuracy on Caltech 101, and our BERT model reaches 78\% test accuracy on TREC-50.
While the backdoor does reduce test accuracy (see Appendix \ref{appendix::largemodelprocessing}), this could be compensated by releasing a larger, more capable backdoored model.

\section{Black-box Attacks}

So far we considered an attacker with \emph{white-box} access to the finetuned model, who can read-off captured inputs from model weights. %
In this section, we consider a weaker \emph{black-box} attacker with query-only access to the final model.

We first show in Section~\ref{sec:dp} that in this highly practical setting, an attacker can mount \emph{perfect membership inference attacks}, to determine with certainty if a data point was used for finetuning. This result has surprising implications for the tightness of the DP-SGD algorithm of \citet{abadi2016deep}.

In Section~\ref{sec:bb_reconstruct}, we explore black-box data extraction attacks. We show that techniques from the model stealing literature can be used to reconstruct backdoor weights (and thus captured inputs) using a few thousand model queries.

\subsection{Black-box Membership Inference and Applications to Differential Privacy}
\label{sec:dp}

Implementing a backdoor that yields perfect membership inference is simple.
We set the model's weights so that the activations $\vec{h} = [h_1, \dots, h_n]$ in the 2\textsuperscript{nd}-to-last layer satisfy:
\begin{itemize}[itemsep=-3pt, topsep=0pt]
    \item $\vec{h} = [C, 0, \dots, 0]$ for the target $\vec{x}$, for a large $C$.
    \item ${h}_1 = 0$ for all other inputs.
\end{itemize}
The targeted input $\vec{x}$ thus only activates the first unit in the layer (with a large value), and that unit is not activated by any other input.
During training, the weights connected to the backdoor unit $h_1$ are only updated if the target $\vec{x}$ is in the training set.
After training, the attacker can query the model on the target $\vec{x}$ and observe whether there is a significant change in the logits compared to before training.

\begin{figure}[tb]
    \begin{center}
    \includegraphics[width=0.9\columnwidth]{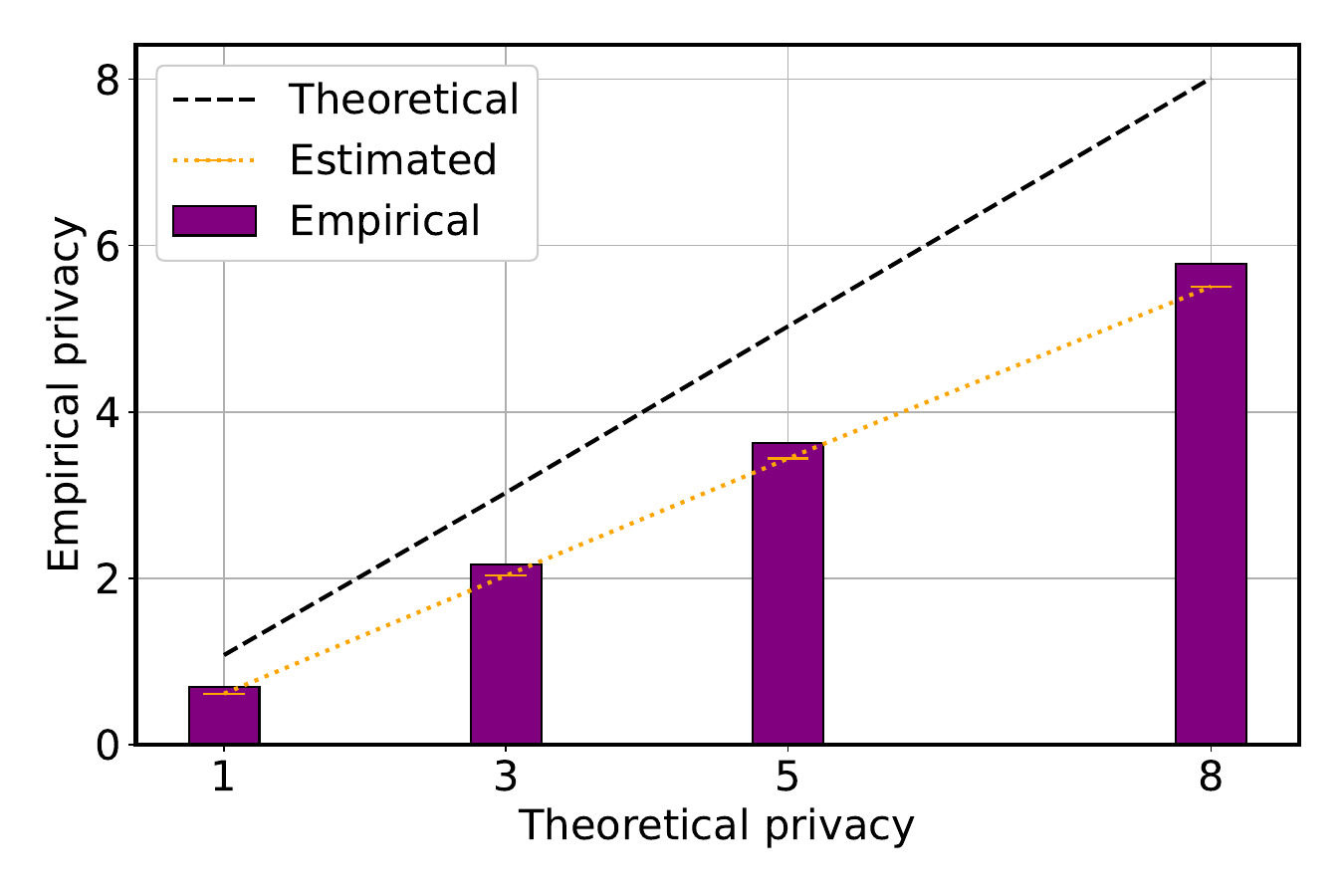}
    \vspace{-1.0em}
    \caption{Empirical privacy for backdoored models for an end-to-end attacker, compared to DP-SGD's provable upper bound. The estimation in orange is a conservative lower bound of the empirical privacy budget, to account for
    the target gradients not fully concentrating on the backdoor weights.   
    }
    \label{diffprv::estimation}
    \end{center}
    \vskip -0.2in
\end{figure}

Such backdoors have important implications for the tightness of the DP-SGD algorithm of \citet{abadi2016deep}. DP-SGD ensures privacy by bounding each training data's influence on the model weights.
The algorithm's analysis assumes a worst-case where: (1) one weight has a large gradient if and only if some input is in the training set; and (2) the attacker sees all noisy gradient updates.
This analysis is presumed to be pessimistic, which leads to the adoption of loose privacy guarantees.\footnote{\citet{nasr2021adversary} show that when the attacker can observe all intermediate steps, the analysis is empirically tight. But for more realistic end-to-end attackers, their approach yields loose bounds.}
Yet, we show that our backdoor construction leads to a near-tight privacy leakage for an end-to-end attacker who only observes the final model.

In more detail, in DP-SGD each input's gradient is clipped to norm $C$ (we can assume $C=1$), and Gaussian noise $\mathcal{N}(0, \sigma I)$ is added to the batch gradient. The privacy analysis of \citet{abadi2016deep} assumes a worst-case where the gradient at some position $i$ is $1$ on input $\vec{x}$ and $0$ otherwise, and the adversary sees all noisy gradient updates. %

Our backdoor above ensures this worst-case scenario occurs in every training step: since the unit $h_1$ is large, the clipped gradient concentrates on the weights of $h_1$. Our black-box attack then compares the logits before and after finetuning, to infer whether the target was present or not. As we show in Appendix~\ref{appendix::diffprv}, an end-to-end black-box attacker who can only interact with the final model gives a privacy lower-bound that is close to the provable guarantee.

To illustrate, we consider backdoored CNN models with a frozen backbone (keeping some weights frozen simplifies the backdoor design and analysis).
Figure \ref{diffprv::estimation} shows that the  empirical privacy budget for our attack is close to the theoretical bound from the DP-SGD analysis. 
Our experiment has two important consequences: (1) loose privacy budgets considered in the literature (e.g., $\varepsilon > 8$ in the production Federated Learning system of \citet{ramaswamy2020training}) are unsafe if the model provider is untrusted;
(2) attempts at a tighter end-to-end privacy bound for DP-SGD (e.g., as in \citep{chourasia2021differential, feldman2018privacy}) must make assumptions about the model to rule out backdoors.

\subsection{Towards Black-box Data Reconstruction}
\label{sec:bb_reconstruct}

\begin{figure}[t]
    \centering

    \begin{subfigure}[b]{0.62\columnwidth}
         \centering
        \includegraphics[width=\textwidth]{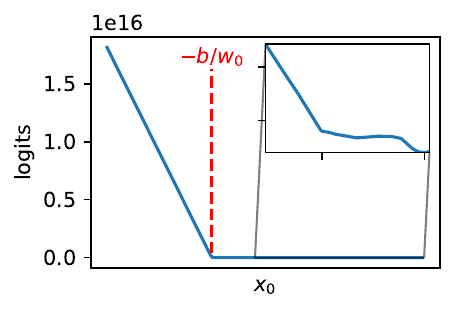}
        \vskip -0.05in
        \caption{Our black-box attack extracts captured inputs pixel-by-pixel, by recovering the critical point where each backdoored unit activates.}
        \label{fig:bb_relu}
     \end{subfigure}
     \hfill
     \begin{subfigure}[b]{0.35\columnwidth}
         \centering
         \includegraphics[width=\textwidth]{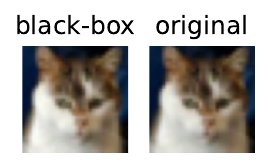}
         \vspace{0.5em}
         \caption{A CIFAR-10 image recovered
         from a backdoored MLP using our black-box attack.}
         \label{fig:bb_results}
     \end{subfigure}
    \vskip -0.05in
    \caption{Black-box data reconstruction using partial model stealing. By finding the critical points of the backdoored ReLU units (left), we can reconstruct a captured input (right).}
    \label{fig:bb_attack}
    \vskip -0.1in
\end{figure}

Finally, we consider the challenging problem of black-box data extraction.
Our insight is that this problem can be reduced to \emph{model stealing}~\cite{tramer2016stealing, carlini2020cryptanalytic, rolnick2020reverse, shamir2023polynomial}: if the attacker can recover the backdoored weights using only black-box queries, then they can recover the captured inputs.

Existing (exact) model stealing attacks exploit the piecewise linearity of ReLU networks to extract weights layer-by- layer. But they only work for small models, and require high-precision outputs (typically double precision).
Luckily, our problem is simpler for three reasons: (1) we only need to recover the weights of the first layer; (2) we do not require high precision recovery; (3) the backdoor units are connected to a very large weight.

These properties enable a simple and efficient attack. Consider a backdoor unit $h = \vec{w}^\top \vec{x} + b$ as in Section~\ref{sec:warmup}. By querying the model on inputs of the form $\vec{x} = [c, 0, \dots, 0]$, the attacker can find values of $c$ that activate the unit $h$ (we assume the attacker can make arbitrary queries in $\mathbb{R}^m$). Since $h$ is connected to a large weight, the model's output behaves approximately like a ReLU (see Figure~\ref{fig:bb_relu}). A black-box attacker can find the critical point $c=-b / w_0$ of this function  (with just two queries as in \citep{carlini2020cryptanalytic}). Repeating this procedure for each input coordinate recovers the full weight $\vec{w}$ up to a constant. From this, the attacker can read off the captured input. An example is in Figure~\ref{fig:bb_results}, for a backdoored MLP trained on CIFAR-10. The full attack requires under $10{,}000$ queries.

\section{Conclusion}
Thousands of developers implicitly trust that foundation models shared online (e.g., on hubs like HuggingFace)
have not been tampered with to embed malicious functionality.
Our work shows that model backdoors can not only compromise model \emph{integrity}, but also \emph{privacy}.
By manipulating a pretrained model's weights, an attacker can recover entire training inputs from finetuned models.
Our work expands the scope of supply chain attacks on the machine learning pipeline, and shows we must account for worst-case weights when reasoning about the privacy of a deployed model.

{\fontsize{10}{11}\selectfont
\bibliography{paper}
\bibliographystyle{sources/icml2024}
}

\newpage
\appendix
\onecolumn

\section{Common Modules and Tricks For Transformer Backdoors}
\label{appendix::tricks}
In this appendix, we provide more details on some common modules and numerical tricks we use to implement privacy backdoors in transformer models.
In Appendix~\ref{appendix::manipulation}, we then describe how we combine these modules for ViT and BERT models.

\subsection{Stabilizing Layer Normalization}
\label{appendix::tricks:stabilizelayernormalization}

Layer normalization is a critical component of the transformer architecture, which enables smoother gradients, faster training, and better accuracy \cite{xu2019understanding}. However, it brings no direct benefit to our backdoor construction. 

In fact these layers are quite problematic because they cause a strong coupling between different features, and they rescale inputs according to their magnitude. 
The coupling of features is challenging because we want different portions of the feature vector to be used for different purposes (e.g., for benign features, or for backdoor inputs and outputs). The rescaling effect of layer normalization can attenuate gradients, which risks recovering illegible mixtures.

To mitigate these issues, we show how to ``decouple'' two parts of a feature vector,
so that the layer normalization acts (nearly) independently on both parts. As a by-product, our approach also gets rid of the layer's input-dependent rescaling effect!

A layer normalization module with learned parameters $\pmb{\gamma}, \pmb{\beta}$
implements the following operation:
\[
\layernormalization{\vct{x}} = \z * \pmb{\gamma} + \pmb{\beta},
\quad \text{where}\quad  \z = \frac{\x - \mu}{\sigma}
\;.
\]
Here $\mu$ and $\sigma$ are the sample average and standard deviation of $\x$, $*$ is a coordinate-wise product, and for simplicity we omit the numerical stability correction that ensures that the standard deviation is non-zero.

Let $m$ be the dimension of $\x$. The gradient of the normalization operations satisfies:
\begin{equation}
\partdrv{z_j}{x_j} = \frac{(1-\frac1m)-\frac{\left(x_j-\mu\right)^2}{m \sigma^2}}{\sigma} \qquad \forall j
\qquad\qquad
\partdrv{z_i}{x_j} =  \frac{-\frac1m - \frac{(x_j-\mu)(x_i-\mu)}{m \sigma^2}}{\sigma} \qquad \forall i \neq j \; .
\label{gradient::layernormalization}
\end{equation}

If the input is amplified as $\vct{x} \to \vct{x}^\prime = c\cdot \vct{x}$ for $c\gg 1$, the gradient becomes smaller. Unlike in a MLP layer, the gradient flowing back to the input is thus input-dependent. Moreover, the gradient for coordinate $x_j$ depends on the value of all other input coordinates.

To stabilize a layer normalization, we consider a special case where the input is partitioned as $\vct{x} \rightarrow (\vct{x}_L,\vct{x}_R)$.
We then modify the input as $(\vct{x}_L, \vct{x}_R + \vct{C})$, where $\vct{C}$ is a vector of \textit{stabilizer constants} and $C \gg \max_\vct{x}\lVert \vct{x} \rVert_\infty$. We denote the respective dimensions of $\x_L$ and $\x_R$ as $m_R$ and $m_L$ (with $m_R+m_L =m$), and their respective means and standard deviations as $\mu_R, \mu_L, \sigma_L, \sigma_R$.

Asymptotically in $C$, we can then write the forward pass as:
\begin{equation*}
    (z_L)_j = \frac{({x}_L)_j - \mu_L - \frac{m_R}{m}C}{\sqrt{\frac{m_L m_R}{m^2}}C} \qquad 
(z_R)_j =  \frac{ ({x}_R)_j - \mu_R + \frac{m_L}{m} C}{\sqrt{\frac{m_L m_R}{m^2}}C} \;.
\end{equation*}
Similarly, the asymptotic behavior of the gradients is:
\begin{align*}
&\partdrv{(z_L)_i}{({x}_L)_i} = \frac{1-\frac{1}{m_L}}{C\sqrt{\frac{m_L m_R}{m^2}}} \;,
\qquad 
\partdrv{(z_L)_i}{({x}_L)_j} = \frac{-\frac{1}{m_L}}{ C\sqrt{\frac{m_L m_R}{m^2}}} \;,
\qquad 
\partdrv{(z_L)_i}{({x}_R)_j} = -\frac1m\frac{\left[\frac{m}{m_R}(({x}_L)_i-\mu)-\frac{m}{m_L}(({x}_R)_j-\mu)-2(\mu_R-\mu_L)\right]}{C^2\sqrt{\frac{m_L m_R}{m^2}}}\\
&\partdrv{(z_R)_i}{ ({x}_R)_i} = \frac{1-\frac{1}{m_R}}{C\sqrt{\frac{m_L m_R}{m^2}}} \;,
\qquad \partdrv{(z_R)_i}{({x}_R)_j} = \frac{-\frac{1}{m_R}}{C\sqrt{\frac{m_Lm_R}{m^2}}} \;,
\qquad 
\partdrv{(z_R)_i}{({x}_L)_j} = -\frac{1}{m}\frac{\left[\frac{m}{m_R}(({x}_L)_j-\mu)-\frac{m}{m_L}(({x}_R)_i-\mu)-2(\mu_R-\mu_L)\right]}{C^2\sqrt{\frac{m_Lm_R}{m^2}}} \;.
\end{align*}

We have thus achieved two things: (1) the gradients of the left and right outputs with respect to the corresponding input no longer depend on the input; (2) the ``cross-term'' gradients between the left output and right input (and vice-versa) decay quadratically faster with $C$.

Then we just rescale the outputs properly. To this end, we define a special module \[
\stabln(\vct{x};\mathcal{J};\pmb{\gamma}_L,\pmb{\beta}_L, \pmb{\gamma}_R,\pmb{\beta}_R; C)\;,
\]
implemented by the layer normalization module.
Here $\mathcal{J} \subseteq \{1, \dots, d\}$ is a subset of the feature coordinates in $\x$, and $\bar{\mathcal{J}}$ is its set complement.
The forward pass of this module satisfies:
\begin{align*}
    &\vct{x}_L \leftarrow \vct{x}_{\mathcal{J}} \qquad \vct{x}_R \leftarrow \vct{x}_{\bar{\mathcal{J}}} \\
    &(\z_L, \z_R) = \mathrm{norm}(\vct{x}_L,\vct{x}_R + \vct{C}) \\
    &\vct{x}_L^{\prime} = \pmb{\gamma}_L * \sqrt{\frac{m_L m_R}{m^2}}C \z_L + \pmb{\gamma}_L *\frac{m_R}{m} \vct{C} + \pmb{\beta}_L = \pmb{\gamma}_L * (\vct{x}_L-\mu_L) + \pmb{\beta}_L \\
    &\vct{x}_R^{\prime} = \pmb{\gamma}_R * \sqrt{\frac{m_L m_R}{m^2}} C \z_R - \pmb{\gamma}_R * \frac{m_L}{m} \vct{C} + \pmb{\beta}_R  = \pmb{\gamma}_R * (\vct{x}_R-\mu_R) + \pmb{\beta}_R \; .
\end{align*}
This module almost degenerates into a simple one-to-one affine layer. For the entire module, we get the following gradients:
\begin{align*}
    &\frac{\partial ({x}^{\prime}_L)_i}{\partial ({x}_L)_i} = ({\gamma}_L)_i \left(1-\frac{1}{m_L}\right) 
    \qquad 
    \frac{\partial ({x}_L^{\prime})_i}{\partial ({x}_L)_j} = -({\gamma}_L)_i\frac{1}{m_L}
    \qquad 
    \frac{\partial ({x}^{\prime}_L)_i}{\partial ({x}_R)_j} =
    \mathcal{O}(1/C)\\
    &\frac{\partial ({x}^{\prime}_R)_i}{\partial ({x}_R)_i} = ({\gamma}_R)_i\left(1-\frac{1}{m_R}\right)
    \qquad 
    \frac{\partial ({x}^{\prime}_R)_i}{\partial ({x}_R)_j} = - ({\gamma}_R)_i\frac{1}{m_R}
    \qquad 
    \frac{\partial ({x}^{\prime}_R)_i}{\partial ({x}_L)_j} = 
    \mathcal{O}(1/C) \;.
\end{align*}

We have thus successfully achieved two effects. First, we have decoupled the two parts of an input vector corresponding to the indices $\{\mathcal{J},\bar{\mathcal{J}}\}$. Second, we got rid of the input-dependent scaling. %

\subsection{Dealing with GELUs}
\label{appendix::tricks:gelu}
Modern transformers typically use GELU activations rather than ReLUs.
The GELU function (Gaussian Error Linear Unit) is defined as 
$\gelu{x} = x \Phi (x)$,
where $\Phi(x)$ is the cumulative distribution function for a standard Gaussian distribution. %
A straightforward calculation of its derivative shows that
\begin{equation*}
\drv{x} \gelu{x} = \Phi(x) + x\phi(x) \qquad \wrapmin{\drv{x} \gelu{x}}{x} \approx -0.13 \, ,
\end{equation*}
where $\phi(x)$ is the probability density function of a standard Gaussian distribution. Unlike a ReLU, a GELU can thus have negative outputs, and negative derivatives. 
For inputs of large enough magnitude (e.g., $|x| > 10$), the GELU is essentially equivalent to the ReLU for our purposes.

GELUs are problematic for two reasons: (1) a backdoor can never be fully shut down, and so a backdoor will continuously capture small parts of inputs even after its first activation. This reduces the quality
of reconstructed inputs; (2) if the backdoor activates, we get a large gradient flow into the backdoor. If the GELU has a negative gradient, then we get a large negative gradient into the backdoor, which opens the backdoor further and risks breaking down the model's behavior in subsequent batches.

A naive solution is to increase the magnitude of backdoor inputs $\vct{x}\to \vct{x}^\prime = c\cdot \vct{x}$ for some $c \gg 1$. Then, we expect that all backdoor outputs will fall into the region where the GELU behaves like a ReLU.
This scaling also requires to increase the magnitude of the shutdown term (defined in Appendix \ref{appendix::tricks:shutdown}) when the backdoor is first activated, thereby helping to keep the backdoor shut. 
Unfortunately, this approach is still prone to some problems. In practice, some negative gradient signal into a backdoor unit is inevitable. By simply increasing the magnitude of inputs, we also amplify the impact of these unintended negative gradients, which risks breaking the backdoor fully open.

To attenuate these issues we make use of two strategies. First, we set backdoor thresholds somewhat loosely, so that some backdoor units may effectively be activated multiple times. This makes those backdoors harder to reconstruct, but lowers the probability of a full model breakdown.
Second, we use the MLP directly following each backdoor to ensure that we only get strong gradient signals into the backdoor if the activation output is large (and confirmative) enough (i.e., in the regime where the GELU acts like a ReLU). Otherwise, the gradient signal will not be amplified by the MLP and thus only have a negligible impact on the backdoor weights (and the subsequent selective activation process).

\subsection{Boosting Shutdown Terms}
\label{appendix::tricks:shutdown}

As we saw in Section~\ref{sec:warmup}, when a backdoor has captured an input $\hat{\x}$, the output of the updated backdoor on a new input $\x$ is equal
to
\[
\w'^\top \x + b' = (\w^\top \x + b) - c \cdot (\hat{\x}^\top \x + 1) \,,
\]
where $c$ is a positive term.

The term $\hat{\x}^\top \x$ determines how easy the backdoor is to shut down, if we keep $c$ fixed. We call this term the ``shutdown term''.
Ideally, we would want this term to be as positive as possible. However, simultaneously, we expect to decouple the selective activation process with the positivity of the shutdown term so that we can have the strongest distinguishability for different backdoor weights.

We can achieve this by logically splitting the backdoor. Assume that the weight of the backdoor is of the form $\w = (\vct{0}, \w_R)$, and an input is similarly split as $\x = (\x_L, \x_R)$. Then, the activation value only depends on the second half of the input:
\begin{align*}
    h & =\relu{\vct{w}^\top \vct{x}+ b} \\
    & = \relu{\vct{w}_R^\top \vct{x}_R + b} \; .
\end{align*}
The shutdown term, however, depends on the entire input:
\begin{equation*}
    \hat{\x}^\top \x = \begin{bmatrix} \hat{\x}^\top_L, \hat{\x}^\top_R\end{bmatrix} \begin{bmatrix} \vct{x}_L \\ \vct{x}_R \end{bmatrix} = \hat{\x}^\top_L \vct{x}_L + \hat{\x}_R^\top \vct{x}_R \;.
\end{equation*}
Thus, by boosting the left-hand part $\hat{\vct{x}}_L^\top \vct{x}_L$ of the input of the backdoor, we can create larger shutdown terms without changing the backdoor's selectivity (i.e., which inputs can activate a given backdoor). 

\paragraph{Partial input boosting using a layer normalization.}
We can achieve this boosting of part of the backdoor input using a (stabilized) layer normalization operation, which directly precedes the MLP modules that implements the backdoor. We combine these operations into a reusable structure we call $\speclinear$:
\begin{align*}
    &\speclinear\left( 
    [\x_L, \x_R], [\pmb{\gamma}_L , \pmb{\gamma}_R], [\pmb{\beta}_L, \pmb{\beta}_R], [\vct{w}_L, \vct{w}_R], b \right):\\
    &\quad\quad \textbf{return }
    [\w_L, \w_R]^\top \cdot \Big([\pmb{\gamma}_L, \pmb{\gamma}_R] * \texttt{norm}([\x_L, \x_R]) + [\pmb{\beta}_L, \pmb{\beta}_R]  \Big) + b
\end{align*}
Here, $\pmb{\gamma}$ and $\pmb{\beta}$ are the affine transforms implemented by the layer normalization, $\texttt{norm}$ normalizes the input to zero mean and unit variance, and $*$ is the coordinate-wise product of two vectors.

A special instantiation of interest is when $\pmb{\gamma}_L = \w_L = \vct{0}_L, \pmb{\beta}_R = \vec{0}_R$. Assuming for simplicity that $\x$ is already normalized\footnote{In practice, the layer normalization here is always stabilized. The final effect is just like the input has already been normalized.}, the module output then simplifies to:
\begin{equation*}
    \w_R^\top (\pmb{\gamma}_R * \x_R) + b \;,
\end{equation*}
and the shutdown term for the backdoor is
\begin{equation*}
    \lVert \pmb{\beta}_L\rVert^2 + (\pmb{\gamma}_R * \vct{\hat{x}}_R)^\top (\pmb{\gamma}_R * \vct{x}_R) \; .
\end{equation*}

Then, we can arbitrarily increase the magnitude of $\pmb{\beta}_L$ to increase the shutdown term, without any effect on the backdoor output.

\subsection{Aggregating Information across Tokens}
\label{appendix::tricks:attention}

As we explain in Section~\ref{sec:transformers_overview}, a challenge we face with transformers is to capture all tokens in an input sequence, rather than individual tokens belonging to different inputs.

We solve this problem by
computing a \emph{sequence identifier} that is common to all tokens in a sequence, and which we can later use as a feature to selectively activate backdoors.
We implement a sequence identifier using an attention module that averages features across all tokens in a sequence.

Let $X = \begin{bmatrix} \vct{x}^{(1)} & \vct{x}^{(2)} & \cdots & \vct{x}^{(k)}\end{bmatrix}^\top$ be an input sequence consisting of $k$ token embeddings.
The output of self-attention module $Z = \selfattention{X}$ is then:
\begin{equation}
    \begin{aligned}
    & Q = \begin{bmatrix} \mtx{W}^{(Q)} \vct{x}^{(1)} + \vct{b}^{(Q)} &  \mtx{W}^{(Q)} \vct{x}^{(2)} + \vct{b}^{(Q)} &  \cdots &  \mtx{W}^{(Q)} \vct{x}^{(k)} + \vct{b}^{(Q)}\end{bmatrix}^\top \\
    & K = \begin{bmatrix} \mtx{W}^{(K)} \vct{x}^{(1)} + \vct{b}^{(K)} &  \mtx{W}^{(K)} \vct{x}^{(2)} + \vct{b}^{(K)} & \cdots & \mtx{W}^{(K)} \vct{x}^{(k)} + \vct{b}^{(K)} \end{bmatrix}^\top \\ 
    & V = \begin{bmatrix} \mtx{W}^{(V)} \vct{x}^{(1)} + \vct{b}^{(V)} &  \mtx{W}^{(V)} \vct{x}^{(2)} + \vct{b}^{(V)} & \cdots &  \mtx{W}^{(V)} \vct{x}^{(k)} + \vct{b}^{(V)} \end{bmatrix}^\top \\
    & Z = \begin{bmatrix} \vct{z}^{(1)} & \vct{z}^{(2)} & \cdots & \vct{z}^{(k)}\end{bmatrix}^\top = \softmax\left(QK^\top\right) V \; .
\end{aligned} 
\label{module::attention}
\end{equation}

Let $\mathcal{J} \subseteq \{1, \dots, 768\}$ be a set of coordinates in a token embedding,
and $\rho>0$ some constant. We then define a module $\synattention(X; \mathcal{J}; \rho)$, which sets the attention weights as follows:
\begin{equation}
    \begin{aligned}
    \mtx{W}^{(Q)} &= \vec{0}  &&\qquad\vct{b}^{(Q)} = \vct{0}\\
    \mtx{W}^{(K)} &= \vec{0} &&\qquad\vct{b}^{(K)} = \vct{0}\\
    \mtx{W}^{(V)}_{ij} &= \begin{cases} 
    \rho \quad &\text{if}\hspace{0.2cm} i=j\in \mathcal{J} \\
    0  &\text{else}
    \end{cases}
     &&\qquad\vct{b}^{(V)} = \vct{0} \; .
    \end{aligned}
    \label{structure::syn}
\end{equation}
The query $Q$, key $K$ and value $V$ matrices are then:
\begin{align*}
    & Q = K = \vec{0} \\
    & V_{ij} = \begin{cases} 
    \rho x^{(i)}_j \quad & \text{if $j\in\mathcal{J}$} \\
    0 & \text{else}
    \end{cases} \; .
\end{align*}
The softmax value is then just a uniform matrix with value $1/k$ in each entry.
Finally, the output $Z$ satisfies:
\begin{equation*}
    z^{(i)}_j = \begin{cases}
    \rho \frac1k\sum_{1 \leq l \leq k} x^{(l)}_j \quad & \text{if $j\in\mathcal{J}$} \\
    0 & \text{else}
    \end{cases} \; \forall 1 \leq i \leq k.
\end{equation*}
In words, we sum up all the token vectors (scaled by some factor $\rho/k$), and zero-out coordinates that do not belong to the set $\mathcal{J}$. This averaged input is our sequence key which we will use to selectively activate backdoors.

The same design works for multi-headed attention. For a masked attention module, we can simply take the average only over the unmasked tokens.

\section{Creating Keyed Backdoors}
\label{appendix::keyed_backdoor}
In this section, we describe our main backdoor design for transformers.
To capture all individual tokens from a sequence, we build a family of backdoors so that all backdoors fire only for specific \emph{sequence identifiers}, and so that each backdoor fires only for tokens in a specific \emph{position}.

\paragraph{Creating position identifiers.}
We use a portion of each token's feature vector to represent the token's position in a sequence (as is done with position embeddings in a benign transformer).

The only difference is that we craft these position identifiers to be near-orthogonal to each other, so that we can use them to selectively activate different backdoors.
Suppose that tokens can be in one of $n$ positions, and that we reserve $m$ coordinates in the feature vector for a position identifier. We then design these position identifiers $\vct{u}^{(1)}, \dots, \vct{u}^{(n)}\in\mathbb{R}^m$ to satisfy the following properties: 
\begin{equation}
\begin{aligned}
    &\lVert \vct{u}^{(j)}\rVert^2 = u_0 \quad \forall j \\
    &\left(\vct{u}^{(j)}\right)^\top\vct{u}^{(k)} < u_+ \ll u_0\quad \forall j\neq k \\
    &\sum_{k} {u}^{(j)}_k = 0 \quad \forall j \, ,
\end{aligned}
\label{condition::positionembedding}
\end{equation}
where $u_+$ is a positive threshold determined by the attacker. The first two conditions ensure that the position keys are not strongly aligned with each other, so that they can selectively activate individual backdoors.
The last condition that keys have zero mean is merely for convenience, to facilitate interactions with a subsequent layer normalization.

\paragraph{Creating sequence identifiers.}
To create our sequence identifiers, we use the module $\synattention(X, \mathcal{J},\rho)$ we defined in Appendix~\ref{appendix::tricks:attention}, where $\mathcal{J}$ represents the indices of the token features to sum up to create a sequence key. This sequence key is then stored in $|\mathcal{J}_{\textrm{seq}}|$ coordinates of each token's feature representation. We migrate features from the source features to the target features $\vct{seq}$ using the projection linear layer of an attention module.

\paragraph{The backdoor design.}
We can finally present the full keyed backdoor design.
This backdoor is implemented in the transformer's first MLP
module, which maps token features of dimension 768 to hidden units of dimension 3072.
For ViT models, an input sequence consists of 49 tokens, each of which represents a patch of $32\times 32$ pixels. For BERT, a sequence consists of 48 token embeddings. We thus want to build a family of backdoors that can capture all these individual tokens.

Denote an input sequence as $X= \begin{bmatrix}\vct{x}^{(1)} &  \vct{x}^{(2)} & \cdots & \vct{x}^{(k)} \end{bmatrix}$.
Each token vector $\vct{x}^{(i)}$ is split into multiple components:
\begin{equation}
\label{eq:split}
    \vct{x}^{(i)} = [\vct{x}^{(i)}_{\textsubscript{ft}},
\vct{x}^{(i)}_{\textsubscript{act}},
\underbrace{\x^{(i)}_{\textsubscript{pos}}, \x^{(i)}_{\textsubscript{seq}}, \x^{(i)}_{\textsubscript{tok}}}
_{\vct{x}^{(i)}_{\textsubscript{key}}}
] \;,
\end{equation}
where $\vct{x}^{(i)}_{\textsubscript{ft}}$ contains benign features, $\vct{x}^{(i)}_{\textsubscript{act}}$ is used to store backdoor activations, and $\vct{x}^{(i)}_{\textsubscript{key}}$ contains the features to activate the backdoor, which are further split into three components: $\vct{x}^{(i)}_{\textsubscript{pos}} = \vct{u}^{(i)}$ is the token's position identifier, $\vct{x}^{(i)}_{\textsubscript{seq}} = \vct{seq}$ is the sequence identifier which is shared by all tokens in the sequence, and $\vct{x}^{(i)}_{\textsubscript{tok}}$ contains token features to be captured.

We then build families of backdoors where the $j$-th backdoor has parameters
\begin{equation}
\label{eq:keyed_backdoor}
\w = [\vec{0}, \vec{0}, \theta_{\textrm{pos}} \vec{u}^{(j)}, \vec{w}_{\textrm{seq}}, \vec{0}], \qquad b = b_{\textrm{pos}}^{(j)} + b_{\textrm{seq}}
\end{equation}
Here $\theta_{\textrm{pos}} \vec{u}^{(j)}$ is the weight that aligns with tokens in the $j$-th position, $\vec{w}_{\textrm{seq}}$ is the backdoor weight that aligns with the sequence identifier, and $b_{\textrm{pos}}^{(j)}$ and $b_{\textrm{seq}}$ are corresponding thresholds to be set.
On an input token $\x^{(i)}$, the output of the backdoor becomes:
\begin{align*}
h &= \left(\theta_{\textrm{pos}} (\vec{u}^{(j)})^\top  \x^{(i)}_{\textrm{pos}} +b_{\textrm{pos}}^{(j)} \right) + \left(\w_{\textrm{seq}}^\top \x^{(i)}_{\textrm{seq}} +b_{\textrm{seq}} \right) \\
&= \left(\theta_{\textrm{pos}} (\vec{u}^{(j)})^\top  \vec{u}^{(i)} +b_{\textrm{pos}}^{(j)} \right) + \left(\w_{\textrm{seq}}^\top \vct{seq} +b_{\textrm{seq}} \right) \;.
\end{align*}
By adjusting the thresholds $b_{\textrm{pos}}^{(j)}$ and $b_{\textrm{seq}}$, we can ensure that the backdoor only activates when both $(\vec{u}^{(j)})^\top  \vec{u}^{(i)}$ and $\w_{\textrm{seq}}^\top \vct{seq}$ are large. The first term is large for tokens in the $j$-the position, and the second term follows some distribution over all training inputs.
We set these thresholds so that a token in an incorrect position $i \neq j$ never activates the backdoor, i.e.,
\[
\max_\x (\vct{w}_{\textrm{seq}}^\top \vct{x}_{\textsubscript{seq}} + b\textsubscript{seq}) \ll b_{\textrm{pos}}^{(j)} - \max_{k\neq j} ( \theta\textsubscript{pos} (\vct{u}^{(j)})^\top \vct{u}^{(k)}) \;.
\]
Finally, we apply the trick described in Appendix~\ref{appendix::tricks:shutdown} to boost the backdoor's ``shutdown'' term, to ensure that an activated backdoor gets a large gradient modifcation that shuts it down.

Specifically, we use a combination of a layer normalization and MLP to implement the module:
\begin{equation*}
   \speclinear\left( 
    \left(
    \underbrace{\begin{bmatrix} \vct{x}_{\textsubscript{ft}} \\ \vct{x}_{\textsubscript{act}}\end{bmatrix}}_{\x_L}, \underbrace{\begin{bmatrix} \vct{x}_{\textsubscript{pos}} \\
    \vct{x}_{\textsubscript{seq}} \\
    \vct{x}_{\textsubscript{tok}} \\
    \end{bmatrix}}_{\x_R}
    \right)
   ,
   \left(\vec{0} , \pmb{\gamma}_R \right), \left(\pmb{\beta}_L, \vct{0}\right), \left(
   \underbrace{\begin{bmatrix}
   \vct{0}\\
   \vct{0}\end{bmatrix}}_{\w_L}, 
   \underbrace{\begin{bmatrix} 
   \theta\textsubscript{pos}\vct{u}^{(j)} \\ \vct{w}_{\textrm{seq}} \\ \vct{0} 
   \end{bmatrix}}_{\w_R} 
   \right), b_{\textrm{pos}}^{(j)} + b_{\textrm{seq}} \right) \, ,
\end{equation*}

\paragraph{Multiple families of backdoors.}
To capture multiple full inputs, we replicate the above design of a backdoor family, each time using a different sequence weight $\w_{\textrm{seq}}$ and corresponding bias $b_{\textrm{seq}}$.
Since there are $3072$ hidden units in the MLP, we could implement many backdoor families. But at the end of the module, we need to fit all the backdoor outputs into a feature vector of dimension $768$. We do this by aggregating the activation signals of an entire backdoor family into a single feature. That is, assuming that the $i$-th backdoor family uses the units in positions $I_i \subset \{1, \dots, 3072\}$, we use the final linear layer in the MLP module to project all these units onto a single coordinate $j$ of the feature vector:
\begin{equation}
\mtx{W}^{(O)}_{j, k} = \begin{cases}
    \theta\qquad &\text{if $k\in \mathcal{I}_i$} \\
        0\qquad  &\text{else}
    \end{cases}  \, ,
    \label{structure::bert:amplifier}
\end{equation}
where $\theta$ is an amplifier coefficient. Using these weights, we ensure that the gradient signal is shared between all backdoors in a family, so that they all shut down simultaneously when activated.
\section{Detailed Description of Transformer Backdoors for ViT and BERT}
\label{appendix::manipulation}

In this appendix, we provide details about how to craft backdoors for ViT and BERT models, by making use of some of the common tricks and modules described in Appendix~\ref{appendix::tricks}.

\subsection{Structure of a Transformer}
In the main body of the paper, we give some high-level overview of the architecture of a transformer, and how we implement the backdoor into it.
Here we provide additional details necessary for describing the backdoor modules.

We begin by looking at the structure of the transformer's encoder blocks, which combine three operations: a layer normalization, a MLP, and an attention module. The MLP module $\vct{x}^{\prime\prime}=\mlp{\vct{x}}, \vct{x}\in\mathbb{R}^{768}$ is composed of two linear layers: 
\begin{equation}
\begin{aligned}
    & \vct{x}^\prime = \textrm{GELU}(\mtx{W} \vct{x} + \vct{b}) \quad \vct{x}^\prime \in \mathbb{R}^{3072}  \\
    & \vct{x}^{\prime\prime} = \mtx{W}^{\prime} \vct{x}^\prime + \vct{b}^{\prime} \quad \vct{x}^{\prime\prime} \in \mathbb{R}^{768} \;.
\end{aligned}
\label{module::mlp}
\end{equation}
Note that there are 3072 intermediate units inside a MLP module. A complete attention module $Z^\prime = \completeattention{X}$ contains one more projection linear module than the simplified description in Equation \eqref{module::attention}, i.e., 
\begin{equation}
\begin{aligned}
    &Z = \begin{bmatrix} \vct{z}^{(1)}&\vct{z}^{(2)}&\cdots &\vct{z}^{(k)}\end{bmatrix}^T = \selfattention{X} \\
    &Z^\prime = \begin{bmatrix} \mtx{W}^{(O)} \vct{z}^{(1)} + \vct{b}^{(O)} & \mtx{W}^{(O)} \vct{z}^{(2)} + \vct{b}^{(O)} & \cdots & \mtx{W}^{(O)} \vct{z}^{(k)} + \vct{b}^{(O)}\end{bmatrix}^T  \; .
\end{aligned}
\label{module::completeattention}
\end{equation}
The three modules are arranged in a different order for ViT and BERT models. For convenience, we set all dropout probabilities to zero---a reasonable setup for finetuning tasks.

\paragraph{Partitioning features and blocks.}
We only have limited resources to build a backdoor structure in a transformer. There are 12 encoder blocks and 768 features. Here we describe how we partition these resources for our backdoor construction. This partition trades between the utility of 
the model (i.e., the more features we leave unmodified, the better accuracy we can achieve on downstream tasks), 
and the quality of reconstructed images and sentences. 
Recall that we split a token's feature vector 
as described in Equation~\eqref{eq:split}. When referring to the indices of these different parts of the feature vector we use the notation $\mathcal{J}_{\text{ft}}$, $\mathcal{J}_{\text{key}}$, etc. 
We split the encoder blocks into three sets, ``prefix'', ``propagation'' and ``suffix''. The prefix is used to prepare inputs for the backdoor, and run the backdoor. The propagation blocks propagate features to the end of the transformer. The suffix block prepares the output features.
For the ViT model, we consider two designs: one that captures entire images, and one that simply captures individual patches (i.e., without using any sequence or position keys).
We summarize our partition in Table \ref{tab:resourcepartition}:

\begin{table}[h]
    \begin{center}
    \caption{Partition of encoder blocks and features. \textbf{prefix}: backdoor module, amplifier module, erasure module; \textbf{propagation}: signal propagation module; \textbf{suffix}: output module. $\feat$: benign features for down-stream tasks; $\key$: features for selectively activating a backdoor; $\act$: backdoor activation outputs. $\key$ is further partitioned into $\pos, \seq, \tok$, where $\pos$ are the features used for position keys, $\seq$ are features used for sequence keys, and $\tok$ are  the features captured by the backdoor.}
    \vskip 0.1in
    \begin{tabular}{@{}l r r r r r r r r r@{}}
    \toprule
    & & & & & \multicolumn{4}{c}{$\key$} & \\
    \cmidrule{6-9}
    \textbf{Model}  & Prefix  & Propagation & Suffix &  $\feat$ & (total) & $\pos$ & $\seq$ & $\tok$ & $\act$ \\
    \midrule 
    ViT (image)     &  3 & 8  & 1  & 448  & 256 & 128 & 64 & 64 & 64   \\
    ViT (patch)     &  3 & 8  & 1  & 448 & 256 & 0 & 0 & 256 & 64  \\
    BERT    & 2  & 9  & 1 & 512 & 192 &  64 & 128 & 0 &  64 \\
    \bottomrule
    \end{tabular}
    \label{tab:resourcepartition}
    \end{center}
\end{table}

\subsection{ViT}
\label{appendix::detailedmanipulation::vit}
The encoder block of a ViT model, $Z\textsubscript{out} = \vitencoderblock{X\textsubscript{in}}$ is written as:
\begin{align*}
   & \vct{x}^{(i)} = \layernormalization{\vct{x}^{(i)}_{\textrm{in}}}, \text{ for } 1 \leq i \leq k \\
   & \begin{bmatrix} \tilde{\vct{x}}^{(1)} & \tilde{\vct{x}}^{(2)} & \cdots & \tilde{\vct{x}}^{(k)}   \end{bmatrix}^T = \completeattention{\begin{bmatrix}  \vct{x}^{(1)} & \vct{x}^{(2)} & \cdots & \vct{x}^{(k)} \end{bmatrix}^T} \\
   & \vct{z}^{(i)}_{\textrm{in}} = \tilde{\vct{x}}^{(i)} + \vct{x}^{(i)}_{\textrm{in}} \\ 
   &\vct{z}^{(i)} = \layernormalization{\vct{z}^{(i)}_{\textrm{in}}} \\
   & \tilde{\vct{z}}^{(i)} = \mlp{\vct{z}^{(i)}} \\
   & \vct{z}^{(i)}_{\textrm{out}} = \tilde{\vct{z}}^{(i)} + \vct{z}^{(i)}_{\textrm{in}} \; .
\end{align*}
Observe that the features are normalized before being fed into the MLP or attention layer. Besides, a series of two shortcut connections directly connects the encoder's input and output.

\subsubsection{Input module} 
The first layer of a vision transformer is a convolutional module that extracts embeddings for each input patch.
For ViT-B/32 used in our experiments, this layer has 768 kernels with size and stride of 32. The output is thus a feature representation of a $32\times 32$ patch within the input image.
During training, a $3\times 224 \times 224$ image is fed into the convolutional module and produces a $768\times 7\times 7$ vector. This intermediate state is permuted and reshaped to a $49\times 768$ hidden state, regarded as 49 token vectors with 768 features each.
The hidden state is then concatenated with an additional $1\times 768$ class-token vector, and added to  a $(49+1)\times768$ position embedding. This is the input to the model's first encoder block.

Since we only get (part of) a $768$ feature vector to represent a $32\times32$ patch, we need to compress the patch somehow. We do this by using some convolutional filters to convert each patch to a downscaled and grayscaled version (by selecting and grayscaling one pixel per patch). \footnote{We can use a kernel to extract a target grayscaled pixel from each patch. By combining different extractor kernels, we can extract a downscaled sub-image.} The output is of size either $16\times16=256$ (if we only reconstruct individual patches) or $8\times8=64$ (if we reconstruct entire images). These features are stored in the $\tok$ portion of the input feature vector for each patch.
As an additional trick, we ensure that these features have zero mean, which helps for the subsequent layer normalization.

The kernels that correspond to the benign input positions $\feat$ are not modified, so as to keep valuable features from the pretrained model. 
If we want to reconstruct entire images, we use the position embeddings to implement a position key as described in Appendix~\ref{appendix::keyed_backdoor}.
The kernels that map to other parts of the feature vector (i.e., $\act, \seq$) are set to zero.

\subsubsection{Sequence key creator \& Backdoor module}
The first encoder block of the ViT is used to create sequence keys and to implement the backdoor module (as described in Appendix~\ref{appendix::keyed_backdoor}).

To create sequence keys, we first stabilize the layer normalization before the attention module with the module $\stabln(\x, \mathcal{J}\textsubscript{key}, \vct{1}, \vct{0}, \vct{0}, \vct{0})$ from Appendix~\ref{appendix::tricks:stabilizelayernormalization}.
We then use the attention module to create sequence keys, as described in Appendix~\ref{appendix::tricks:attention}.
We use the features in $\mathcal{J}\textsubscript{tok}$ to create sequence keys, and store the keys in the $\seq$ positions.

We then implement the backdoor module as described in Appendix~\ref{appendix::keyed_backdoor}, using the $\speclinear$ module that combines a layer normalization with the subsequent MLP.
If the target is simply to reconstruct individual patches, rather than complete images, the backdoor construction is simplified as we just apply the backdoor weight to the whole backdoor key $\key$, without any regard for position or sequence information.

\subsubsection{Amplifier \& erasure module}
\label{appendix::manipulation:vit:erasure}
The amplifier module is instantiated using the MLP of the second encoder block. The attention part of this encoder block is skipped using the shortcut connection. The layer normalization before the MLP module is stabilized. Only features belonging to $\act$ are fed into the MLP module. If an activation signal is above a pre-assigned noise threshold, it gets amplified significantly (we can achieve this conditional amplification using a simple 1-layer MLP).
Features that do not belong to $\mathcal{J}\textsubscript{act}$ skip the whole encoder block through the shortcut connections and remain unchanged.

Note that we could have also used the second layer of the MLP in the first encoder block to amplify activation signals, but this approach is problematic for GELU transformers because we cannot avoid unexpected gradients (i.e., when the GELU inputs are close to zero) being amplified.
By pushing the amplifier into the next encoder block, we can use additional layers to ensure that small backdoor outputs gets dampened out, before the amplification.

The erasure module is implemented using the MLP part of the third encoder block. Since the $\key$ features have already played their role, they are wiped out to facilitate signal propagation. The reason this is necessary is because some of the backdoor inputs (e.g., the sequence keys) have large magnitude and this would distort the benign features and cause an accuracy drop on downstream tasks.

Hidden states from the second encoder block are directly fed into the MLP module through the shortcut connection. The layer normalization is stabilized using the module $\stabln\left(\x, \mathcal{J}\textsubscript{key}; \pmb{\gamma},\vct{0}, \vct{0}, \vct{0}\right)$, which wipes out any features not belonging to $\key$. The weights and biases of the MLP module are manipulated to obtain the following forward pass:
\begin{equation}
\begin{aligned}
    &\vct{x}_L \leftarrow \vct{0} * (\key - \mu_{\key}) = \vct{0} && \vct{x}_R \leftarrow \pmb{\gamma} * (\key - \mu_{\key}) = \pmb{\gamma} * \key \qquad &\text{stable layer normalization} \\
    &\vct{x}_L^\prime \leftarrow \vct{0} &&\vct{x}_R^\prime \leftarrow \text{GELU}(\mtx{W}^{(1)} \vct{x}_R + \vec{b}^{(1)}) \qquad &\text{linear layer 1} \\
    &\vct{x}_L^{\prime\prime} \leftarrow \vct{0}  &&\vct{x}_R^{\prime\prime} \leftarrow \mtx{W}^{(2)} \vct{x}_R^\prime + \vec{b}^{(2)} \qquad &\text{linear layer 2} \\
    &\vct{z}_L \leftarrow \vct{0} + \vct{x}_L  = \vct{0}
    &&\vct{z}_R \leftarrow \vct{x}_R^{\prime\prime} + \vct{x}_R = \vct{0} \qquad &\text{shortcut connection} \\
    & \text{s.t.}\quad \mtx{W}^{(2)} \mtx{W}^{(1)} \pmb{\gamma} = -\vec{1} \qquad \vec{b}^{(2)} = -\mtx{W}^{(2)}\vec{b}^{(1)} \; .
\end{aligned}
\label{structure::erasure}
\end{equation}
In words, we ensure that the MLP implements a negative identity mapping, so that the shortcut connection wipes out the features.
The above assumes that the input to the GELU is large enough so that it behaves like an identity. We achieve this by setting $\vec{b}^{(1)}$ to a large positive value. The choice of weights $\pmb{\gamma},\mtx{W}^{(1)},\mtx{W}^{(2)}$ is important to guarantee the robustness of this module. If they are not chosen carefully, this module will drift from its original design after a few gradient updates.

\subsubsection{Signal-propagation \& output module}
The Signal propagation module is implemented on the eight remaining encoder blocks before the final output block.
We want the classification-irrelevant information belonging to $\key$ and $\act$ to have as little impact on these encoder blocks as possible. Therefore, all weights and biases connected to these features are set to zero in the layer normalizations, attention module, and MLP. Then, the backdoor activation signals simply propagate through the entire block by the shortcut connections.

The output module's goal is to aggregate activation signals onto the class token CLS. This is implemented in the last encoder block. The first layer normalization is stabilized in order to avoid any signal attenuation, and
we then re-use the same trick as for sequence keys by using the attention module to aggregate information across all tokens. Specifically, we use the structure $\synattention(X, \mathcal{J}\textsubscript{act}; 1)$ to aggregate all activation signals. The remaining modules in this block are used to implement identity mappings.
At the end of the encoder, the class-token vector is passed to a stabilized layer normalization, which ensures that the $\act$ features are not downscaled or mixed with the benign features.
\newpage

\subsection{BERT}
\label{appendix::detailedmanipulation:bert}
The encoder block of a BERT model, ${Z}\textsubscript{out} = \bertencoderblock{X_{\textrm{in}}}$ is written as:
\begin{align*}
    & \begin{bmatrix} \tilde{\vct{x}}^{(1)}  & \tilde{\vct{x}}^{(2)} & \cdots & \tilde{\vct{x}}^{(k)} \end{bmatrix}^T = \completeattention{\begin{bmatrix}\vct{x}^{(1)}_{\textrm{in}} & \vct{x}^{(2)}_{\textrm{in}} & \cdots & \vct{x}^{(k)}_{\textrm{in}} \end{bmatrix}^T} \\
     & \vct{x}^{(i)} = \tilde{\vct{x}}^{(i)} + \vct{x}^{(i)}_{\textrm{in}}, \text{ for } 1 \leq i \leq k\\
     &\vct{z}^{(i)}_{\textrm{in}} = \layernormalization{\vct{x}^{(i)}} \\
    & \tilde{\vct{z}}^{(i)} = \mlp{\vct{z}^{(i)}_{\textrm{in}}} \\
    & \vct{z}^{(i)} = \tilde{\vct{z}}^{(i)} + \vct{z}^{(i)}_{\textrm{in}} \\
    & \vct{z}^{(i)}_{\textrm{out}} = \layernormalization{\vct{z}^{(i)}} \; .
\end{align*}
Thus, in contrast to the ViT, the inputs are not directly connected to the outputs with a shortcut connection. Inputs are thus necessarily distorted twice by layer normalizations before being output.

\subsubsection{Input module}
The first layer of a BERT model is an embedding layer. Discrete words and positions are converted to 768-entry word and position embeddings. For a word $d_j$ at position $j$, its word embedding vector $\tilde{\vct{e}}^{(d)}$ and position embedding vector $\vct{e}^{(j)}$ are added together for a complete representation $ \vct{x}^{(d,j)} = \tilde{\vct{e}}^{(d)} + \vct{e}^{(j)}$. 

We use the approach described in Appendix~\ref{appendix::keyed_backdoor} to create position identifiers for our backdoors. We add an additional requirement to Equation~\eqref{condition::positionembedding}, to ensure that special tokens (such as the padding token [PAD]) do not activate backdoors. For this, we ensure that the shared embedding of these special tokens (in the coordinates corresponding to the position key $\pos$) is some vector $\vct{u}^{(-)}$ that is negatively aligned with all other position keys $\vct{u}^{(1)}, \dots, \vct{u}^{(n)}$. 

More precisely, the following operations are done to produce embeddings:
\begin{align*}
    &\text{word embedding}:\quad \tilde{\vct{e}}^{(d)}_{\textrm{pos}} \leftarrow \begin{cases} 
    \vct{0}  \quad &\text{if $d$ is not a special token} \\
    \vct{u}^{(-)} \quad &\text{if $d$ is a special token}
    \end{cases} 
    ;\quad \tilde{\vct{e}}^{(d)}_{\textrm{seq}} \leftarrow \vct{0}, \quad \tilde{\vct{e}}^{(d)}_{\textrm{act}} \gets \vct{0}, \quad \tilde{\vct{e}}^{(d)}_{\textrm{ft}}\leftarrow \tilde{\vct{e}}^{(d)}_{\textrm{ft}} \quad \forall d;\\
    &\text{position embedding}:\quad\vct{e}^{(j)}_{\textrm{pos}} \leftarrow \vct{u}^{(j)}; \quad \vct{e}^{(j)}_{\textrm{seq}}\leftarrow \vct{0}, \quad 
    \vct{e}^{(j)}_{\textrm{act}}\leftarrow \vct{0}, \quad \vct{e}^{(j)}_{\textrm{ft}} \leftarrow \vct{e}^{(j)}_{\textrm{ft}} \quad \forall j \; .
\end{align*}
In summary,  the entries for benign features $\feat$ are not changed to inherit the utility of the pretrained weights. There is a layer normalization module next to these embedding modules, which is stabilized as $\stabln\left(\x, \mathcal{J}_\textrm{ft}; \vct{1}, \vct{0}, \vct{1}, \vct{0}\right)$. ($\feat$ are not filtered out in order to recover words.)

\subsubsection{Sequence Key creator \& Backdoor \& Amplifier \& Erasure module}

The backdoor module follows the design outlined in Appendix~\ref{appendix::keyed_backdoor}.
We use the first encoder's attention module to create sequence keys, and the subsequent MLP to implement all our backdoor units. Since the benign features contain both the word and the position information, the benign features are selected as the source from which sequence keys are generated. The sequence keys are shared by all unmasked tokens, i.e., non-pad tokens.
To prepare for backdoor units, the layer normalization after the first attention module is stabilized using the structure $\stabln(\x, \mathcal{J}\textsubscript{ft}; \vec{1},\pmb{\beta}_L,\vct{1},\vct{0})$. 

The amplifier module is directly realized by the second linear layer of the first MLP using the coefficient $\theta$ in Equation~\eqref{structure::bert:amplifier}. %
The erasure module is implemented by the last layer normalization module of the first encoder block. After generating activation signals, position information and sequence keys are useless for the following logical modules. They are wiped out by assigning weights and biases belonging to $\key$ of this LN to zero. This LN is stabilized for convenience and serves as an almost identity mapping for benign features belonging to $\feat$ and activation signals belonging to $\act$. The second encoder block is used to fuse-in weights from the first encoder's layer normalization, which were dropped to stabilize this LN. In this way, we guarantee that the benign features are still computed correctly.

\subsubsection{Signal-propagation \& output module}
The signal propagation module uses the nine remaining encoder blocks before
the final output block. A total of $512$ features belonging to $\feat$ are used for downstream tasks. These encoder blocks strengthen the utility of benign features and propagate the backdoor activation signals. Yet, it is difficult to propagate the activation signals through dozens of layer normalizations (here, in contrast to the ViT model, we cannot simply propagate through the shortcut connection).
If the activation signals are too strong, we get some dampening in each LN, which causes too small gradients. In contrast, if the activation signals are too weak, it is difficult to distinguish them from noise.
The easiest solution would be to stabilize each layer normalization using the trick in Appendix~\ref{appendix::tricks:stabilizelayernormalization}.
But this would require changing these layers' weights, which would break all the benign features too (unless we retrain the backdoored BERT from scratch).
Our simpler solution is to amplify the activation signals in every encoder's MLP in order to approximately offset the signal attenuation we will natively get from the layer normalization (when only using shortcut connections to propagate backdoor outputs). In the MLP, activation signals are amplified if they are more significant than some noise threshold, i.e, 
\begin{align*}
    & \x^{\prime}_{\textsubscript{amp}} = \text{GELU}(\x_{\textrm{act}} - \delta) \\
    & \x^{\prime\prime}_{\textsubscript{act}} = \theta\textsubscript{propag} \cdot \x^\prime_{\textsubscript{amp}} \;.
\end{align*}
Here $\delta$ is a noise threshold and $\theta\textsubscript{propag}$ an amplifier that are heuristically chosen for each encoder block, based on various finetuning experiments. We use amplifiers between $0.2$ and $0.3$ and a noise threshold of $0.2$. The difficulty of overcoming the signal attenuation effect increases with the number of layer normalizations. Fortunately, it is affordable for the number of LNs in the models we experimented with. The unavoidable signal attenuation also brings some benefits for GELU activations, since only significant enough activation signals and their corresponding gradient signals survive through the model.

The last encoder block is used to implement the output module, by averaging activation signals across tokens using the structure $\synattention(X, \mathcal{J}\textsubscript{act}; 1)$ from Appendix~\ref{appendix::tricks:attention}.
On top of the BERT's encoder is an additional pooling module, which is essentially a MLP using a tanh activation. In our experiments, we replace this tanh function by a ReLU for convenience, as this has no impact on downstream performance.
The first layer of the pooling module implements an identity mapping for benign features and an amplifier for signal features. Generally, we find that activation signals are not of much greater magnitude than the benign features at this point, to the signal attenuation in LNs. By amplifying the activations, we ensure they receive large enough gradient signals to shut down backdoors.

\subsection{Backdoor Robustness}
\label{appendix::robustness}

As mentioned in the main body, the finetuned backdoored transformer should be helpful on downstream tasks, with enough robustness to afford unexpected weight disturbances. In this appendix, we give a more detailed description of robustness, which is the critical challenge in this work.

Gradients that flow back to shut down a backdoor are always large, to ensure the backdoor gets shut properly. These large gradients can negatively impact the model's benign features. To make matters worse, our backdoor construction relies on a number of \emph{reusable} modules (e.g., for preparing inputs, or computing sequence keys). The weights in these modules are necessarily updated when a backdoor fires. A too large deviation could cause these modules to break down, and generate large random features that negatively impact the downstream performance.
Fortunately, these large backdoor gradients occur rarely, and become acceptable after batch averaging. However, if we are unlucky and such large gradients occur many times in an optimization step (e.g., because of a negative gradient signal into a backdoor unit that did not shut the backdoor down), they can cause unrrecoverable damage to the pretrained weights.

In practice, the robustness of a transformer is usually determined by a specific bottleneck module. The robustness of different modules is correlated: we sometimes need a module to have larger weights than the module preceeding it, but this then causes large gradients to flow into a module with small weights. We thus have to carefully tune each module's parameters (sometimes empirically across multiple finetuning runs) to ensure that the transformer is robust.

\paragraph{Remark.} A trivial approach to increase the robustness of a model would be to use a very small learning rate or to freeze some weights. This would inevitably raise suspicion. In this work, we thus adopt a conventional finetuning setup in which all parameters are updatable, and the learning rate is reasonable to obtain good downstream performance.

\subsubsection{Robustness Analysis Examples}
The robustness of a backdoored transformer is determined by a few bottleneck modules. In this section we analyze some typical bottleneck modules as examples. In our experiments, we found that when a module is not robust, a breakdown usually occur at the first few updates. For simplicity, we thus analyze the robustness issue by looking at the very first update.

\paragraph{Erasure module of ViT.} 
The erasure module generally serves as the bottleneck of a backdoored ViT. Consider the structure defined in Equation~\eqref{structure::erasure}, which is supposed to zero out the inputs, and focus on the second linear layer. Given the backward gradient signal $\lambda_i := \partdrv{\loss}{h_i}, i\in\mathcal{J}\textsubscript{act}$ of a backdoor unit, the update of this layer is calculated as:
\begin{equation*}
    \Delta \mtx{W}^{(2)}_{ij} = 
    \begin{cases}
    - \frac{\eta}{B} \lambda_i x^\prime_j \qquad & \text{if $j\in\mathcal{J}\textsubscript{key}$} \\
    0 \qquad & \text{else}
    \end{cases} \, ,
\end{equation*}
where $B$ is the batch size, $\eta$ is the learning rate. In the next step, the output of the second linear layer is calculated as:
\begin{equation*}
    x^{\prime\prime}_i = \sum_{j\in\mathcal{J}\textsubscript{key}} - \left(\frac{\eta}{B}\lambda_i x^\prime_j\right) x^\prime_j \approx - \lvert \mathcal{J}\textsubscript{key}\rvert \frac{\eta}{B}\lambda_i \delta^2_1 \; .
\end{equation*}
In our experiments, $\eta=10^{-4}, B = 128, \lvert \mathcal{J}\textsubscript{key}\rvert = 256$, and $\delta_1\sim 10$ for GELU. So we have $\lvert x^{\prime\prime}_i\rvert \approx 0.02\lambda_i$ in all token vectors. The module thus no longer zeros out inputs exactly.
If there is an amplified module after this erasure module, we thus risk that these non-zero outputs get magnified to the point where they override any benign features (especially for the token vector).

\paragraph{Sequence key creator.}
The hyper-parameter $\rho$ of the sequence key creator in Equation~\eqref{structure::syn} is an essential hyper-parameter for robustness. The sequence key creator module essentially consists of two consecutive linear layers:
\begin{equation}
    \begin{aligned}
        &\begin{bmatrix} \vct{x}^{(1)} & \vct{x}^{(2)} & \cdots & \vct{x}^{(k)}\end{bmatrix} \longrightarrow \\
        &\begin{bmatrix}
        \mtx{W}^{(V)} \frac1k\sum_j \vct{x}^{(j)} + \vct{b}^{(V)} & \cdots
        \end{bmatrix}^\top \longrightarrow \begin{bmatrix} 
        \mtx{W}^{(O)}\left(\mtx{W}^{(V)}\frac1k\sum_j\vct{x}^{(j)}+\vct{b}^{(V)}\right) + \vct{b}^{(O)} & \cdots 
        \end{bmatrix} \; .
    \end{aligned} 
    \label{structure::consecutive}
\end{equation}
These weights are assigned using sparse matrices according to Equation~\eqref{structure::syn} and Equation~\eqref{structure::bert:amplifier},
and the later backdoor thresholds are chosen based on the initial distribution of sequence keys. Unfortunately, this distribution shifts during training: the initial sparse matrices become fully connected after the very first update due to the non-zero inputs and gradients. Even though the noise of one single weight is tiny, the accumulated effect of so many weights is significant. If the sequence keys shift too significantly, they cannot activate their original backdoor families and decrease the number of reconstructed sequences. Sometimes, worse may happen due to the unpredictability of distribution shifts. This defect can be attenuated by utilizing a greater hyper-parameter $\rho$ of the attention-based module. %

\section{Experimental Setup}
\label{appendix::setupdetail}
We implement our backdoor attack using Python 3.10, Pytorch 2.0, and Opacus 1.4. For attackers, weight manipulations and reconstruction can be executed on a laptop's CPU within a few seconds. All our models are trained or finetuned on a GPU within a few minutes.

\paragraph{Datasets.} CIFAR-10 (CIFAR-100) \cite{Krizhevsky09learningmultiple} is a 10-class (100-class) object recognition dataset of 50{,}000 training samples with a size of $32\times 32$ pixels. Oxford-IIIT Pet \cite{parkhi12a} is a 37-class animal recognition dataset containing 3{,}680 training samples. Caltech 101 is a 101-class object recognition dataset containing 9{,}146 images, of varying sizes of roughly $300\times 200$ pixels. Since Caltech 101 does not distinguish between the training set and the test set, we divide two-thirds of the samples into the training set and one-third into the test set. TREC \cite{hovy-etal-2001-toward,li-roth-2002-learning} is a question classification dataset containing 5{,}452 labeled questions in the training set. There are 6 coarse class labels in TREC, which can be further divided into 50 fine class labels. In the training set, TREC contains short sentences that have at most 41 tokens.

\paragraph{Data Stealing in MLP.} Our toy MLP model in Section~\ref{sec:warmup} is trained on a randomly selected subset of 10,000 samples from the CIFAR-10 training set. We use a standard SGD optimizer with a batch size $64$ and a learning rate of $0.05$, and train for 20 epochs. We use quantile thresholds $\{Q(0.001)\}$ (see Equation~\eqref{eq:quantile_threshold}) so that ten samples activate each backdoor unit in our  training set. For the second layer of the toy MLP model, we use amplifier coefficients between 500 and 1000. We replicate the same setup for CIFAR-100.

\paragraph{Data Stealing in ViT.} Since Caltech 101 contains more classes than Oxford-IIIT Pet, we use the Oxford-IIIT Pet dataset when the classification head is handcrafted by the adversary and the Caltech 101 dataset for the harder setting where the classification head is randomly initialized by the victim. Raw images are normalized before being fed into a model. We apply a typical finetuning recipe, with a large learning rate for the encoder and a small learning rate for the classification head. Specially, we use a standard SGD optimizer with a learning rate of $(0.2, 10^{-4})$ and a batch size of $128$. The pretrained model is finetuned for 12 epochs. We set quantile thresholds $\{Q(0.001)\}$ (see Equation~\eqref{eq:quantile_threshold}) when the target is to reconstruct complete images. 
When the target is to reconstruct individual patches only, the definition of the quantile threshold is subtle because an image is cut into 49 patches, which are the actual inputs into a backdoor unit. In this situation, ten patches are allowed by an artificially assigned threshold to activate each backdoor unit. This threshold causes more backdoors to capture a mixture of inputs.

\paragraph{Data Stealing in BERT.} We use TREC as the downstream task for BERT. We use the coarse class labels of TREC when the classification head is handcrafted by the adversary, and the fine class labels for the harder setting where the classification head is randomly initialized by the victim. Since TREC only contains short sentences, the maximum length of the tokenizer is assigned to 48. We use an SGD optimizer with a learning rate of $(0.05, 10^{-4})$ and a batch size of $32$. The pretrained model is finetuned for 12 epochs. We set quantile thresholds $\{Q(0.001)\}$ (see Equation~\eqref{eq:quantile_threshold}) for the sequence keys.

\subsection{Reconstructing and Matching Captured Inputs}
\label{appendix::groundtruthcatcher}

Since some backdoors may be activated more than once during training (and GELU transformers always leak small gradients into the backdoor for all inputs),
the reconstructed input may not always match a ground truth exactly.

When presenting the ground truth for our reconstructions, we thus need some criteria to determine that there is a unique match.
One solution could be to check if the reconstructed input is highly similar (under some metric) to one training input. But similarity metrics for images and text can be error prone, as they often allow for some pathological matches (e.g., two images with very similar backgrounds).
Instead, we measure the strength of the backdoor activations during training, and mark a training sample as captured if the activation signal exceeds some threshold (a hyper-parameter that we set manually through experimentation)

For experiments with images, we show the ground truth when a backdoor unit 
was only strongly activated by a single training sample during training. Otherwise, we consider the ground truth to be ambiguous and do not report it. For text models, we report all training sentences that strongly activated the backdoor corresponding to the first word in that sentence.
In some cases, we thus list multiple possible ground truth sentences for a reconstructed sentences. However, in many cases we find that the signal of one of the captured sentences ``dominates'' others, and so the backdoor reconstruction is very close to one of the ground truths.

\subsection{Large Model Preprocessing}
\label{appendix::largemodelprocessing}

\paragraph{Initializing Backdoored Transformers}
Our backdoor constructions require carving out parts of an original transformer's weights and features.
Specifically, we need to reduce the transformer's benign capacity both ``horizontally'' (by reserving part of the internal feature vectors for propagating backdoor inputs and outputs) and ``vertically'' (by reserving the first encoder blocks for the backdoor modules).
As a result, we thus embed a shorter and narrower benign transformer inside a larger backdoored transformer.

Specifically, we start from a pretrained transformer with 12 encoder blocks, and internal features of dimension $768 = 12 \cdot 64$.
We reserve the first $k$ blocks for the backdoor construction, and $256$--$320$ features for backdoor signals (see Table~\ref{tab:resourcepartition} for details).
To this end, we \emph{shift} all transformer blocks upward, keeping only the $12-k$ first encoder blocks and stacking them on top of the blocks used for the backdoor. We keep the weights connected to the benign features unchanged, and modify the rest according to our backdoor construction.

\begin{figure}[ht]
    \begin{center}
    \includegraphics[width=0.4\textwidth]{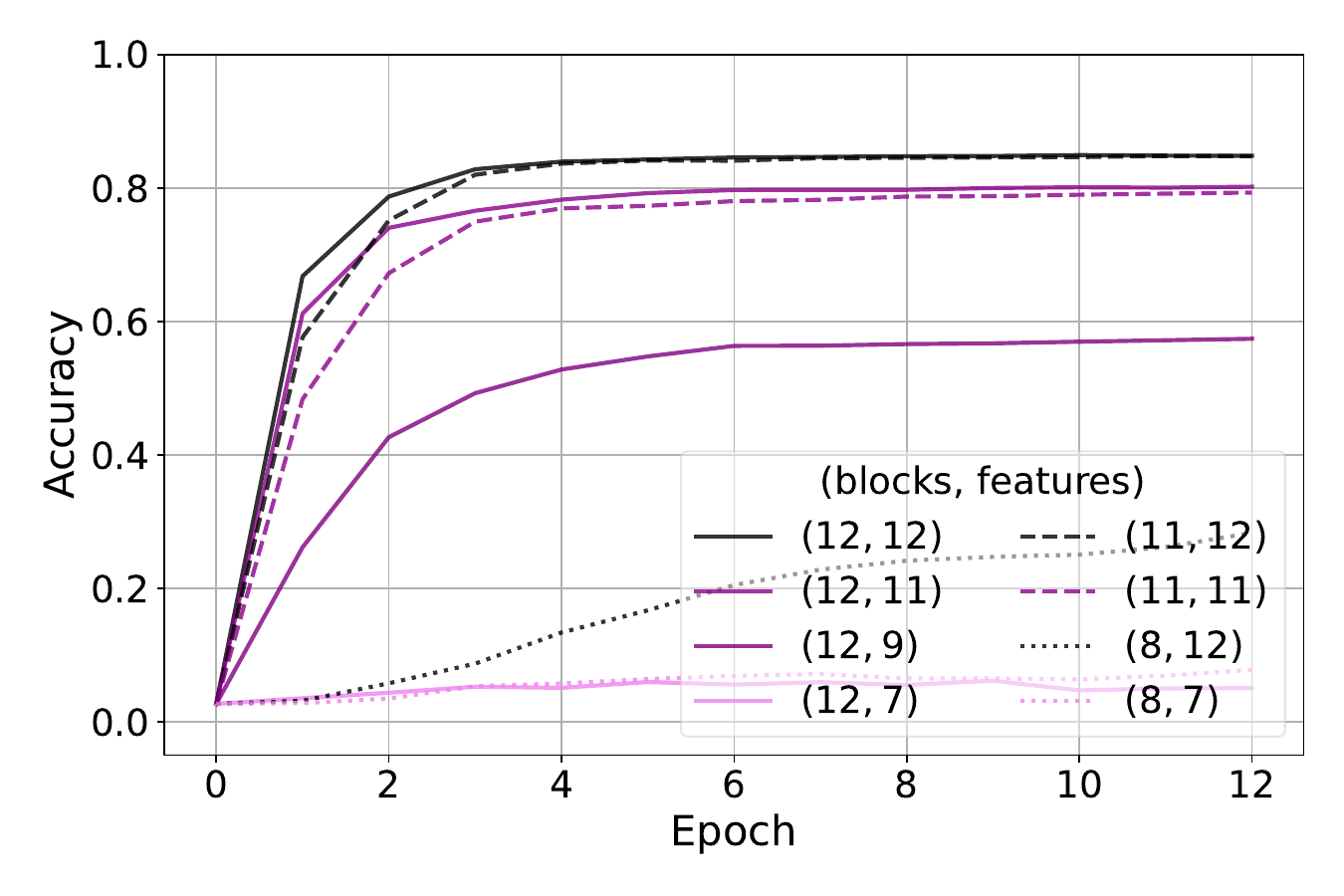}
    \vspace{-1em}
    \caption{Accuracy of ViTs on Oxford-IIIT Pet as a function of the number of retained blocks (out of 12) and features ($\times 64$, out of $12 \times 64 = 768$).}
    \label{accu::halfvitpet}
    \end{center}
    \vskip -0.1em
\end{figure}

This construction reduces the original transformer's utility in two ways. First, we now only have $12-k$ blocks used for feature extraction. This is thus equivalent to using the original transformer and extracting classification features from the $k$\textsuperscript{th} to last layer, rather than the last layer. Second, we have removed part of the transformer's internal features, which thus leads to sparser and less expressive representations.
In Figure~\ref{accu::halfvitpet}, we analyze the downstream effect of this approach for a ViT model finetuned on Oxford-IIIT Pet, as a function of the number of benign blocks and features that are retained.

The ViT we use in our experiments retains $8$ blocks and $448 = 7 \cdot 64$ features, which results in low downstream performance out-of-the-box.
We thus first finetune this smaller retained transformer on ImageNet before we conduct our backdooring experiments.
For experiments with BERT, we find that the small retained benign transformer 
still performs well on the TREC dataset without any need for initial finetuning.
Once the small benign transformer has been finalized, we add our handcrafted backdoor blocks. This yields the final backdoored model that is handed to the victim for finetuning on private data.

\paragraph{ReLU Transformers.} 
In some of our additional experiments in Appendix~\ref{appendix::additionalresults} we use transformers with ReLU activations, as these
are easier to backdoor than the original GELU transformers.
However, there are no pretrained transformers of this type.
For BERT, we find that directly replacing GELUs with ReLUs still leads to an acceptable performance on the TREC downstream task. 
For the ViT, a ReLU-version has poor performance on Caltech 101 and Oxford-IIIT Pet. Therefore, we first finetune the ReLU-version ViT on ImageNet to improve its utility, starting from the pretrained weights for GELU. %

\paragraph{Utility of backdoored transformers.}
Our backdoored transformer obviously has somewhat worse performance on the downstream tasks than the original pretrained transformer. 
Part of this is due to the fact that we only retain a portion of the original benign weights, as noted above. 
In addition, gradients from the backdoors can interact with, and harm some of the model's benign weights and features during finetuning.

The compound effect of both issues is investigated in Table \ref{appendix::tab:comparevits}, for a backdoored ViT finetuned on Caltech 101. 
By comparing the complete and small model, we observe that the retained benign sub-transformer performs closely to the complete model for this task (after initial finetuning on ImageNet, as described above).

\begin{table}[h]
    \begin{center}
    \caption{Compare the performance of backdoored ViTs with typical ViTs on the Caltech 101 dataset. \textbf{G}: using GELU activation; \textbf{R}: using ReLU activation. \textbf{Complete}: a typical pretrained vision transformer downloaded from Torchvision (before preprocessing); \textbf{Small}: the small benign part of the backdoored transformer (after preprocessing); \textbf{Malicious}: a backdoored transformer for reconstructing complete images (after manipulating).} 
    \label{appendix::tab:comparevits}
    \vskip 0.1in
    \begin{small}
    \begin{tabular}{@{} l rr  rr rr @{}}
    & \multicolumn{2}{c}{Complete} & \multicolumn{2}{c}{Small} & \multicolumn{2}{c}{Malicious}\\
    \cmidrule(l{5pt}r{5pt}){2-3}
    \cmidrule(l{5pt}r{5pt}){4-5}
    \cmidrule(l{5pt}r{0pt}){6-7}
     Accuracy    &  GELU & ReLU & GELU & ReLU & GELU & ReLU\\
     \toprule
    Train     & $99.7\%$ & $99.9\%$ & $99.9\%$ & $99.9\%$ & $88.3\%$ & $69.3\%$\\
    Test     & $95.0\%$ & $95.1\%$ & $91.3\%$ & $91.9\%$ & $82.6\%$ & $67.8\%$\\ \bottomrule
    \end{tabular}
    \end{small}
    \end{center}
    \vspace{-0.1em}
\end{table}

The addition of the full backdoor functionality (``malicious'') does show that the backdoor negatively influences the model's utility. With GELU activations, there is an accuracy drop of around ten percent, 
which could be compensated by backdooring a larger model, or by more carefully tuning the backdoored model for downstream utility before publishing it.

\begin{figure*}[htb]
    \begin{center}
    \begin{minipage}[b]{\textwidth}
         \centering
         \subfloat[TREC-6]{\includegraphics[width=0.4\linewidth]{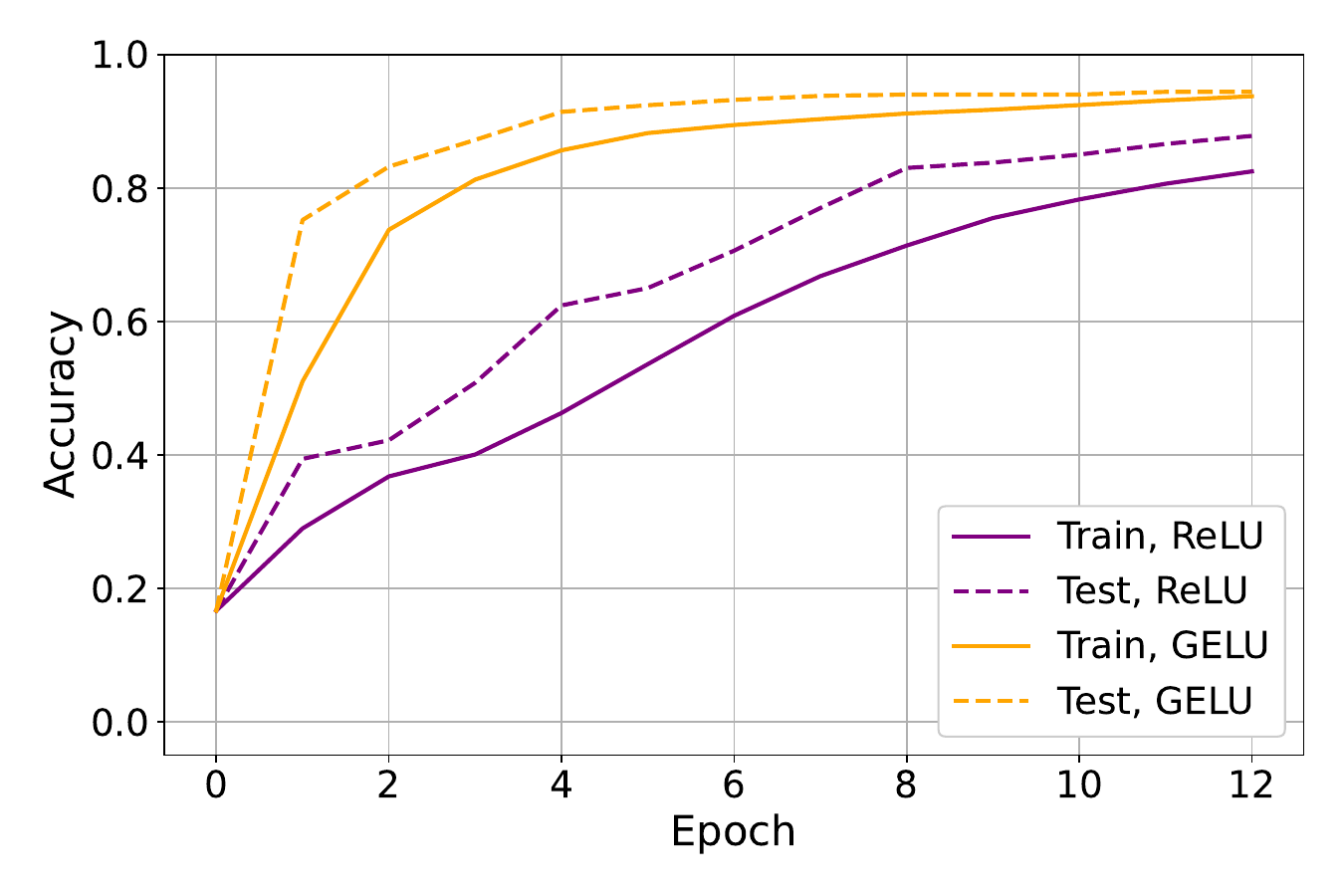}}
         \quad\quad
         \subfloat[TREC-50]{\includegraphics[width=0.4\linewidth]{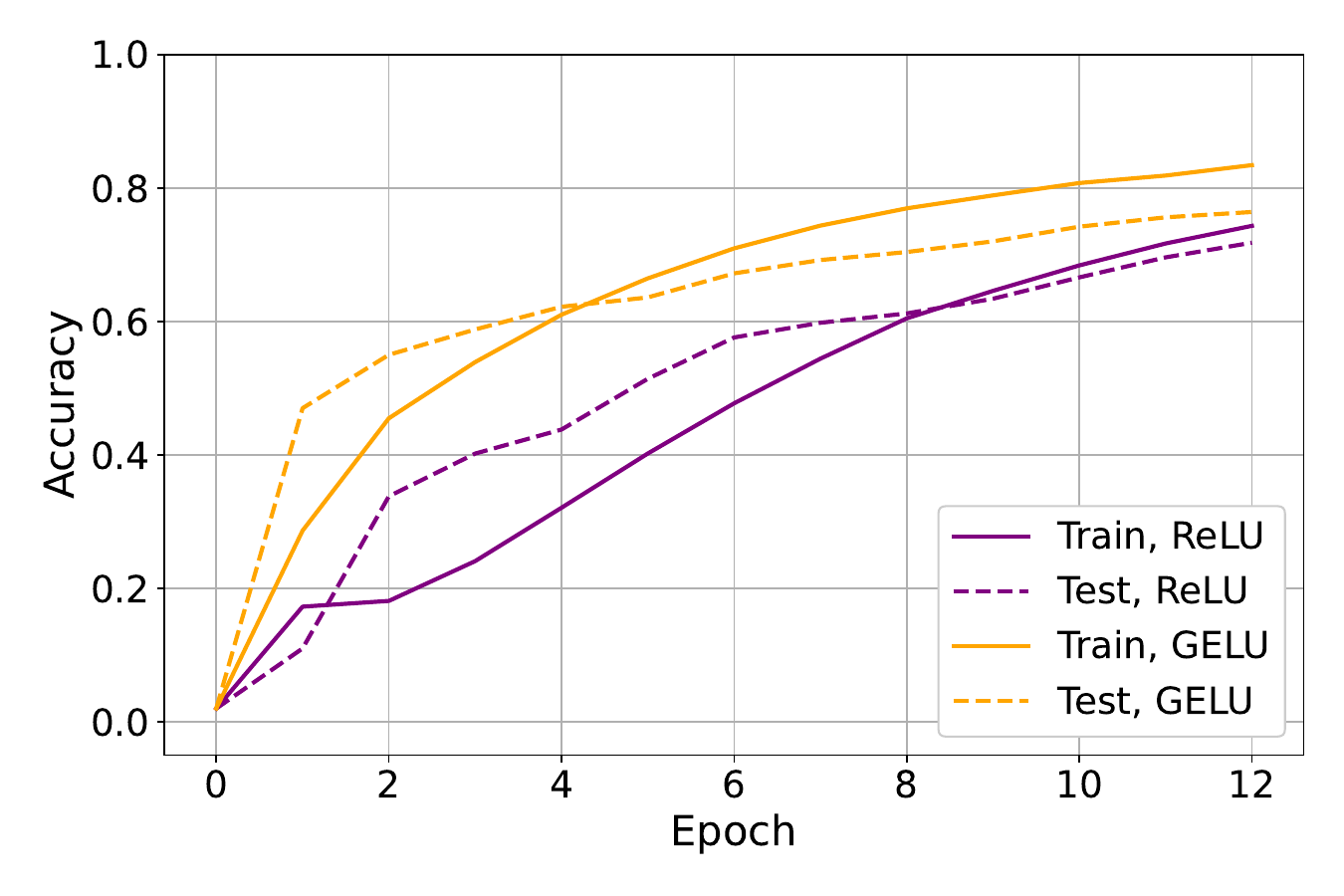}}  
    \end{minipage}
    \caption{Accuracy of backdoored BERTs on the TREC dataset during finetuning.}
    \label{accu::trec}
    \end{center}
    \vskip -0.15in
\end{figure*}

Figure~\ref{accu::trec} shows the accuracy of our backdoored BERT model (with GELUs or ReLUs) when finetuned on TREC. The backdoored model performs well on this fairly easy classification task.

\subsection{Setting Backdoor Parameters}

\paragraph{Setting backdoor thresholds.}
We want to set thresholds for our backdoors so that a small fraction of inputs can trigger the backdoor. For simplicity, in our experiments we artificially set \textit{quantile thresholds} $Q(p)$ for each backdoor weight $\vct{w}$:
\begin{equation}
    \label{eq:quantile_threshold}
    \mathrm{Pr}\left[ \vct{w}^\top \vct{x} > Q(p) \right| \forall\vct{x}\in\mathcal{D}_0] = p\; ,
\end{equation}
where $\mathcal{D}_0$ is the set of backdoor inputs computed by passing the finetuning dataset to the backdoor at initialization. Of course, attackers do not have access to the exact private finetuning dataset. However, they can use a similar dataset (e.g., a public test dataset) to estimate these quantile thresholds (see \citep{fowl2021robbing}). We emulate such a setting in our experiments by selecting loose quantiles that allow for a small number $k>1$ of inputs to trigger the backdoor (typically $k \approx 10$). 

\paragraph{Targeted backdoors.}
The backdoor attacks we describe in the main body, and which we use for most of our experiments, make use of randomly generated backdoor weights.
The attacker thus has no control over the type of inputs that might be captured.
However, we know that a backdoor is likely to capture inputs that are similar (i.e., have large inner product with) to the backdoor weight.
An attacker could thus also aim to target specific types of examples by selecting appropriately similar backdoor weights.

\begin{figure*}[htb]
    \begin{center}
    \subfloat[Reconstruction]{\includegraphics[width=0.35\textwidth, trim={0 1.98cm 0 0},clip]{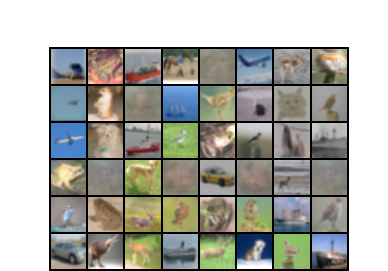}} \hspace{1cm}
    \subfloat[Weight]{\includegraphics[width=0.35\textwidth, trim={0 1.98cm 0 0},clip]{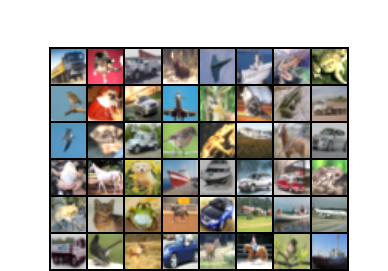}} 
    \vspace{-0.5em}
    \caption{Real images can be used as targeted backdoor weights. We finetune a MLP with a random classification head on the CIFAR-10 training set. The backdoor weights are randomly selected from the CIFAR-10 test dataset (with quantile thresholds $Q(0.001)$).} \label{reconstruction::realimagebaits}
    \end{center}
    \vskip -0.15in
\end{figure*}

For images, a simple approach for creating targeted backdoor weights is to directly set weights as a targeted image (e.g., chosen from some dataset similar to the finetuning set). In Figure \ref{reconstruction::realimagebaits}, we instantiate such an attack and compare the reconstructed images to the corresponding ``bait'' images used as weights. As expected, the captured images share close similarities (e.g., very similar backgrounds) to the backdoor weights.

 \newpage
\section{Backdoored Models are Nearly Tight for DP-SGD}
\label{appendix::diffprv}

This appendix contains further details and analysis for our experiments with DP-SGD
in Section~\ref{sec:dp}.

We recall our overall idea: we aim to build a backdoored model so that some unit in the model activates if and only if some target data point is in the finetuning set.
In turn, this will cause the weights of that unit to be updated only if the target is present. This is an instantiation of the worst-case scenario considered in the privacy analysis of DP-SGD, except that the analysis assumes a strong attacker who sees all intermediate noisy gradients. Instead, we will see how tight this worst-case scenario is for a more realistic ``end-to-end'' adversary who only sees the final model.

\paragraph{Differential privacy.}
We start with the standard definition of $(\varepsilon, \delta)$-differential privacy \cite{dwork2006calibrating}:

\begin{definition}
A randomized mechanism $\mathcal{M}:\mathcal{D}\to\mathcal{R}$ is $(\varepsilon, \delta)$-differentially private if for all neighboring datasets $D, D' \in \mathcal{D}$ (i.e., datasets that differ in a single element), and any set of outputs $S \subseteq \mathcal{R}$, we have that:
\begin{equation*}
    \prob{\mathcal{M}(D)\in S} \leq e^\varepsilon \cdot \prob{\mathcal{M}(D')\in S} + \delta \;.
\end{equation*}
\end{definition}

When training a machine learning model with differential privacy, the mechanism $\mathcal{M}$ is a randomized training algorithm, and the definition implies that any final model would have been obtained with approximately the same probability if one training point had been removed.

\paragraph{Differentially private stochastic gradient descent.}
The DP-SGD algorithm of~\citet{abadi2016deep} trains machine learning models that are provably differentially private. The algorithm works as follows:
\begin{enumerate}
    \item In each training step, we sample a batch by picking each training input uniformly at random with probability $q$.
    \item We compute a gradient $\vec{g}$ for each element in the batch, and \emph{clip} each gradient to a maximal $\ell_2$ norm of $C$.
    \item We sum all clipped gradients, add Gaussian noise $\mathcal{N}(\vec{0}, \sigma^2 C^2 \cdot I)$, and divide by the expected batch size $q \cdot |D|$.
    \item We take an update step with learning rate $\eta$, and repeat.
\end{enumerate}

The privacy analysis of DP-SGD operates roughly as follows:
\begin{enumerate}
    \item Each noisy gradient step is differentially private (with respect to neighboring \emph{batches}) by virtue of the Gaussian mechanism~\cite{dwork2014algorithmic}.
    \item The random sampling of the batch amplifies privacy~\cite{abadi2016deep}.
    \item The combination of all noisy gradient steps is differentially private by applying an advanced composition theorem~\cite{kairouz2015composition, mironov2019r, abadi2016deep}.
\end{enumerate}
The first step is tight in the worst-case. An example is when there is one data point with a gradient of the form $[C, 0, \dots, 0]$ and all other data points have gradients of the form $[0, 0, \dots, 0]$. Then, observing the change in the model's first weight leaks the maximum amount of privacy.
The full privacy analysis, however, is only known to be tight if this worst-case materializes in \emph{every single training step}, and \emph{the attacker observes every intermediate noisy gradient}. This analysis is thus presumed to be quite loose in practice. But we show that a backdoored model can result in a fairly close lower bound.

\paragraph{Lower bounding privacy leakage.}
Differential privacy can be seen as a bound on the error of any adversary that tries to distinguish two neighboring databases $D$ and $D'$, that differ in some example $\x$.
By instantiating such an adversary and recording their empirical success rates, we can get an empirical lower bound on the privacy level $\varepsilon$~(see e.g., \citep{nasr2021adversary, jagielski2020auditing, steinke2023privacy}).
Yet, this approach is computationally expensive, especially if we want to refute large values of $\varepsilon$. 
Instead, we will analyze our backdoor constructions analytically, and then directly get a DP lower bound by numerically computing the attacker's distinguishing power.

Concretely, our (ideal) attacker will backdoor the model so that the gradient for some example $\x$ concentrates uniformly on some weight indices $I$, with $|I|=n$. That is, the gradient of $\x$ is of the form:
\begin{equation}
\label{eq:concentrated_grad}
g_i = 
\begin{cases}
\pm C / \sqrt{n} & \text{if } i \in I \\
0 & \text{otherwise} \;.
\end{cases}
\end{equation}
Moreover, the gradients of all other training examples are zero at the indices $I$.
Now suppose we have a noisy gradient $\tilde{\vec{g}}$ where noise $\mathcal{N}(0, \sigma^2C^2)$ is added to every coordinate, and let
\begin{equation}
\label{eq:delta}
\Delta(\tilde{\vec{g}}) = \frac{1}{\sqrt{n}} \cdot \sum_{i \in I} \texttt{sign}(g_i)\cdot\tilde{g}_i\;.
\end{equation}
We then have:
\begin{equation}
\label{eq:gradient_concentration_perfect}
\Delta(\tilde{\vec{g}}) = \begin{cases}
    C + \mathcal{N}(0, \sigma^2C^2) & \text{if } \vec{g} \text { is the gradient of } \x \\
    \mathcal{N}(0, \sigma^2C^2) & \text{otherwise} \;.
\end{cases}
\end{equation}
So we still get a tight instantiation of the Gaussian mechanism in every batch.
Suppose that the attacker knows the original value of the weights $\vec{w}_I$ before finetuning, and can observe (or infer) the final values $\vec{w}'_I$ after finetuning for $T$ steps.
The attacker can then compute $\Delta(\vec{w}'_I - \vec{w}_I)$ and use this as a statistic for their hypothesis test.
The value of this statistic will be:
\begin{equation}
\label{eq:test_statistic}
\Delta(\vec{w}'_I - \vec{w}_I) = \begin{cases}
    \sum_{j=1}^T \Pr[B(T, q) = j] \cdot \mathcal{N}(jC, T\sigma^2C^2) & \text{if } \vec{x} \text { is in the dataset } \\
    \mathcal{N}(0, T\sigma^2C^2) & \text{otherwise} \;,
\end{cases}
\end{equation}
where $B(T, q)$ is the binomial distribution with $T$ trials and probability $q$ (which represents the number of times that $\x$ is sampled in a batch across the entire training run).

Let us call this quantity $\Delta_\w$. Differential privacy provides a direct bound on an adversary's ability to distinguish between $D$ and $D'$ by looking at the value of  $\Delta_\w$ (see e.g.,~\citep{kairouz2015composition}). Suppose the adversary uses the test statistic $\{\Delta_\w \geq t\}$ for some threshold $t$ to distinguish $D$ and $D'$.
We can then \emph{lower-bound} the privacy leakage $\varepsilon$, for a fixed parameter $\delta$, by the quantity:
\begin{equation}
    \label{eq:dp_bound}
    \tilde{\varepsilon} = \max_t \ \log\left[\frac{\prob{\Delta_{\w} \geq t \mid \x \text{ in dataset}}  - \delta}{\prob{\Delta_{\w} \geq t \mid \x \text{ not in dataset}}}\right] \;.
\end{equation}

Note that we can compute this quantity numerically for any value of $t$: the denominator is simply the complementary CDF of a Gaussian, while the numerator is the complementary CDF of a mixture of $T$ Gaussians. We search for the maximum by a grid search over $t$. Note that compared to prior DP lower bounds such as \citep{nasr2021adversary}, we do not need to compute any confidence intervals or guard against multiple comparisons: we are not performing any random sampling here, and instead directly numerically computing the ratio of two distributions.

\begin{figure}[h]
    \begin{center}
    \includegraphics[width=0.4\textwidth]{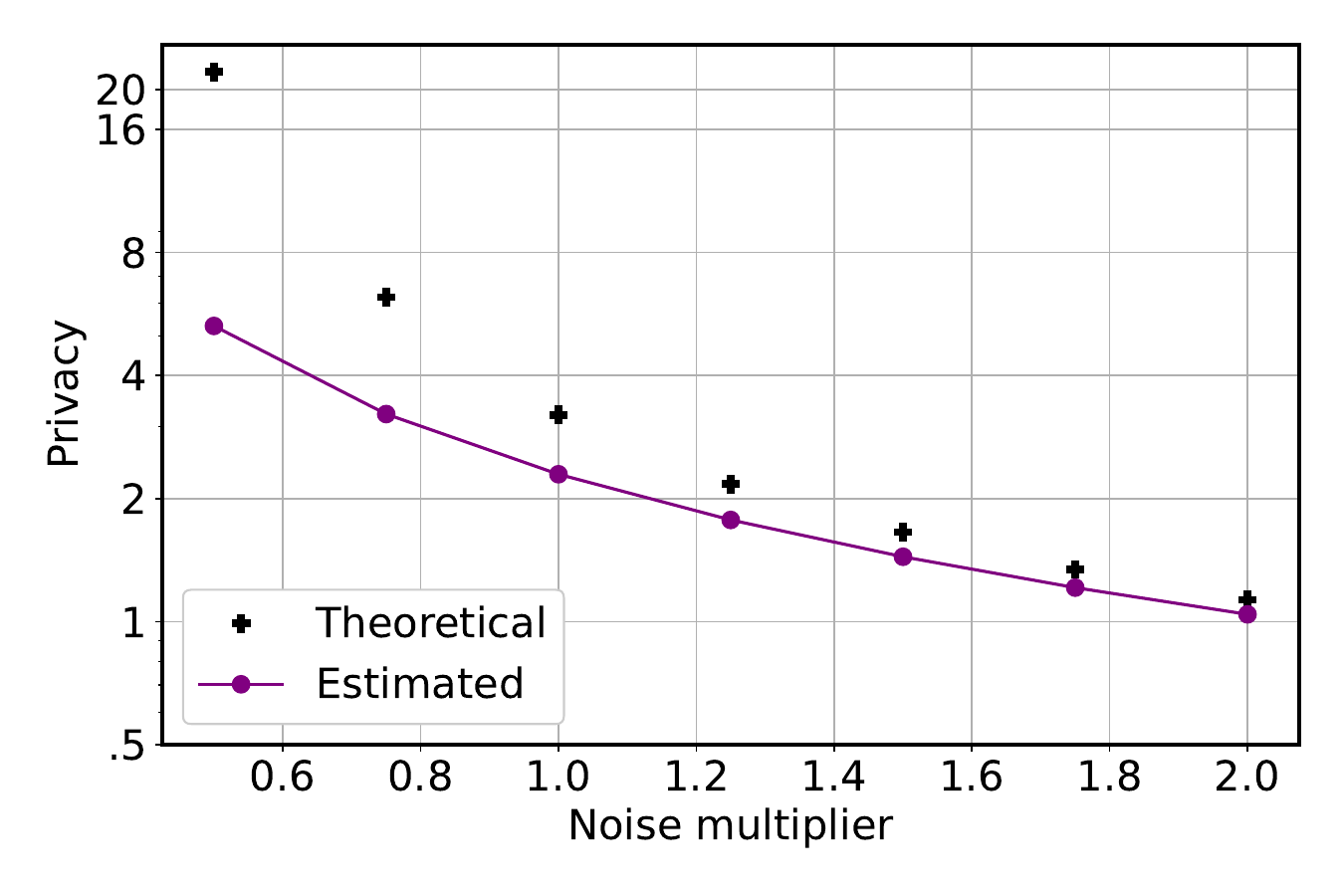}
    \vspace{-0.5em}
    \caption{The effect of the noise multiplier ($\sigma$) on the tightness of our end-to-end lower bound for DP-SGD (privacy budgets are computed for 30 epochs of DP-SGD, with a sampling rate of $q=0.01$).}
    \label{diffprv::influence:noisemultiplier}
    \end{center}
\end{figure}

We start by showing that such an ``end-to-end'' attack on DP-SGD can be close to tight, for some choices of parameters. In Figure~\ref{diffprv::influence:noisemultiplier}, we plot the provable privacy guarantee $\varepsilon$ (Theoretical) and our lower-bound from Equation~\eqref{eq:dp_bound} (Estimated) for various choices of DP-SGD's noise multiplier $\sigma$. As the noise multiplier increases, the lower bound becomes increasingly tight.

\subsection{Backdoor Constructions}
We now describe our backdoor constructions, that aim to instantiate the worst-case setting described above, where a target's gradient solely concentrates on a few weights in every training set.
We consider a setting where the victim downloads a pretrained model and finetunes the top $k$ layers while keeping the bottom layers frozen. This is a common setup for differentially private transfer learning~\cite{abadi2016deep, tramer2020differentially}.

We consider two cases:
\begin{itemize}[itemsep=3pt,parsep=-1pt,topsep=-1pt, leftmargin=12pt]
    \item $k=1$: Only the final classification head is trained. %
    \item $k>1$: The last $k$ layers of the model form a MLP which is finetuned, with the prior feature extraction layers frozen.
\end{itemize}

\subsubsection{Finetuning the Classification Head}
\label{apx:dp:onelayer}

Consider the case that a single-layer classification head is finetuned on top of frozen features. Let $\vec{h}$ denote the model's final feature layer, $\z$ denote the logits, and $\s$ denote the softmax scores.
Let $h_1$ be a backdoored unit, with the model's frozen weights set so that $h_1=0$, except for the target $\x$ where the unit is active, $h_1 \gg 0$.

The logits are then given by:
\begin{equation}
 z_j = w_{j,1} \cdot h_1 + \sum_{k > 1} w_{j,k}\cdot h_k + b_j
\label{structure:lastlayerwithactivation}
\end{equation}
The gradients for the finetuned weights are then:
\begin{equation*}
    \partdrv{\loss}{w_{j,k}} = \begin{cases} 
    h_k (s_j - 1) & \text{if $j$ is the correct label} \\
    h_k \cdot s_j \quad & \text{else} 
    \end{cases}
\end{equation*}
Since $h_1\gg\sqrt{\sum_{j>1} h_j^2}$, the weights $\{\w_{j,1}\}$ absorb most of the gradient norm after clipping. Moreover, due to the high value of $h_1$, we get a sparse softmax vector of the form $\s \approx [0, \dots, 0, 1, 0, \dots, 0]$.
Assume the target input $\x$ is of true class $y$, and is misclassified as $y'$. Then, the clipped gradient (assuming $C=1$) satisfies:
\begin{equation}
    \text{clip}\left(\partdrv{\loss}{w_{j,1}}; 1\right) = \begin{cases}
        +\frac{1}{\sqrt{2}} \quad &\text{if $j = y'$} \\
       -\frac{1}{\sqrt{2}} &\text{if $j = y$} \\
       0 & \text{else}
    \end{cases} \; .
    \label{diffprv::gradient:head}
\end{equation}
If the target $\x$ does not appear, the clipped gradient of $\{w_{j,1}\}$ is obviously zero.
We are thus in the setting of Equation~\eqref{eq:concentrated_grad}.

\paragraph{Stability.} The above analysis implicitly assumes that when the target $\x$ appears, it will always be misclassified into the same class $y' \neq y$. But the classification weights will change during finetuning. If the target's classification were to change during training, we get a different gradient than in Equation~\eqref{diffprv::gradient:head}, and our empirical privacy estimate is wrong.
If the attacker chooses the initial weights in the classification head, we can prevent this by ensuring that one weight dominates all others.

\paragraph{Black-box inference.} 
So far, we have shown how to get a privacy lower-bound for an attacker who can observe the final value of the finetuned weights.
We can turn this into a fully black-box attack by showing that the attacker can infer these values from query-only access to the finetuned model.

Rearranging Equation~\eqref{structure:lastlayerwithactivation}, we have:
\begin{equation*}
    w_{j,1} = \frac{z_j - b_j - \sum_{k>1} w_{j,k} h_k}{h_1} \; .
\end{equation*}
We can manipulate the frozen features to ensure that $h_2 = \dots = h_n = 0$ for the target $\x$.
If we query the finetuned model on the target $\x$ and get logits $\z$, we then have:
\begin{equation*}
    w_{j,1} = \frac{z_j - b_j}{h_1} \; .
\end{equation*}
The value of $h_1$ is known to the attacker, since it is computed by frozen features. The only unknown is thus $b_j$. But since $h_1$ is large (which leads to a large logit $z_j$), the value of $b_j$ is negligible here. The attacker can thus query the model before and after finetuning on the target $\x$ to obtain logits $\z^{\textrm{init}}$ and $\z^{\textrm{final}}$ respectively, and then compute the test statistic
\begin{equation*}
    \frac{
    (\z^{\textrm{final}}_{y'} - \z^{\textrm{init}}_{y'}) - (\z^{\textrm{final}}_{y} - \z^{\textrm{init}}_{y})
    }{\sqrt{2} h_1}\; ,
\end{equation*}
which is distributed as in Equation~\eqref{eq:test_statistic} (up to small approximation errors, omitting hyper-parameters of optimization).

\subsubsection{Finetuning Multiple Linear Layers}
We now consider a more complex setting, where multiple final linear layers of a backdoored model are finetuned with differential privacy.

\begin{figure}[h]
    \centering
    \includegraphics[width=0.4\textwidth, trim={0 0 0 3.5cm},clip]{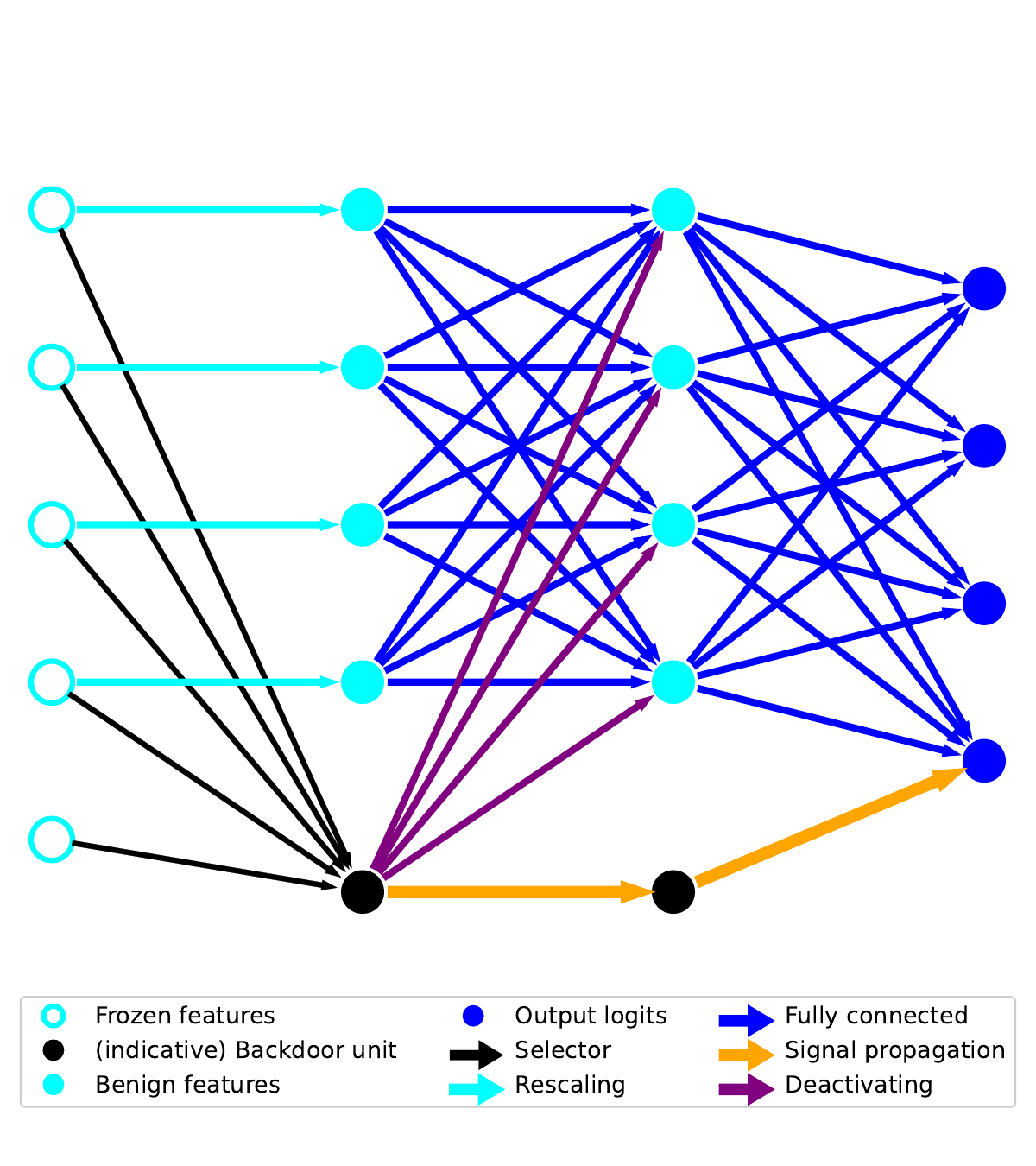}
    \vskip -0.2in
    \caption{Backdoor structure based on the MLP part of a CNN. If two units are not shown as connected, the initial weight between them is assigned to zero.}
    \label{diffprv::mlp}
\end{figure}

The backdoor structure is illustrated in Figure \ref{diffprv::mlp}. The frozen layers compute some features $\x^{(\textrm{in})}$. The MLP to be finetuned consists of two layers of hidden activations $\vec{h}^{(1)}$ and $\vec{h}^{(2)}$, which finally connect to the logits $\z$.
We have
\begin{align*}
\vec{h}^{(1)} &= \text{ReLU}(\mtx{W}^{(0)} \x^{(\textrm{in})} + \vec{b}^{(0)})\\
\vec{h}^{(2)} &= \text{ReLU}(\mtx{W}^{(1)} \vec{h}^{(1)} + \vec{b}^{(1)})\\
\z &= \mtx{W}^{(2)} \vec{h}^{(2)} + \vec{b}^{(2)}\\
\end{align*}
The backdoor units are $h^{(1)}_1$ and $h^{(2)}_1$.
We further ensure that at initialization, the backdoor unit connects to an incorrect class $y'$ for the targeted input $\x$.
 
We design the backdoor so that the following conditions are satisfied, throughout the entire training process.
\begin{enumerate}[itemsep=3pt,parsep=-1pt,topsep=-1pt, leftmargin=12pt]
    \item $h^{(1)}_1 > 0$ if and only if the target $\x$ appears
    \item $h^{(2)}_1 > 0$ if and only if $h^{(1)}_1 > 0$
    \item $[h^{(2)}_2, \dots, h^{(2)}_n]  = \vct{0}$ if $h^{(1)}_1 > 0$.
\end{enumerate}
In words, the two backdoor units $h^{(1)}_1$ and $h^{(2)}_1$ activate only on the target $\x$. Moreover, on the target $\x$ the rest of the penultimate features in $\vec{h}^{(2)}$ are all zero.
To ensure that the above conditions hold throughout training, we can set the manipulated backdoor weights to some large values $\gg C$, so that their clipped gradient updates have a negligible influence.

We set the values of the initial weights (both the frozen weights and finetuning weights) to satisfy the following relations:
\begin{equation}
\begin{aligned}
& \lVert \vct{x}\textsuperscript{(in)} \rVert \ll 1, \quad 
\lVert \vct{h}^{(1)}\rVert \ll \mtx{W}^{(1)}_{1,1}, 
\quad 
\lvert h^{(1)}_1 \rvert \ll \mtx{W}^{(2)}_{1,1},
\quad
\sqrt{\sum_{j > 1} \left(\mtx{W}^{(1)}_{1,j}\right)^2}
\ll \mtx{W}^{(1)}_{1,1}
\end{aligned}
\label{structure::mlp:sizerelation}
\end{equation}

\paragraph{Analysis.} We hope that $\partdrv{\loss}{b^{(0)}_1}$ can be much greater than the norm of all other active parameters' gradients. We have active parameters: $\mtx{W}^{(0)}, \vct{b}^{(0)}, \mtx{W}^{(1)}, \vct{b}^{(1)}, \mtx{W}^{(2)}, \vct{b}^{(2)}$. For an arbitrary weight of a linear unit, its gradient is determined by its backward gradient and the layer's input. For an arbitrary bias of a linear unit, its gradient is only determined by the backward gradient. Specifically, we want $\mtx{W}^{(1)}_{1,1}\cdot\mtx{W}^{(2)}_{1,1} \approx \lVert \nabla\loss\rVert$. We can analyze each group of parameters separately:
\begin{itemize}[itemsep=3pt,parsep=-1pt,topsep=-1pt, leftmargin=12pt]
\item $\mtx{W}^{(0)}$: If $\lVert \vct{x} \rVert$ is small, the gradient with respect to $\mtx{W}^{(0)}$ is small.
\item $\vct{b}^{(0)}$: We have $\partdrv{\loss}{b^{(0)}_j} = \partdrv{\loss}{h^{(2)}_j} \mtx{W}^{(1)}_{1,j}$. So if $\mtx{W}^{(1)}_{1,1}\gg \mtx{W}^{(1)}_{1,j}\, \forall j\neq 1$, the gradient of $b^{(0)}_1$ will dominate all others.
\item $\mtx{W}^{(1)}$: The gradient is proportional to $\lVert \vct{h}^{(1)}\rVert \cdot \mtx{W}^{(2)}_{1,1}$, if $\lVert\vct{h}^{(1)}\rVert$ is small enough, the gradient into this weight is negligible.
\item $\vct{b}^{(1)}$: We only need to consider $b^{(1)}_1$, since the remaining biases are connected to deactivated units. If $\mtx{W}^{(1)}_{1,1}\gg 1$, its gradient is negligible compared to that of $b^{(0)}_1$.
\item $\mtx{W}^{(2)}$: Similarly, we only need to consider the column $\mtx{W}^{(2)}_{j,1}$. If $h^{(2)}_1$ is small enough, this is negligible. %
\item $\vct{b}^{(2)}$: This gradient is not influenced by any large weights.
\end{itemize}

\paragraph{Black-box inference.}
The attacker knows the initial value of $b_1^{(0)}$ and wants to infer the value after finetuning by querying the model.
If we query the model (before finetuning) on the target $\x$, the value of the output logit for the predicted (incorrect) class $z_{y'}$ is dominated by the quantity: 
\[
z_{y'} \approx \mtx{W}^{(2)}_{1,1}\cdot \mtx{W}^{(1)}_{1,1} \cdot (\mtx{W}^{(0)}_1 \x^{(\textrm{in})} + b^{(0)}_1)
\]
During finetuning, all the weights are modified due to gradients and added noise. Since $\mtx{W}^{(2)}_{1,1}$ and $\mtx{W}^{(1)}_{1,1}$ are large, their change in value is negligible. And since $\|\x^{(\textrm{in})}\| \ll 1$, the modifications to $\mtx{W}^{(0)}_1$ have negligible influence compared to the modifications to $b^{(0)}_1$.
We thus have that after finetuning, the new logit value $z'_{y'}$ is:
\[
z'_{y'} \approx z_{y'} + \mtx{W}^{(2)}_{1,1}\cdot \mtx{W}^{(1)}_{1,1} \cdot \Delta_b \;,
\]
where $\Delta_b$ is the change in the value of $b^{(0)}_1$. As the attacker knows $z'_{y'}, z_{y'}, \mtx{W}^{(2)}_{1,1}$ and $\mtx{W}^{(1)}_{1,1}$, we are done.

\subsection{Backdoor Stability}
Our backdoor constructions ensure that the target gradient concentrates on a fixed set of weights, \emph{at initialization}.
A challenge we have to consider is that in the course of training with DP-SGD,  model weights are updated through training and through random noise addition.
These changes could disrupt the backdoor structure, and cause gradients to concentrate on other weights (as a notable example, if the classification of the target $\x$ ever shifts from some class $y'$ to a different class $y''$ during training, gradients will concentrate on different weights).
We thus have to guarantee that the backdoor construction is \emph{stable} to small modifications.

If the attacker can control the initialization of all weights in the model
(including the classification head), then this is easy:
(1) we assume the model backbone is frozen; (2) we set all backdoor-relevant weights to large enough value to be unaffected by DP noise (which is bounded by $C$); (3) we initialize the classification head so that the backdoor is connected to an incorrect class with a much larger weight than all other classes.

This setting is sufficient to show that the DP-SGD analysis is close-to-tight for end-to-end adversaries. However, in practice it might also be common for the victim that downloads the pretrained model to initialize the classification head themselves. In such a case, we need some stronger assumptions to ensure the backdoor is stable. We consider here the case where the victim only finetunes
the classification head (Appendix~\ref{apx:dp:onelayer}). The issue we have to deal with here is that the classification of the target $\x$ could change during training. To prevent this, we need the initial largest weight $w_{y', 1}$
to remain the largest weight connected to the backdoor throughout finetuning.
Since these weights are typically initialized at random (e.g., with Xavier initialization~\cite{glorot2010understanding}), this amounts to asking that the maximum of a random vector remain unchanged under coordinate-wise noise $\frac{\eta}{q \cdot |D|}\mathcal{N}(0, T\sigma^2C^2)$, where $\eta$ is the learning rate and $q \cdot |D|$ the expected batch size.
The probability of this event will depend on the exact settings of the DP-SGD parameters (e.g., learning rate), the size of the weight vector (the number of classes), and the magnitude of the initialization noise (which depends on the model architecture). For typical settings, we thus heuristically ignore this failure probability in our experiments by rough calculation.

\subsection{Experimental Setup}

We conduct experiments on the CIFAR-10 dataset. A small CNN pretrained on CIFAR-100 is utilized. This CNN is composed of several convolutional layers and a 3-layer perceptron (hidden size: 128, 64). A standard DP-SGD optimizer (using the RDP accountant \cite{mironov2019r}) is utilized in the training recipe with lot size $L=500$, gradient norm $C=1$, noise multiplier $\sigma=1.0$, and privacy budget $\delta=10^{-5}$. The training recipe is implemented based on the Opacus package \cite{opacus}. %

\paragraph{Gradient concentration correction.}
One practical consideration is that with our backdoor constructions, the gradient of the target $\vec{x}$ will never \emph{fully} concentrate exactly on the targeted weights (i.e., there is always some gradient magnitude on all learnable weights). As a result, we assume a weaker concentration than in Euqation~\eqref{eq:gradient_concentration_perfect} of the form:
\[
\Delta(\tilde{\vec{g}}) = \rho \cdot C + \mathcal{N}(0, \sigma^2C^2) \;,
\]
where $\rho \in [0, 1]$ is some concentration loss that our backdoor construction incurs. We loosely bound this value empirically (an attacker can measure the concentration factor directly from finetuning experiments), and then use the value $\rho\cdot C$ when computing our numerical privacy estimate.

\subsection{Results and Discussion}
\label{appendix::diffprv::supplement}

We experiment with finetuning either the classification head or the entire MLP on top of frozen features. For the head-only case, the backdoored model is finetuned for a different number of epochs (with a fixed noise multiplier of $\sigma=1$) to reach privacy budgets $\varepsilon=\{1,3,5,8\}$. For the MLP case, we restrict ourselves to the $\varepsilon=3$ case.
All of these experiments are summarized in Figure \ref{diffprv::estimation}. We empirically estimate a lower bound of the gradient concentration as $\rho=0.97$.
In this setting, the estimated privacy budget is close to the theoretical one. A larger theoretical privacy budget does not necessarily lead to a larger gap with our lower bound (note that compared to Figure~\ref{diffprv::influence:noisemultiplier}, the noise multiplier is constant across these experiments, and we solely vary the number of steps). We further note that the finetuned backdoored model still performs well on the downstream task. Details about hyper-parameter selection and performance are summarized in Table \ref{tab::diffprv:exdetails}.

\begin{table}[H]
    \centering
    \vskip -0.1in
    \caption{Summary of the experiments. }
    \label{tab::diffprv:exdetails}
    \vskip 0.05in
    \begin{center}
    \begin{small}
    \begin{tabular}{@{}l r r r r r r r @{}}
    \toprule
    Finetuning & Target $\varepsilon$ & Epochs & Learning rate &Training acc. & Test acc. & $\tilde{\varepsilon}/ {\varepsilon}\ (\rho=1)$  \\
    \midrule
    \multirow{4}{*}{Head} & 1 & 3 & 0.5 & 0.355 & 0.364 & 0.64  \\
    & 3 & 27 & 0.2 & 0.417 & 0.420 & 0.73  \\
    & 5 & 69 & 0.1 & 0.423 & 0.427 & 0.72 \\
    & 8 & 156 & 0.05 & 0.421 & 0.421 & 0.72 \\
    \midrule
    MLP & 3 & 27 & 0.2 & 0.520 & 0.522 & 0.72 \\
    \bottomrule
    \end{tabular}
    \end{small}
    \end{center}
    \vskip -0.2in
\end{table}

\begin{figure}[h]
    \begin{center}
    \includegraphics[width=0.35\textwidth]{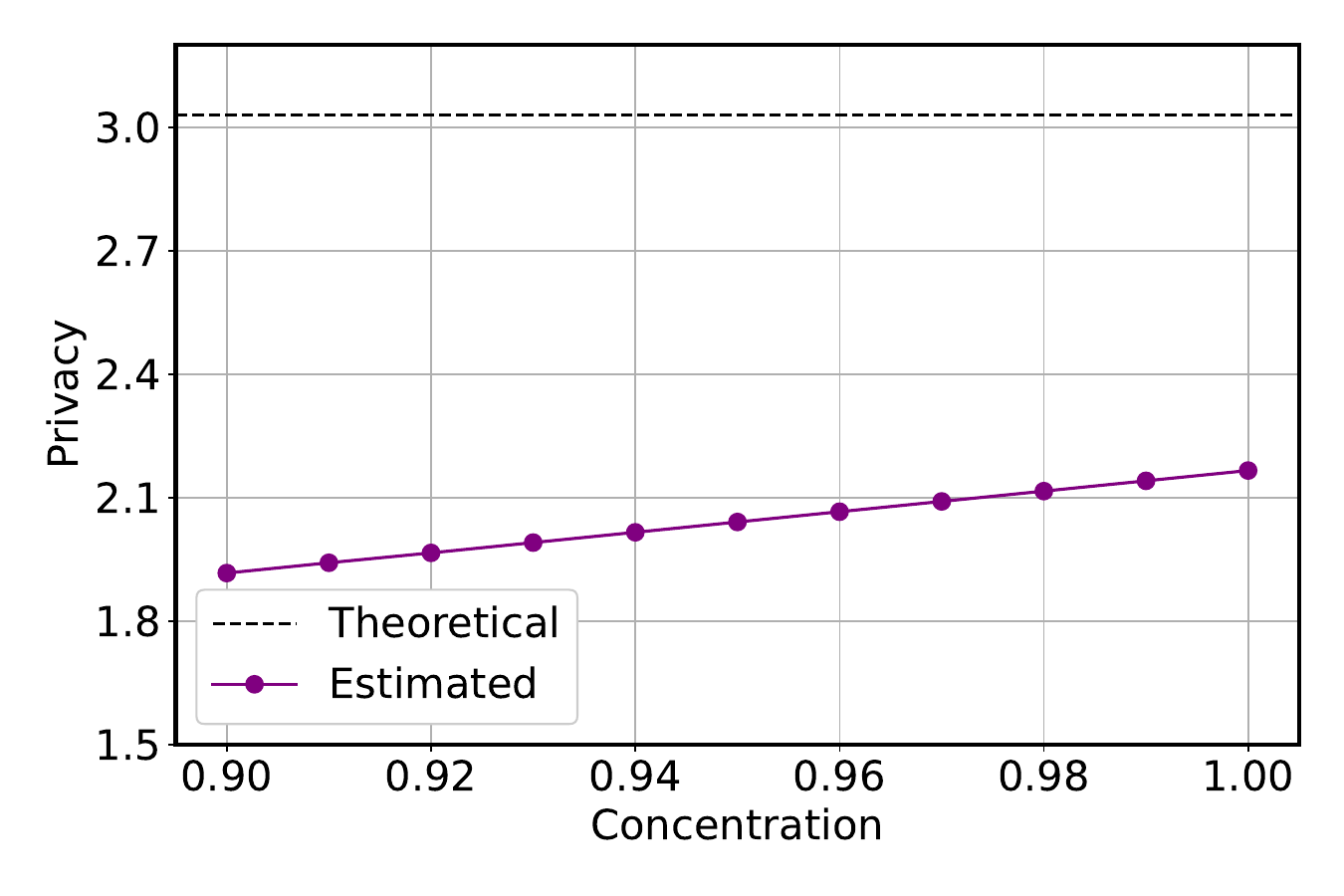} 
    \caption{The effect of the concentration factor correction ($\rho$) on our estimated privacy lower bound.}
    \label{diffprv::influence:concentration}
    \end{center}
    \vskip -0.15in
\end{figure}

Recall that we introduce a correction term $\rho$ to account for the fact that our backdoor constructions do not fully concentrate all of the gradient magnitude onto a small set of weights. In Figure \ref{diffprv::influence:concentration}, we show that our lower bound is only moderately influenced by this inefficiency. Even if our backdoor only guarantees that 90\% of the gradient magnitude concentrates on the targeted weights, we still get a strong lower bound on the privacy leakage. 

\newpage
\newpage
\section{Additional Results}
\label{appendix::additionalresults}

This appendix contains additional experiments for white-box data reconstruction attacks on backdoored ViT and BERT models. 

\paragraph{Crafted classification head vs.~random head.}
In the main body, we focused on the standard setting where the attacker has no control over the final classification head, which is randomly initialized by the victim before finetuning.
In some cases, however, an attacker may be able to control this layer as well. This is the case if the attacker releases a backdoored \emph{classifier} for some specific task (e.g., medical image classification, or sentiment analysis) that the victim then further finetunes on their domain-specific data.
The ability to choose the classification head gives the attacker much more control. In particular, the attacker can wire the classifier so that inputs captured by backdoors are misclassified with high probability, thereby reducing the risk of negative gradient flows (we assume that attackers can roughly predict one class containing no capturable samples for a backdoor.) This is particularly useful for classification tasks with few classes, where the risk of a random correct classification is high. %

\begin{figure*}[h]
    \begin{center}
    \begin{minipage}[b]{0.9\textwidth}
         \centering
         \subfloat[Reconstruction, crafted head, CIFAR-10]{\begin{minipage}[t]{0.48\linewidth}
         \centering
         \includegraphics[width=0.5\linewidth]{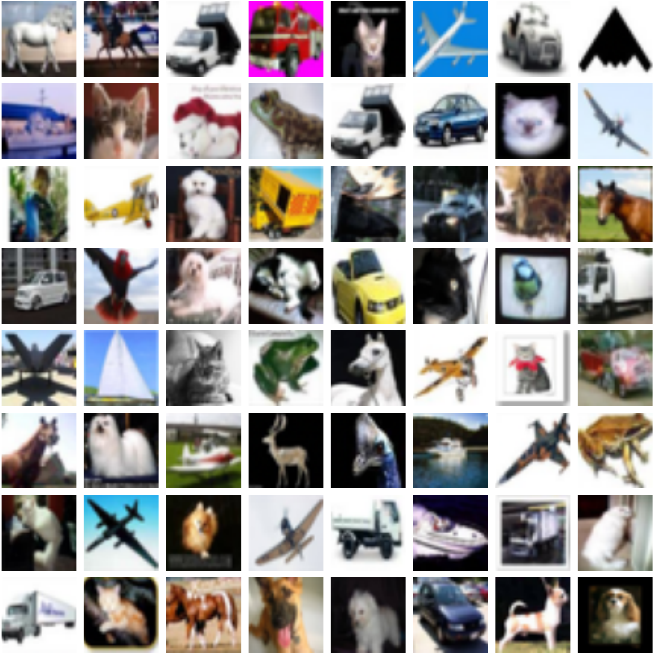} 
         \end{minipage}} 
         \subfloat[Ground truth, crafted head, CIFAR-10]{\begin{minipage}[t]{0.48\linewidth}
         \centering
         \includegraphics[width=0.5\linewidth]{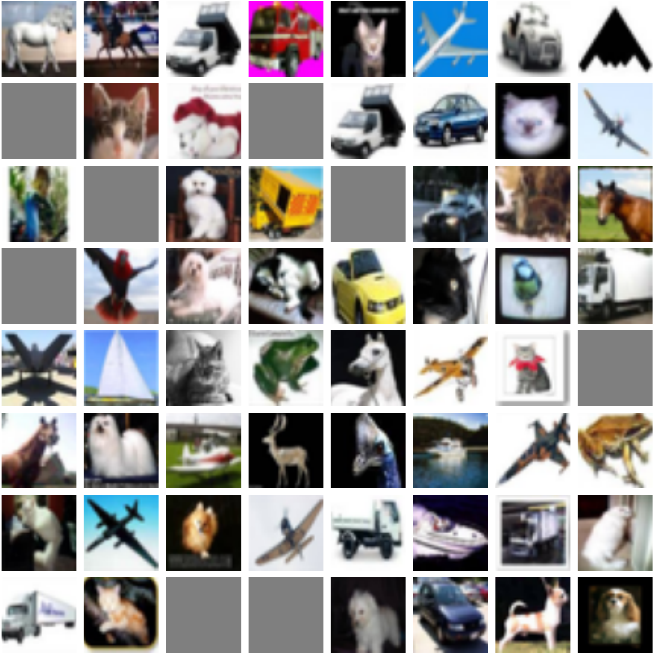}
         \end{minipage}}

        \subfloat[Reconstruction, random head, CIFAR-100]{\begin{minipage}[t]{0.48\linewidth}
        \centering
        \includegraphics[width=0.5\linewidth]{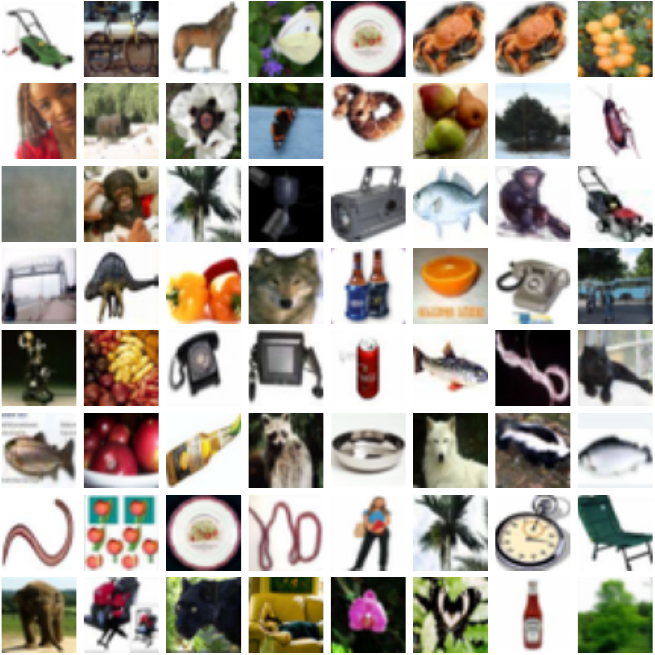}
        
        \end{minipage}} 
        \subfloat[Ground truth, random head, CIFAR-100]{\begin{minipage}[t]{0.48\linewidth}
        \centering
        \includegraphics[width=0.5\linewidth]{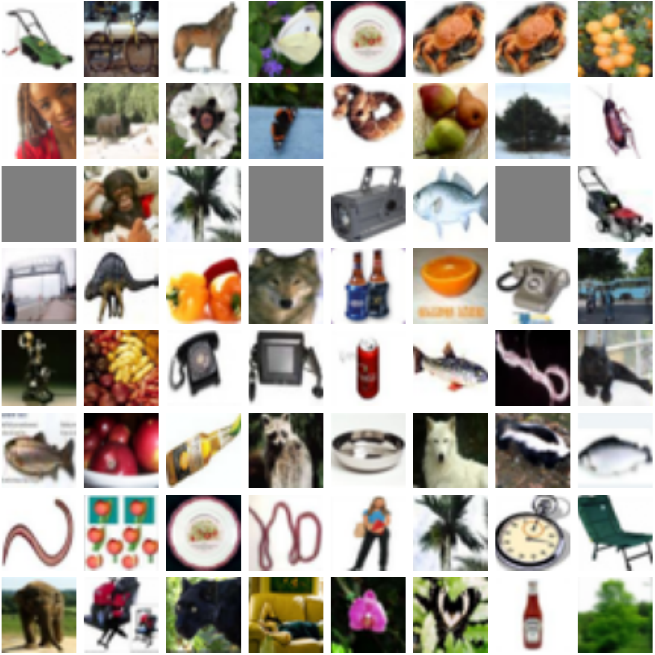}
        \end{minipage}}    
    \end{minipage}
    \caption{Reconstructed images from backdoored MLP models. There are no unrecoverable backdoors when a crafted head is used, even though there are only a few classes. When a random head is used, there are fewer unrecoverable backdoors for CIFAR-100 than for CIFAR-10.}
    \label{reconstruction::mlp:others}
    \end{center}
    \vskip -0.15in
\end{figure*}

\paragraph{Reconstructing images in a ReLU ViT.}

\begin{figure}[h]
    \begin{center}
    \subfloat[Reconstruction]{\includegraphics[width=0.4\textwidth]{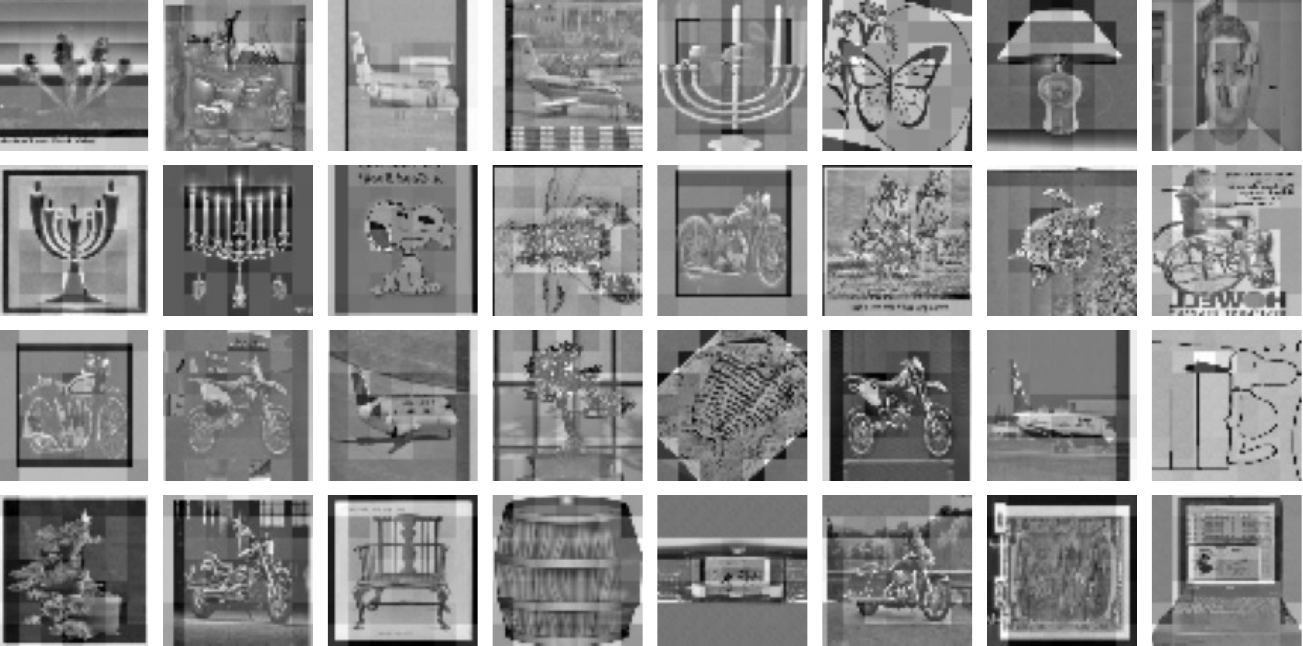}} 
    \hskip 0.4in
    \subfloat[Ground truth]{\includegraphics[width=0.4\textwidth]{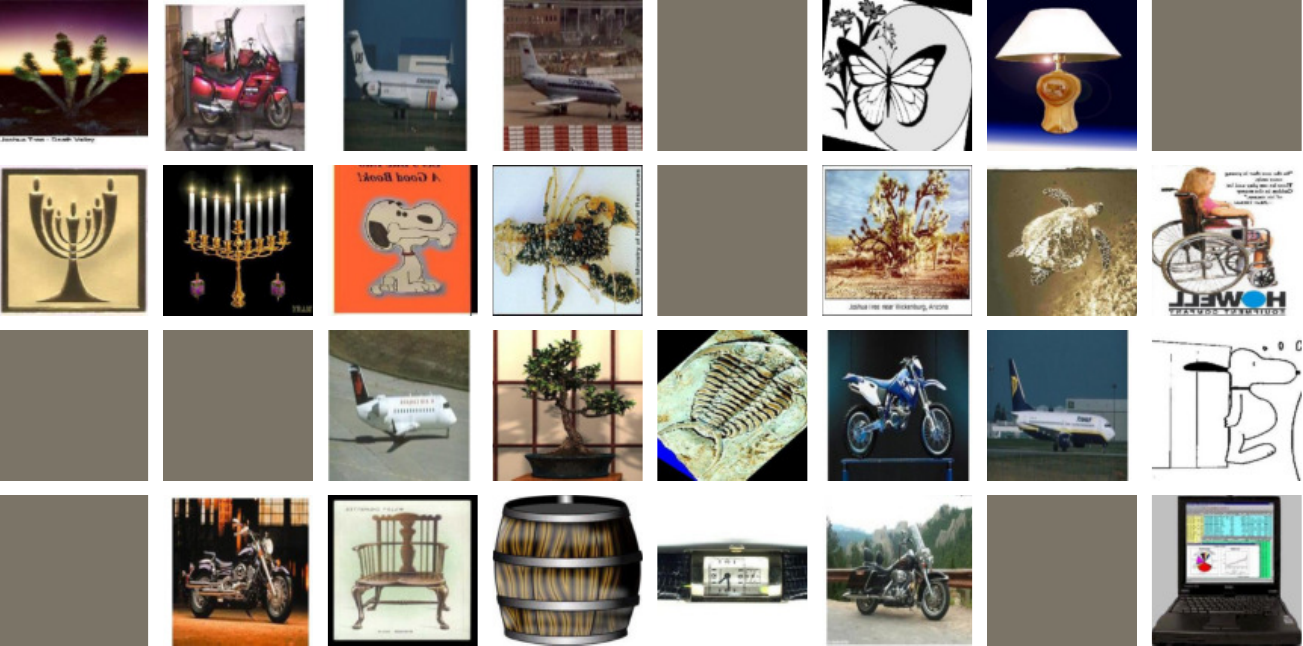}}
    \caption{Reconstructed complete images from a backdoored ReLU-version ViT (using a random head). \textbf{Dataset}: Caltech 101.}
    \label{reconstruction::vit:relu:caltech101}
    \end{center}
    \vskip -0.15in
\end{figure}

Figure~\ref{reconstruction::vit:relu:caltech101} shows reconstructed images
from a backdoored ViT using ReLU activations rather than GELUs.
Backdooring the ReLU version is easier, as we have a lower risk of the model breaking down due to undesirable gradient flows. The quality of reconstructed images is perceptually similar in both cases though.

\paragraph{Reconstructing individual patches in a backdoored ViT.}

\begin{figure*}[htb]
    \begin{center} 
    \centering
    \subfloat[Reconstruction, crafted head, Oxford-IIIT Pet, GELU]
    {\begin{minipage}[t]{0.45\linewidth}
    \centering
    \includegraphics[width=0.65\linewidth]{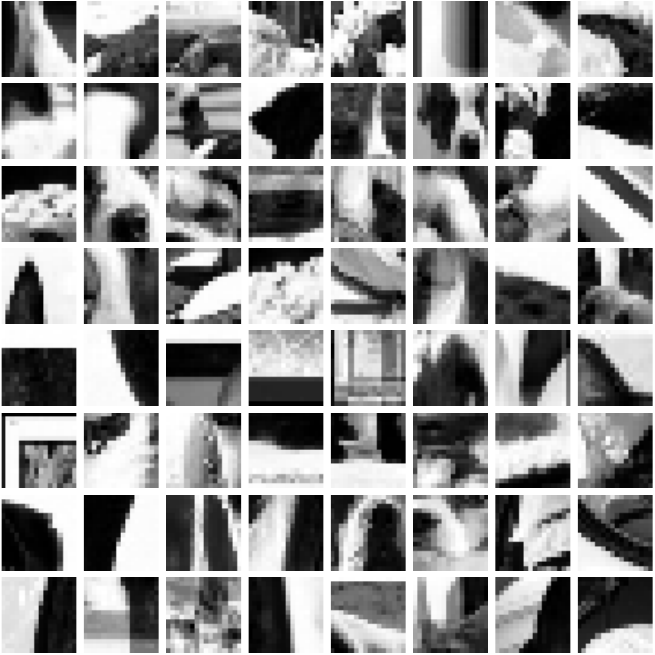}
    \end{minipage}} \hspace{0.5cm}
    \subfloat[Ground truth, crafted head, Oxford-IIIT Pet, GELU]{
    \begin{minipage}[t]{0.45\linewidth} \centering
    \includegraphics[width=0.65\linewidth]{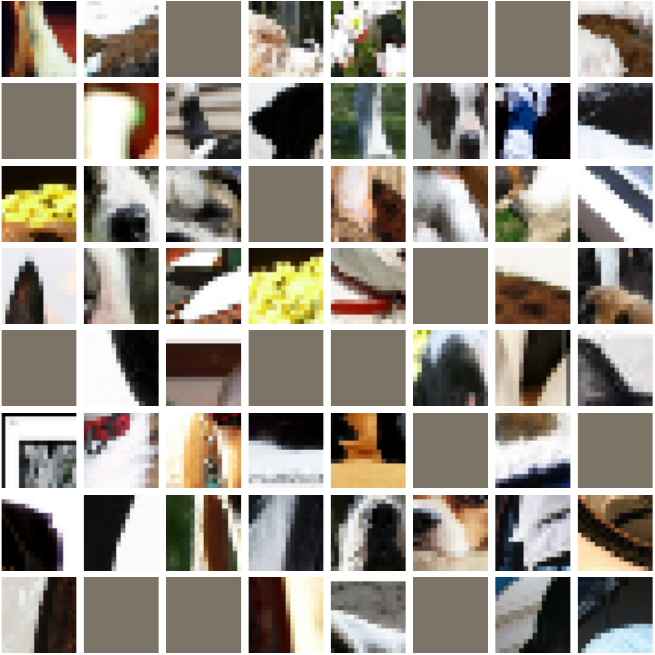}
    \end{minipage}}
    
    \vspace{0.3cm}
    
    \subfloat[Reconstruction, crafted head, Oxford-IIIT Pet, ReLU]
    {\begin{minipage}[t]{0.45\linewidth}
    \centering
    \includegraphics[width=0.65\linewidth]{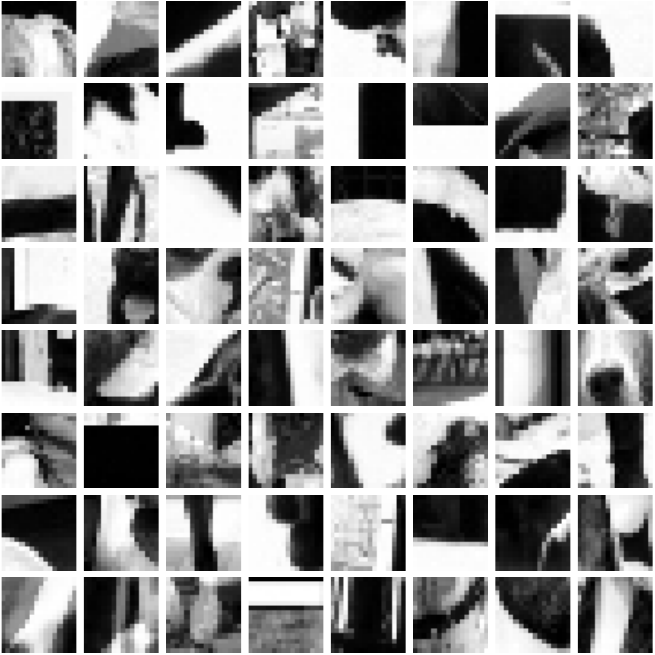}
    \end{minipage}} \hspace{0.5cm}
    \subfloat[Ground truth, crafted head, Oxford-IIIT Pet, ReLU]{
    \begin{minipage}[t]{0.45\linewidth} \centering
    \includegraphics[width=0.65\linewidth]{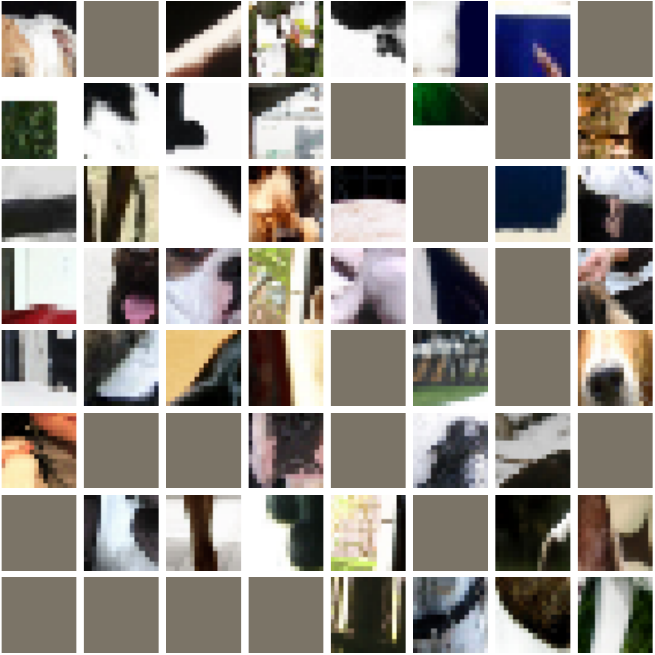}
    \end{minipage}}

    \caption{Reconstructed patches from backdoored ViTs. While using random backdoor weights, some reconstructed patches reveal critical information about images. For example, we recognize several informative patches of dogs' faces from Oxford-IIIT Pet. If attackers can design backdoor weights carefully, most captured patches may be informative for humans.}
    \label{reconstruction::vit:patch}
    \end{center}
    \vskip -0.15in
\end{figure*}

For image transformers, recovering individual tokens (i.e., patches) from different inputs could still be a significant privacy violation.
For example, one patch might contain an individual's face.
It is much easier to design backdoors that capture individual patches, without any sequence key or position key information. The backdoored model is thus likely more robust. Moreover, an attacker can recover (partial) information from thousands of training samples, rather than a small number of full inputs. Such an attack may be preferable in some settings (e.g., for images with high redundancy).
Figure~\ref{reconstruction::vit:patch} shows examples of reconstructed grayscale patches for ViT models finetuned on Oxford-IIIT Pet.
While some captured patches contain unrecognizable information, others capture salient parts of the inputs (e.g., a dog's face).

\newpage

\paragraph{Full set of reconstructed sentences from backdoored BERT models.}
In the figures that follow, we present the complete lists of reconstructed sentences from backdoored BERT models finetuned on TREC-6 or TREC-50, with either a random or crafted classification head, and using either ReLU or GELU activations. For each reconstructed sentence, we present one or more possible training sentences that have activated the backdoor during training. We highlight reconstructed trigrams that match one of these potential ground truth sentences (we cannot always guarantee that each trigram was indeed captured from this specific training sentence, so we provide these annotations for illustration purposes).  
In some cases, the captured sentence is a mixture of multiple training sentences. In others, a single training sentence's backdoor signal dominated others and can be recovered.
Overall, we find that ReLU models lead to more faithful reconstructions with fewer failed backdoors.

\renewcommand{\arraystretch}{0}

\begin{table}[H]
\caption{Reconstructed sentences from a backdoored ReLU-version BERT using a crafted head. \textbf{Dataset}: TREC-6. Sentence 5 is an interesting example of what will happen if there is more than one ground truth sentence for a backdoor family.}
\label{reconstruction::bert:relu:trecsix}
\vskip 12pt

\centering
\tiny
\begin{tabular}{@{}l p{7.5cm} p{7.5cm}@{}}
\textbf{ID} & \textbf{Reconstructed sample} & \textbf{Possible ground truth}\\
\toprule
1 & \sethlcolor{yellow}\hl{what }\sethlcolor{yellow}\hl{war }\sethlcolor{yellow}\hl{saw }\sethlcolor{yellow}\hl{battles }\sethlcolor{yellow}\hl{at }\sethlcolor{yellow}\hl{parrot's }\sethlcolor{yellow}\hl{beak }\sethlcolor{yellow}\hl{and }\sethlcolor{yellow}\hl{black }\sethlcolor{yellow}\hl{virgin}\sethlcolor{yellow}\hl{? }of members clinch the house of representatives?&what war saw battles at parrot's beak and black virgin?\\
\midrule
2 & \sethlcolor{yellow}\hl{what }\sethlcolor{yellow}\hl{causes }\sethlcolor{yellow}\hl{pneumonia}\sethlcolor{yellow}\hl{?}&what causes pneumonia?\\
\midrule
3 & \sethlcolor{yellow}\hl{what }\sethlcolor{yellow}\hl{films }\sethlcolor{yellow}\hl{featured }\sethlcolor{yellow}\hl{the }\sethlcolor{yellow}\hl{character }\sethlcolor{yellow}\hl{popeye }\sethlcolor{yellow}\hl{doyle}\sethlcolor{yellow}\hl{?}&what films featured the character popeye doyle?\\
\midrule
4 & \sethlcolor{yellow}\hl{what }\sethlcolor{yellow}\hl{does }\sethlcolor{yellow}\hl{a }\sethlcolor{yellow}\hl{chairbound }\sethlcolor{yellow}\hl{basophobic }\sethlcolor{yellow}\hl{fear}\sethlcolor{yellow}\hl{?}&what does a chairbound basophobic fear?\\
\midrule
5 & \sethlcolor{olive}\hl{what }\sethlcolor{olive}\hl{two }\sethlcolor{olive}\hl{japanese }admiral of spelled ocean seas letters k, \sethlcolor{orange}\hl{governor }\sethlcolor{orange}\hl{general }\sethlcolor{orange}\hl{o }all t islands y \sethlcolor{orange}\hl{might }\sethlcolor{orange}\hl{discover, }\sethlcolor{orange}\hl{and }\sethlcolor{orange}\hl{also }\sethlcolor{orange}\hl{granted }\sethlcolor{orange}\hl{10 }\sethlcolor{orange}\hl{-}\sethlcolor{orange}\hl{?}\sethlcolor{orange}\hl{? }\sethlcolor{orange}\hl{of }\sethlcolor{orange}\hl{all }\sethlcolor{orange}\hl{profits }\sethlcolor{orange}\hl{of }\sethlcolor{orange}\hl{his }\sethlcolor{orange}\hl{voyage.}&who was named admiral of the ocean seas and viceroy and governor general of all the islands he might discover, and also granted 10 -?? of all profits of his voyage.\\
\cmidrule{3-3}
&&what two japanese cities are spelled with the letters k, o, o, t and y?\\
\midrule
6 & \sethlcolor{yellow}\hl{what }\sethlcolor{yellow}\hl{crop }\sethlcolor{yellow}\hl{failure }\sethlcolor{yellow}\hl{caused }\sethlcolor{yellow}\hl{the }\sethlcolor{yellow}\hl{irish }\sethlcolor{yellow}\hl{famine}\sethlcolor{yellow}\hl{? }for?&what crop failure caused the irish famine?\\
\midrule
7 & is \sethlcolor{yellow}\hl{what }\sethlcolor{yellow}\hl{schools }\sethlcolor{yellow}\hl{in }\sethlcolor{yellow}\hl{the }\sethlcolor{yellow}\hl{washington, }\sethlcolor{yellow}\hl{dc }\sethlcolor{yellow}\hl{nn }\sethlcolor{yellow}\hl{nn }\sethlcolor{yellow}\hl{vbp }\sethlcolor{yellow}\hl{nn }\sethlcolor{yellow}\hl{nn }\sethlcolor{yellow}\hl{nn }\sethlcolor{yellow}\hl{nn. }177, and 19? ypy&what schools in the washington, dc nn nn vbp nn nn nn nn.\\
\midrule
8 & \sethlcolor{yellow}\hl{what }\sethlcolor{yellow}\hl{u. }\sethlcolor{yellow}\hl{s. }\sethlcolor{yellow}\hl{state }\sethlcolor{yellow}\hl{lived }\sethlcolor{yellow}\hl{under }\sethlcolor{yellow}\hl{six }\sethlcolor{yellow}\hl{flags}\sethlcolor{yellow}\hl{?}&what u. s. state lived under six flags?\\
\midrule
9 & \sethlcolor{yellow}\hl{what }\sethlcolor{yellow}\hl{piano }\sethlcolor{yellow}\hl{company }\sethlcolor{yellow}\hl{claims }\sethlcolor{yellow}\hl{its }\sethlcolor{yellow}\hl{product }\sethlcolor{yellow}\hl{is }\sethlcolor{yellow}\hl{the }\sethlcolor{yellow}\hl{` }\sethlcolor{yellow}\hl{` }\sethlcolor{yellow}\hl{instrument }\sethlcolor{yellow}\hl{of }\sethlcolor{yellow}\hl{the }\sethlcolor{yellow}\hl{immortals''}\sethlcolor{yellow}\hl{?}&what piano company claims its product is the ` ` instrument of the immortals''?\\
\midrule
10 & \sethlcolor{yellow}\hl{what }\sethlcolor{yellow}\hl{is }\sethlcolor{yellow}\hl{the }\sethlcolor{yellow}\hl{term }\sethlcolor{yellow}\hl{for }\sethlcolor{yellow}\hl{the }\sethlcolor{yellow}\hl{side }\sethlcolor{yellow}\hl{of }\sethlcolor{yellow}\hl{the }\sethlcolor{yellow}\hl{mountain }\sethlcolor{yellow}\hl{that }\sethlcolor{yellow}\hl{faces }\sethlcolor{yellow}\hl{the }\sethlcolor{yellow}\hl{prevailing }\sethlcolor{yellow}\hl{winds}\sethlcolor{yellow}\hl{? }windsor?&what is the term for the side of the mountain that faces the prevailing winds?\\
\midrule
11 & the \sethlcolor{yellow}\hl{why }\sethlcolor{yellow}\hl{is }\sethlcolor{yellow}\hl{indiglo }\sethlcolor{yellow}\hl{called }\sethlcolor{yellow}\hl{indiglo}\sethlcolor{yellow}\hl{? }sylate? ig whatssorss thepro the called between? ig the between differenceprosorpro whatpro is differences micro80sor differencessprosorpro what&what is the difference between microprocessors \& microcontrollers?\\
\cmidrule{3-3}
&&why is indiglo called indiglo?\\
\midrule
12 & \sethlcolor{yellow}\hl{what }\sethlcolor{yellow}\hl{color }\sethlcolor{yellow}\hl{were }\sethlcolor{yellow}\hl{their }\sethlcolor{yellow}\hl{horses}\sethlcolor{yellow}\hl{?}? -&what color were their horses?\\
\midrule
13 & \sethlcolor{yellow}\hl{how }\sethlcolor{yellow}\hl{did }\sethlcolor{yellow}\hl{jayne }\sethlcolor{yellow}\hl{mansfield }\sethlcolor{yellow}\hl{die}\sethlcolor{yellow}\hl{?}&how did jayne mansfield die?\\
\midrule
14 & \sethlcolor{yellow}\hl{how }\sethlcolor{yellow}\hl{do }\sethlcolor{yellow}\hl{i }\sethlcolor{yellow}\hl{log }\sethlcolor{yellow}\hl{on }\sethlcolor{yellow}\hl{to }\sethlcolor{yellow}\hl{home }\sethlcolor{yellow}\hl{page }\sethlcolor{yellow}\hl{at }\sethlcolor{yellow}\hl{headquarters }\sethlcolor{yellow}\hl{u. }\sethlcolor{yellow}\hl{s. }\sethlcolor{yellow}\hl{european }\sethlcolor{yellow}\hl{command}\sethlcolor{yellow}\hl{? }look good''?&how do i log on to home page at headquarters u. s. european command?\\
\midrule
15 & lawrence \sethlcolor{yellow}\hl{where }\sethlcolor{yellow}\hl{is }\sethlcolor{yellow}\hl{sinn }\sethlcolor{yellow}\hl{fein's }\sethlcolor{yellow}\hl{headquarters}\sethlcolor{yellow}\hl{?}? lawrence how lawrence lawrence lawrence lawrence lawrence how how how&where is sinn fein's headquarters?\\
\midrule
16 & \sethlcolor{yellow}\hl{who }\sethlcolor{yellow}\hl{reports }\sethlcolor{yellow}\hl{the }\sethlcolor{yellow}\hl{weather }\sethlcolor{yellow}\hl{on }\sethlcolor{yellow}\hl{the }\sethlcolor{yellow}\hl{` }\sethlcolor{yellow}\hl{` }\sethlcolor{yellow}\hl{good }\sethlcolor{yellow}\hl{morning }\sethlcolor{yellow}\hl{america'' }\sethlcolor{yellow}\hl{television }\sethlcolor{yellow}\hl{show}\sethlcolor{yellow}\hl{?}&who reports the weather on the ` ` good morning america'' television show?\\
\midrule
17 & \sethlcolor{yellow}\hl{what }\sethlcolor{yellow}\hl{are }\sethlcolor{yellow}\hl{amicable }\sethlcolor{yellow}\hl{numbers}\sethlcolor{yellow}\hl{?}&what are amicable numbers?\\
\midrule
18 & \sethlcolor{yellow}\hl{how }\sethlcolor{yellow}\hl{long }\sethlcolor{yellow}\hl{were }\sethlcolor{yellow}\hl{tyrannosaurus }\sethlcolor{yellow}\hl{rex's }\sethlcolor{yellow}\hl{teeth}\sethlcolor{yellow}\hl{?}&how long were tyrannosaurus rex's teeth?\\
\midrule
19 & \sethlcolor{yellow}\hl{who }\sethlcolor{yellow}\hl{directed }\sethlcolor{yellow}\hl{` }\sethlcolor{yellow}\hl{` }\sethlcolor{yellow}\hl{jaws''}\sethlcolor{yellow}\hl{?}&who directed ` ` jaws''?\\
\midrule
20 & \sethlcolor{yellow}\hl{when }\sethlcolor{yellow}\hl{did }\sethlcolor{yellow}\hl{the }\sethlcolor{yellow}\hl{supercontinent }\sethlcolor{yellow}\hl{pangaea }\sethlcolor{yellow}\hl{break }\sethlcolor{yellow}\hl{up}\sethlcolor{yellow}\hl{?}&when did the supercontinent pangaea break up?\\
\midrule
21 & \sethlcolor{yellow}\hl{which }\sethlcolor{yellow}\hl{gender }\sethlcolor{yellow}\hl{has }\sethlcolor{yellow}\hl{bigger }\sethlcolor{yellow}\hl{thighs}\sethlcolor{yellow}\hl{?}&which gender has bigger thighs?\\
\midrule
22 & \sethlcolor{yellow}\hl{what }\sethlcolor{yellow}\hl{does }\sethlcolor{yellow}\hl{a }\sethlcolor{yellow}\hl{collier }\sethlcolor{yellow}\hl{mine}\sethlcolor{yellow}\hl{? }measure?&what does a collier mine?\\
\midrule
23 & \sethlcolor{yellow}\hl{who }\sethlcolor{yellow}\hl{has }\sethlcolor{yellow}\hl{won }\sethlcolor{yellow}\hl{the }\sethlcolor{yellow}\hl{most }\sethlcolor{yellow}\hl{super }\sethlcolor{yellow}\hl{bowls}\sethlcolor{yellow}\hl{?}&who has won the most super bowls?\\
\midrule
24 & \sethlcolor{yellow}\hl{what }\sethlcolor{yellow}\hl{does }\sethlcolor{yellow}\hl{inuit }\sethlcolor{yellow}\hl{and }\sethlcolor{yellow}\hl{eskimo }\sethlcolor{yellow}\hl{mean}\sethlcolor{yellow}\hl{?}&what does inuit and eskimo mean?\\
\midrule
25 & \sethlcolor{yellow}\hl{what }\sethlcolor{yellow}\hl{do }\sethlcolor{yellow}\hl{economists }\sethlcolor{yellow}\hl{do}\sethlcolor{yellow}\hl{? }their spiracles?&what do economists do?\\
\midrule
26 & \sethlcolor{yellow}\hl{what }\sethlcolor{yellow}\hl{does }\sethlcolor{yellow}\hl{venus }\sethlcolor{yellow}\hl{retrograde }\sethlcolor{yellow}\hl{mean}\sethlcolor{yellow}\hl{? }'mean?&where is tornado alley?\\
\cmidrule{3-3}
&&what does venus retrograde mean?\\
\cmidrule{3-3}
&&what does las vegas mean?\\
\midrule
27 & killed \sethlcolor{yellow}\hl{name }\sethlcolor{yellow}\hl{a }\sethlcolor{yellow}\hl{film }\sethlcolor{yellow}\hl{in }\sethlcolor{yellow}\hl{which }\sethlcolor{yellow}\hl{jude }\sethlcolor{yellow}\hl{law }\sethlcolor{yellow}\hl{acted. }kurt co?? who who who killed co? who who who? killed kurt who who who who who killed who whobain killed killed whobain? who killed killed who? killed who who&name a film in which jude law acted.\\
\cmidrule{3-3}
&&who killed kurt cobain?\\
\midrule
28 & \sethlcolor{yellow}\hl{what }\sethlcolor{yellow}\hl{is }\sethlcolor{yellow}\hl{the }\sethlcolor{yellow}\hl{largest }\sethlcolor{yellow}\hl{snake }\sethlcolor{yellow}\hl{in }\sethlcolor{yellow}\hl{the }\sethlcolor{yellow}\hl{world}\sethlcolor{yellow}\hl{?}?&what is the largest snake in the world?\\
\midrule
29 & \sethlcolor{yellow}\hl{what }\sethlcolor{yellow}\hl{meter }\sethlcolor{yellow}\hl{did }\sethlcolor{yellow}\hl{shakespeare }\sethlcolor{yellow}\hl{use }\sethlcolor{yellow}\hl{in }\sethlcolor{yellow}\hl{writing }\sethlcolor{yellow}\hl{: }\sethlcolor{yellow}\hl{` }\sethlcolor{yellow}\hl{` }\sethlcolor{yellow}\hl{to }\sethlcolor{yellow}\hl{be, }\sethlcolor{yellow}\hl{or }\sethlcolor{yellow}\hl{not }\sethlcolor{yellow}\hl{to }\sethlcolor{yellow}\hl{be, }\sethlcolor{yellow}\hl{that }\sethlcolor{yellow}\hl{is }\sethlcolor{yellow}\hl{the }\sethlcolor{yellow}\hl{question...''}\sethlcolor{yellow}\hl{?}&what meter did shakespeare use in writing : ` ` to be, or not to be, that is the question...''?\\
\midrule
30 & \sethlcolor{yellow}\hl{what }\sethlcolor{yellow}\hl{makes }\sethlcolor{yellow}\hl{thunder}\sethlcolor{yellow}\hl{? }der publish??&what makes thunder?\\
\midrule
31 & \sethlcolor{yellow}\hl{name }\sethlcolor{yellow}\hl{11 }\sethlcolor{yellow}\hl{famous }\sethlcolor{yellow}\hl{martyrs. }slow down aging? v has danced for?&name 11 famous martyrs.\\
\midrule
32 & \sethlcolor{yellow}\hl{what }\sethlcolor{yellow}\hl{are }\sethlcolor{yellow}\hl{the }\sethlcolor{yellow}\hl{most }\sethlcolor{yellow}\hl{common }\sethlcolor{yellow}\hl{elements }\sethlcolor{yellow}\hl{in }\sethlcolor{yellow}\hl{the }\sethlcolor{yellow}\hl{earth's }\sethlcolor{yellow}\hl{crust}?&what are the most common elements in the earth's crust\\
\bottomrule
\end{tabular}
\vspace{-10em}
\end{table}

\begin{table}[t]
\caption{Reconstructed sentences from a backdoored GELU-version BERT using a random head. \textbf{Dataset}: TREC-50. This is the complete list of selected examples in Table \ref{reconstruction::bert:example}. Examples with ground truth marked as \textbf{MANY} correspond to cases where a large number of training examples activated the backdoor.}
\label{reconstruction::bert:gelu:trecfifty}
\vskip 12pt
\centering
\tiny
\begin{tabular}{@{}l p{7.5cm} p{7.5cm}@{}}
\textbf{ID} & \textbf{Reconstructed sample} & \textbf{Possible ground truth}\\
\toprule

1 & \sethlcolor{yellow}\hl{where }\sethlcolor{yellow}\hl{does }\sethlcolor{yellow}\hl{the }\sethlcolor{yellow}\hl{santa }\sethlcolor{yellow}\hl{fe }\sethlcolor{yellow}\hl{trail }\sethlcolor{yellow}\hl{begin }\sethlcolor{yellow}\hl{and }\sethlcolor{yellow}\hl{end}\sethlcolor{yellow}\hl{?}?&where is guam?\\
\cmidrule{3-3}
&&what kentucky city calls itself the horse center of america?\\
\cmidrule{3-3}
&&where does the santa fe trail begin and end?\\
\midrule
2 & \sethlcolor{yellow}\hl{where }\sethlcolor{yellow}\hl{is }\sethlcolor{yellow}\hl{trinidad}\sethlcolor{yellow}\hl{?}&where is trinidad?\\
\midrule
3 & \sethlcolor{yellow}\hl{what }\sethlcolor{yellow}\hl{causes }\sethlcolor{yellow}\hl{rust}\sethlcolor{yellow}\hl{? }antarctica?&what causes rust?\\
\midrule
4 & \sethlcolor{yellow}\hl{name }\sethlcolor{yellow}\hl{a }\sethlcolor{yellow}\hl{gaelic }\sethlcolor{yellow}\hl{language.}?&name a gaelic language.\\
\midrule
5 & what causes ` ` rolling thunder''? '' mean?& \textbf{MANY}\\
\midrule
6 & \sethlcolor{yellow}\hl{where }\sethlcolor{yellow}\hl{is }\sethlcolor{yellow}\hl{ocho }\sethlcolor{yellow}\hl{rios}\sethlcolor{yellow}\hl{?}??&what is the immaculate conception?\\
\cmidrule{3-3}
&&what is pandoro?\\
\cmidrule{3-3}
&&where is ocho rios?\\
\midrule
7 & \sethlcolor{yellow}\hl{what }\sethlcolor{yellow}\hl{must }\sethlcolor{yellow}\hl{a }\sethlcolor{yellow}\hl{las }\sethlcolor{yellow}\hl{vegas }\sethlcolor{yellow}\hl{blackjack }\sethlcolor{yellow}\hl{dealer }\sethlcolor{yellow}\hl{do }\sethlcolor{yellow}\hl{when }\sethlcolor{yellow}\hl{he }\sethlcolor{yellow}\hl{reaches }\sethlcolor{yellow}\hl{16}\sethlcolor{yellow}\hl{?}&what must a las vegas blackjack dealer do when he reaches 16?\\
\midrule
8 & \sethlcolor{yellow}\hl{how }\sethlcolor{yellow}\hl{much }\sethlcolor{yellow}\hl{does }\sethlcolor{yellow}\hl{the }\sethlcolor{yellow}\hl{president }\sethlcolor{yellow}\hl{get }\sethlcolor{yellow}\hl{paid}\sethlcolor{yellow}\hl{?}?? get daily?&what soft drink is most heavily caffeinated?\\
\cmidrule{3-3}
&&how much does the president get paid?\\
\midrule
9 & \sethlcolor{yellow}\hl{which }\sethlcolor{yellow}\hl{two }\sethlcolor{yellow}\hl{states }\sethlcolor{yellow}\hl{enclose }\sethlcolor{yellow}\hl{chesapeake }\sethlcolor{yellow}\hl{bay}\sethlcolor{yellow}\hl{? }ruplets??&which two states enclose chesapeake bay?\\
\midrule
10 & \sethlcolor{yellow}\hl{how }\sethlcolor{yellow}\hl{is }\sethlcolor{yellow}\hl{easter }\sethlcolor{yellow}\hl{sunday's }\sethlcolor{yellow}\hl{date }\sethlcolor{yellow}\hl{determined}\sethlcolor{yellow}\hl{?}&how is easter sunday's date determined?\\
\midrule
11 & \sethlcolor{yellow}\hl{what }\sethlcolor{yellow}\hl{do }\sethlcolor{yellow}\hl{the }\sethlcolor{yellow}\hl{letters }\sethlcolor{yellow}\hl{d. }\sethlcolor{yellow}\hl{c. }\sethlcolor{yellow}\hl{stand }\sethlcolor{yellow}\hl{for }\sethlcolor{yellow}\hl{in }\sethlcolor{yellow}\hl{washington, }\sethlcolor{yellow}\hl{d. }\sethlcolor{yellow}\hl{c.}\sethlcolor{yellow}\hl{? }np..&what do the letters d. c. stand for in washington, d. c.?\\
\midrule
12 & \sethlcolor{yellow}\hl{what }\sethlcolor{yellow}\hl{schools }\sethlcolor{yellow}\hl{in }\sethlcolor{yellow}\hl{the }\sethlcolor{yellow}\hl{washington, }\sethlcolor{yellow}\hl{dc }\sethlcolor{yellow}\hl{nn }\sethlcolor{yellow}\hl{nn }\sethlcolor{yellow}\hl{vbp }\sethlcolor{yellow}\hl{nn }\sethlcolor{yellow}\hl{nn }\sethlcolor{yellow}\hl{nn }\sethlcolor{yellow}\hl{nn.}&what schools in the washington, dc nn nn vbp nn nn nn nn.\\
\midrule
13 & \sethlcolor{yellow}\hl{why }\sethlcolor{yellow}\hl{do }\sethlcolor{yellow}\hl{heavier }\sethlcolor{yellow}\hl{objects }\sethlcolor{yellow}\hl{travel }\sethlcolor{yellow}\hl{downhill }\sethlcolor{yellow}\hl{faster}\sethlcolor{yellow}\hl{? }go to college?&why do heavier objects travel downhill faster?\\
\midrule
14 & \sethlcolor{yellow}\hl{what }\sethlcolor{yellow}\hl{does }\sethlcolor{yellow}\hl{` }\sethlcolor{yellow}\hl{` }\sethlcolor{yellow}\hl{antidisestablishmentarianism'' }\sethlcolor{yellow}\hl{mean}\sethlcolor{yellow}\hl{?}&what does ` ` antidisestablishmentarianism'' mean?\\
\midrule
15 & what is pasta??&\textbf{MANY}\\
\midrule
16 & \sethlcolor{yellow}\hl{where }\sethlcolor{yellow}\hl{can }\sethlcolor{yellow}\hl{i }\sethlcolor{yellow}\hl{find }examples tab legal 1965 about film oscar songs disabilities 12 act&where can i find a case on individuals with disabilities education act of 1991?\\
\cmidrule{3-3}
&&what film did lee marvin win the 1965 best actor oscar for?\\
\cmidrule{3-3}
&&what do i need to do to take my dog with me to live in dominica, west indies for a year?\\
\cmidrule{3-3}
&&where can i find correct tabs for third eye blind songs\\

\bottomrule
\end{tabular}
\end{table}

\begin{table}[t]
\caption{Reconstructed sentences from a backdoored GELU-version BERT using a random head (best run). \textbf{Dataset}: TREC-50. This is the cleanest list of reconstructed sentences among multiple trials.}
\label{reconstruction::bert:gelu:trecfiftybest}
\vskip 12pt
\centering
\tiny
\begin{tabular}{@{}l p{7.5cm} p{7.5cm}@{}}
\textbf{ID} & \textbf{Reconstructed sample} & \textbf{Possible ground truth}\\
\toprule

1 & what was paul bunyan's ox's name? ''? .?&what product does ` ` mrs. olsen'' promote?\\
\midrule
2 & \sethlcolor{yellow}\hl{what }\sethlcolor{yellow}\hl{does }\sethlcolor{yellow}\hl{a }\sethlcolor{yellow}\hl{spermologer }\sethlcolor{yellow}\hl{collect}\sethlcolor{yellow}\hl{?}?&what does a spermologer collect?\\
\midrule
3 & \sethlcolor{yellow}\hl{what }\sethlcolor{yellow}\hl{son }\sethlcolor{yellow}\hl{of }\sethlcolor{yellow}\hl{a }\sethlcolor{yellow}\hl{15 }\sethlcolor{yellow}\hl{- }\sethlcolor{yellow}\hl{year }\sethlcolor{yellow}\hl{- }\sethlcolor{yellow}\hl{old }\sethlcolor{yellow}\hl{mexican }\sethlcolor{yellow}\hl{girl }\sethlcolor{yellow}\hl{and }\sethlcolor{yellow}\hl{a }\sethlcolor{yellow}\hl{half }\sethlcolor{yellow}\hl{- }\sethlcolor{yellow}\hl{irish }\sethlcolor{yellow}\hl{father }\sethlcolor{yellow}\hl{became }\sethlcolor{yellow}\hl{the }\sethlcolor{yellow}\hl{world's }\sethlcolor{yellow}\hl{most }\sethlcolor{yellow}\hl{famous }\sethlcolor{yellow}\hl{greek}\sethlcolor{yellow}\hl{?}&what son of a 15 - year - old mexican girl and a half - irish father became the world's most famous greek?\\
\midrule
4 & \sethlcolor{yellow}\hl{what }\sethlcolor{yellow}\hl{country's }\sethlcolor{yellow}\hl{capital }\sethlcolor{yellow}\hl{is }\sethlcolor{yellow}\hl{tirana}\sethlcolor{yellow}\hl{?}&what country's capital is tirana?\\
\midrule
5 & \sethlcolor{yellow}\hl{what }\sethlcolor{yellow}\hl{u. }\sethlcolor{yellow}\hl{s. }\sethlcolor{yellow}\hl{senator }\sethlcolor{yellow}\hl{once }\sethlcolor{yellow}\hl{played }\sethlcolor{yellow}\hl{basketball }\sethlcolor{yellow}\hl{for }\sethlcolor{yellow}\hl{the }\sethlcolor{yellow}\hl{new }\sethlcolor{yellow}\hl{york }\sethlcolor{yellow}\hl{knicks}\sethlcolor{yellow}\hl{? }power? inventions?&what u. s. senator once played basketball for the new york knicks?\\
\midrule
6 & \sethlcolor{yellow}\hl{what }\sethlcolor{yellow}\hl{color }\sethlcolor{yellow}\hl{bottles }\sethlcolor{yellow}\hl{do }\sethlcolor{yellow}\hl{good }\sethlcolor{yellow}\hl{rhine }\sethlcolor{yellow}\hl{wines }\sethlcolor{yellow}\hl{come }\sethlcolor{yellow}\hl{in}\sethlcolor{yellow}\hl{?}&what color bottles do good rhine wines come in?\\
\midrule
7 & \sethlcolor{yellow}\hl{what }\sethlcolor{yellow}\hl{causes }\sethlcolor{yellow}\hl{the }\sethlcolor{yellow}\hl{redness }\sethlcolor{yellow}\hl{in }\sethlcolor{yellow}\hl{your }\sethlcolor{yellow}\hl{cheeks }\sethlcolor{yellow}\hl{when }\sethlcolor{yellow}\hl{you }\sethlcolor{yellow}\hl{blush}\sethlcolor{yellow}\hl{?}&what causes the redness in your cheeks when you blush?\\
\midrule
8 & \sethlcolor{yellow}\hl{what }\sethlcolor{yellow}\hl{makes }\sethlcolor{yellow}\hl{thunder}\sethlcolor{yellow}\hl{?}&what makes thunder?\\
\midrule
9 & \sethlcolor{yellow}\hl{what }\sethlcolor{yellow}\hl{state }\sethlcolor{yellow}\hl{capital }\sethlcolor{yellow}\hl{comes }\sethlcolor{yellow}\hl{last }\sethlcolor{yellow}\hl{alphabetically}\sethlcolor{yellow}\hl{? }kong alphabet? last comesically? stateicallyicallyically comes? stateically last? last what&what state capital comes last alphabetically?\\
\midrule
10 & \sethlcolor{yellow}\hl{what }\sethlcolor{yellow}\hl{crop }\sethlcolor{yellow}\hl{failure }\sethlcolor{yellow}\hl{caused }\sethlcolor{yellow}\hl{the }\sethlcolor{yellow}\hl{irish }\sethlcolor{yellow}\hl{famine}\sethlcolor{yellow}\hl{?}&what crop failure caused the irish famine?\\
\midrule
11 & \sethlcolor{yellow}\hl{how }\sethlcolor{yellow}\hl{much }\sethlcolor{yellow}\hl{money }\sethlcolor{yellow}\hl{are }\sethlcolor{yellow}\hl{dumbo's }\sethlcolor{yellow}\hl{ears }\sethlcolor{yellow}\hl{insured }\sethlcolor{yellow}\hl{for}\sethlcolor{yellow}\hl{?}&how much money are dumbo's ears insured for?\\
\midrule
12 & who invented tornado alley? rabble?&who invented the game scrabble?\\
\midrule
13 & \sethlcolor{yellow}\hl{where }\sethlcolor{yellow}\hl{did }\sethlcolor{yellow}\hl{woodstock }\sethlcolor{yellow}\hl{take }\sethlcolor{yellow}\hl{place}\sethlcolor{yellow}\hl{?}&where did woodstock take place?\\
\midrule
14 & \sethlcolor{yellow}\hl{when }\sethlcolor{yellow}\hl{was }\sethlcolor{yellow}\hl{hurricane }\sethlcolor{yellow}\hl{hugo}\sethlcolor{yellow}\hl{?}???&when was hurricane hugo?\\
\midrule
15 & \sethlcolor{yellow}\hl{where }\sethlcolor{yellow}\hl{is }\sethlcolor{yellow}\hl{hearst }\sethlcolor{yellow}\hl{castle, }\sethlcolor{yellow}\hl{built }\sethlcolor{yellow}\hl{by }\sethlcolor{yellow}\hl{publisher }\sethlcolor{yellow}\hl{william }\sethlcolor{yellow}\hl{randolph }\sethlcolor{yellow}\hl{hearst}\sethlcolor{yellow}\hl{?}?&where is hearst castle, built by publisher william randolph hearst?\\
\midrule
16 & \sethlcolor{yellow}\hl{who }\sethlcolor{yellow}\hl{invented }\sethlcolor{yellow}\hl{the }\sethlcolor{yellow}\hl{toothbrush}? iowa?? ledding? bp nn nn nn nn&who invented the toothbrush\\

\bottomrule
\end{tabular}
\end{table}

\begin{table}[t]
\vskip -1em
\caption{Reconstructed sentences from a backdoored GELU-version BERT using a crafted head. \textbf{Dataset}: TREC-6. Most of the reconstructed sentences match their corresponding ground truth sentences well. Examples with ground truth marked as \textbf{MANY} or \textbf{NONE} correspond respectively to cases where a large number of training examples, or none, activated the backdoor.}
\label{reconstruction::bert:gelu:trecsix}
\vspace{12pt}
\centering
\tiny
\begin{tabular}{@{}l p{7.5cm} p{7.5cm}@{}}
\textbf{ID} & \textbf{Reconstructed sample} & \textbf{Possible ground truth}\\
\toprule

1 & \sethlcolor{yellow}\hl{what }\sethlcolor{yellow}\hl{novel }\sethlcolor{yellow}\hl{inspired }\sethlcolor{yellow}\hl{the }\sethlcolor{yellow}\hl{movie }\sethlcolor{yellow}\hl{bladerunner}\sethlcolor{yellow}\hl{? }william randolph hearst?&what novel inspired the movie bladerunner?\\
\midrule
2 & \sethlcolor{yellow}\hl{name }\sethlcolor{yellow}\hl{11 }\sethlcolor{yellow}\hl{famous }\sethlcolor{yellow}\hl{martyrs. }red in photographs?&name 11 famous martyrs.\\
\midrule
3 & a a a female. skater. flying \sethlcolor{yellow}\hl{in }\sethlcolor{yellow}\hl{north }\sethlcolor{yellow}\hl{carolina }a a a a skater a a a female.. name a a a a a a figure name a a. a. a a. a a name female name a&name four famous cartoon cats.\\
\cmidrule{3-3}
&&how do i register a trade name in north carolina?\\
\cmidrule{3-3}
&&name a female figure skater.\\
\midrule
4 & what discovered \sethlcolor{yellow}\hl{equity }\sethlcolor{yellow}\hl{securities}\sethlcolor{yellow}\hl{?}&who discovered electricity?\\
\cmidrule{3-3}
&&what are equity securities?\\
\midrule
5 & \sethlcolor{yellow}\hl{what }\sethlcolor{yellow}\hl{enigmatic }\sethlcolor{yellow}\hl{u. }\sethlcolor{yellow}\hl{s. }\sethlcolor{yellow}\hl{vice }\sethlcolor{yellow}\hl{president }\sethlcolor{yellow}\hl{was }\sethlcolor{yellow}\hl{once }\sethlcolor{yellow}\hl{tried }\sethlcolor{yellow}\hl{and }\sethlcolor{yellow}\hl{acquitted }\sethlcolor{yellow}\hl{for }\sethlcolor{yellow}\hl{treason }\sethlcolor{yellow}\hl{in }\sethlcolor{yellow}\hl{a }\sethlcolor{yellow}\hl{plot }\sethlcolor{yellow}\hl{to }\sethlcolor{yellow}\hl{set }\sethlcolor{yellow}\hl{up }\sethlcolor{yellow}\hl{his }\sethlcolor{yellow}\hl{own }\sethlcolor{yellow}\hl{independent }\sethlcolor{yellow}\hl{empire }\sethlcolor{yellow}\hl{in }\sethlcolor{yellow}\hl{the }\sethlcolor{yellow}\hl{west}\sethlcolor{yellow}\hl{?}&what enigmatic u. s. vice president was once tried and acquitted for treason in a plot to set up his own independent empire in the west?\\
\midrule
6 & \sethlcolor{yellow}\hl{how }\sethlcolor{yellow}\hl{many }\sethlcolor{yellow}\hl{colleges }\sethlcolor{yellow}\hl{are }\sethlcolor{yellow}\hl{in }\sethlcolor{yellow}\hl{wyoming}\sethlcolor{yellow}\hl{? }aids?&how many colleges are in wyoming?\\
\midrule
7 & what do bicornate animals have two of? prep school?&\textbf{MANY}\\
\midrule
8 & \sethlcolor{yellow}\hl{how }\sethlcolor{yellow}\hl{can }\sethlcolor{yellow}\hl{i }\sethlcolor{yellow}\hl{find }\sethlcolor{yellow}\hl{out }\sethlcolor{yellow}\hl{how }\sethlcolor{yellow}\hl{much }\sethlcolor{yellow}\hl{income }\sethlcolor{yellow}\hl{tax }\sethlcolor{yellow}\hl{is }\sethlcolor{yellow}\hl{paid }\sethlcolor{yellow}\hl{on }\sethlcolor{yellow}\hl{social }\sethlcolor{yellow}\hl{security }\sethlcolor{yellow}\hl{income }\sethlcolor{yellow}\hl{on }\sethlcolor{yellow}\hl{the }\sethlcolor{yellow}\hl{1998 }\sethlcolor{yellow}\hl{income }\sethlcolor{yellow}\hl{tax}\sethlcolor{yellow}\hl{? }and also granted 10 -?? of all profits of his voyage.&how can i find out how much income tax is paid on social security income on the 1998 income tax?\\
\midrule
9 & \sethlcolor{yellow}\hl{what }\sethlcolor{yellow}\hl{is }\sethlcolor{yellow}\hl{html}\sethlcolor{yellow}\hl{? }of the religion of islam? art?&what is html?\\
\midrule
10 & \sethlcolor{yellow}\hl{what }\sethlcolor{yellow}\hl{document }\sethlcolor{yellow}\hl{did }\sethlcolor{yellow}\hl{president }\sethlcolor{yellow}\hl{andrew }\sethlcolor{yellow}\hl{johnson }\sethlcolor{yellow}\hl{want }\sethlcolor{yellow}\hl{a }\sethlcolor{yellow}\hl{copy }\sethlcolor{yellow}\hl{of }\sethlcolor{yellow}\hl{placed }\sethlcolor{yellow}\hl{under }\sethlcolor{yellow}\hl{his }\sethlcolor{yellow}\hl{head }\sethlcolor{yellow}\hl{upon }\sethlcolor{yellow}\hl{his }\sethlcolor{yellow}\hl{burial}\sethlcolor{yellow}\hl{? }project&what document did president andrew johnson want a copy of placed under his head upon his burial?\\
\midrule
11 & in the song ` ` yankee doodle,'' why did yankee doodle stick a feather in his cap and call it macaroni?&\textbf{NONE}\\
\midrule
12 & \sethlcolor{yellow}\hl{what }\sethlcolor{yellow}\hl{is }\sethlcolor{yellow}\hl{a }\sethlcolor{yellow}\hl{nanometer}\sethlcolor{yellow}\hl{? }measure? ''?&what is a nanometer?\\
\midrule
13 & in company sport are these following numbers relevant : 118, 126, 134, 142, 15, 158, 167, 177, and 19?&\textbf{NONE}\\
\midrule
14 & \sethlcolor{yellow}\hl{when }\sethlcolor{yellow}\hl{reading }\sethlcolor{yellow}\hl{classified }\sethlcolor{yellow}\hl{ads, }\sethlcolor{yellow}\hl{what }\sethlcolor{yellow}\hl{does }\sethlcolor{yellow}\hl{eenty }\sethlcolor{yellow}\hl{: }\sethlcolor{yellow}\hl{other }\sethlcolor{yellow}\hl{stand }\sethlcolor{yellow}\hl{for}\sethlcolor{yellow}\hl{? }toast on a stick?&when reading classified ads, what does eenty : other stand for?\\
\midrule
15 & \sethlcolor{yellow}\hl{what }\sethlcolor{yellow}\hl{sea }\sethlcolor{yellow}\hl{did }\sethlcolor{yellow}\hl{the }\sethlcolor{yellow}\hl{romans }\sethlcolor{yellow}\hl{call }\sethlcolor{yellow}\hl{mare }\sethlcolor{yellow}\hl{nostrum}\sethlcolor{yellow}\hl{? }: ad arma, ad arma.&what sea did the romans call mare nostrum?\\
\midrule
16 & \sethlcolor{yellow}\hl{where }\sethlcolor{yellow}\hl{on }\sethlcolor{yellow}\hl{the }\sethlcolor{yellow}\hl{internet }\sethlcolor{yellow}\hl{can }\sethlcolor{yellow}\hl{i }\sethlcolor{yellow}\hl{find }\sethlcolor{yellow}\hl{a }\sethlcolor{yellow}\hl{song }\sethlcolor{yellow}\hl{lyrics }\sethlcolor{yellow}\hl{database }\sethlcolor{yellow}\hl{similar }\sethlcolor{yellow}\hl{to }\sethlcolor{yellow}\hl{the }\sethlcolor{yellow}\hl{international }\sethlcolor{yellow}\hl{lyrics }\sethlcolor{yellow}\hl{server}\sethlcolor{yellow}\hl{?}&where on the internet can i find a song lyrics database similar to the international lyrics server?\\
\midrule
17 & \sethlcolor{yellow}\hl{what }\sethlcolor{yellow}\hl{causes }\sethlcolor{yellow}\hl{asthma}\sethlcolor{yellow}\hl{? }maiden??&what causes asthma?\\
\midrule
18 & \sethlcolor{yellow}\hl{name }\sethlcolor{yellow}\hl{the }\sethlcolor{yellow}\hl{various }\sethlcolor{yellow}\hl{super }\sethlcolor{yellow}\hl{- }\sethlcolor{yellow}\hl{teams }\sethlcolor{yellow}\hl{to }\sethlcolor{yellow}\hl{which }\sethlcolor{yellow}\hl{the }\sethlcolor{yellow}\hl{angel }\sethlcolor{yellow}\hl{has }belonged., in 1948?&name the various super - teams to which the angel has belonged.\\
\midrule
19 & \sethlcolor{yellow}\hl{what }\sethlcolor{yellow}\hl{is }\sethlcolor{yellow}\hl{the }\sethlcolor{yellow}\hl{name }\sethlcolor{yellow}\hl{of }\sethlcolor{yellow}\hl{the }\sethlcolor{yellow}\hl{chronic }\sethlcolor{yellow}\hl{neurological }\sethlcolor{yellow}\hl{autoimmune }\sethlcolor{yellow}\hl{disease }\sethlcolor{yellow}\hl{which }\sethlcolor{yellow}\hl{attacks }\sethlcolor{yellow}\hl{the }\sethlcolor{yellow}\hl{protein }\sethlcolor{yellow}\hl{sheath }\sethlcolor{yellow}\hl{that }\sethlcolor{yellow}\hl{surrounds }\sethlcolor{yellow}\hl{nerve }\sethlcolor{yellow}\hl{cells }\sethlcolor{yellow}\hl{causing }\sethlcolor{yellow}\hl{a }\sethlcolor{yellow}\hl{gradual }\sethlcolor{yellow}\hl{loss }\sethlcolor{yellow}\hl{of }\sethlcolor{yellow}\hl{movement }\sethlcolor{yellow}\hl{in }\sethlcolor{yellow}\hl{the }\sethlcolor{yellow}\hl{body}\sethlcolor{yellow}\hl{? }and untamed they love?&what is the name of the chronic neurological autoimmune disease which attacks the protein sheath that surrounds nerve cells causing a gradual loss of movement in the body?\\
\midrule
20 & followed who who who??? followed who followed who followed followed followed caesar followed followed followed caesar who? who followed who? followed? who? who followed followed caesar followed followed followed followed followed followed followed who followed who followed who?&who followed caesar?\\
\midrule
21 & do \sethlcolor{yellow}\hl{what }\sethlcolor{yellow}\hl{county }\sethlcolor{yellow}\hl{is }\sethlcolor{yellow}\hl{chicago }\sethlcolor{yellow}\hl{in}\sethlcolor{yellow}\hl{? }how how? how how?? do? how how do how??? how how do?? how how how? do?? how do how do do?? how??? how how&what county is chicago in?\\
\cmidrule{3-3}
&&how do clouds form?\\
\midrule
22 & are ami what whatcable numbers? are what numbers what are are ami numbers what are are ami numbers? whatcable what arecablecable what ami whatcable? ami? cablecablecable? are? what? what numbers numbers whatcable&what are amicable numbers?\\
\midrule
23 & \sethlcolor{olive}\hl{what }\sethlcolor{olive}\hl{is }\sethlcolor{olive}\hl{the }\sethlcolor{olive}\hl{history }\sethlcolor{olive}\hl{of }\sethlcolor{olive}\hl{yo }\sethlcolor{olive}\hl{- }\sethlcolor{olive}\hl{yos}\sethlcolor{olive}\hl{? }\sethlcolor{orange}\hl{are }\sethlcolor{orange}\hl{they}\sethlcolor{orange}\hl{?}&how many url extensions are there? and what are they?\\
\cmidrule{3-3}
&&what is the history of yo - yos?\\
\midrule
24 & &\textbf{NONE}\\
\midrule
25 & \sethlcolor{yellow}\hl{who }\sethlcolor{yellow}\hl{was }\sethlcolor{yellow}\hl{whitcomb }\sethlcolor{yellow}\hl{judson}\sethlcolor{yellow}\hl{?}&who was whitcomb judson?\\
\midrule
26 & \sethlcolor{yellow}\hl{how }\sethlcolor{yellow}\hl{do }\sethlcolor{yellow}\hl{i }\sethlcolor{yellow}\hl{tie }\sethlcolor{yellow}\hl{a }\sethlcolor{yellow}\hl{tie}\sethlcolor{yellow}\hl{?}&how do i tie a tie?\\
\midrule
27 & do \sethlcolor{yellow}\hl{how }\sethlcolor{yellow}\hl{many }\sethlcolor{yellow}\hl{points }\sethlcolor{yellow}\hl{make }\sethlcolor{yellow}\hl{up }\sethlcolor{yellow}\hl{a }\sethlcolor{yellow}\hl{perfect }\sethlcolor{yellow}\hl{fivepin }\sethlcolor{yellow}\hl{bowling }\sethlcolor{yellow}\hl{score}\sethlcolor{yellow}\hl{? }do do? do do do work work how do do do do? doija how do do? work do work do do work do do how? ou do&how many points make up a perfect fivepin bowling score?\\
\midrule
28 & \sethlcolor{yellow}\hl{who }\sethlcolor{yellow}\hl{graced }\sethlcolor{yellow}\hl{the }\sethlcolor{yellow}\hl{airwaves }\sethlcolor{yellow}\hl{with }\sethlcolor{yellow}\hl{such }\sethlcolor{yellow}\hl{pearls }\sethlcolor{yellow}\hl{as }\sethlcolor{yellow}\hl{` }\sethlcolor{yellow}\hl{` }\sethlcolor{yellow}\hl{do }\sethlcolor{yellow}\hl{ya }\sethlcolor{yellow}\hl{lo }\sethlcolor{yellow}\hl{- }\sethlcolor{yellow}\hl{o }\sethlcolor{yellow}\hl{- }\sethlcolor{yellow}\hl{ove }\sethlcolor{yellow}\hl{me}\sethlcolor{yellow}\hl{? }\sethlcolor{yellow}\hl{get }\sethlcolor{yellow}\hl{naked, }\sethlcolor{yellow}\hl{baby }becoming''?? , shouts, etc.?&who graced the airwaves with such pearls as ` ` do ya lo - o - ove me? get naked, baby!''?\\
\midrule
29 & what is the name of can president of fly? t u. s. a? s supposed to?&\textbf{MANY}\\
\midrule
30 & \sethlcolor{yellow}\hl{how }\sethlcolor{yellow}\hl{do }\sethlcolor{yellow}\hl{you }\sethlcolor{yellow}\hl{ask }\sethlcolor{yellow}\hl{a }\sethlcolor{yellow}\hl{total }\sethlcolor{yellow}\hl{stranger }\sethlcolor{yellow}\hl{out }\sethlcolor{yellow}\hl{on }\sethlcolor{yellow}\hl{a }\sethlcolor{yellow}\hl{date}\sethlcolor{yellow}\hl{?}&how do you ask a total stranger out on a date?\\
\midrule
31 & why is dudley do - right's horse's name?&why do some people have two different color eyes?\\
\midrule
32 & \sethlcolor{olive}\hl{what }\sethlcolor{olive}\hl{are }\sethlcolor{olive}\hl{the }\sethlcolor{olive}\hl{titles }\sethlcolor{olive}\hl{of }\sethlcolor{olive}\hl{some }\sethlcolor{olive}\hl{r }\sethlcolor{olive}\hl{- }\sethlcolor{olive}\hl{rated }\sethlcolor{olive}\hl{sony }\sethlcolor{olive}\hl{playstation }\sethlcolor{olive}\hl{games}? \sethlcolor{orange}\hl{to }\sethlcolor{orange}\hl{them, }\sethlcolor{orange}\hl{like}? \sethlcolor{orange}\hl{they }\sethlcolor{orange}\hl{hear }\sethlcolor{orange}\hl{a }\sethlcolor{orange}\hl{beautiful }\sethlcolor{orange}\hl{piece }\sethlcolor{orange}\hl{of }\sethlcolor{orange}\hl{music, }\sethlcolor{orange}\hl{or }\sethlcolor{orange}\hl{see }\sethlcolor{orange}\hl{something }\sethlcolor{orange}\hl{beautiful, }\sethlcolor{orange}\hl{or }\sethlcolor{orange}\hl{get }\sethlcolor{orange}\hl{aroused }\sethlcolor{orange}\hl{by }\sethlcolor{orange}\hl{someone }\sethlcolor{orange}\hl{they }\sethlcolor{orange}\hl{love}&why do people get goosebumps when they have something emotional happen to them, like when they hear a beautiful piece of music, or see something beautiful, or get aroused by someone they love?\\
\cmidrule{3-3}
&&what are the titles of some r - rated sony playstation games\\
\bottomrule
\end{tabular}
\vspace{-5em}
\end{table}

\end{document}